\documentclass[11pt]{article}
\usepackage{amsmath}
\usepackage{pstricks}
\usepackage{amssymb}
\usepackage{mathabx}
\usepackage[vcentermath]{youngtab}
\usepackage{graphicx}
\usepackage{youngtab}
\usepackage{bbm}
\usepackage{multirow}

\def\half{\frac{1}{2}}
\def\bsh{\backslash}

\newfont{\bbbold}{msbm10 scaled \magstep1}

\def\bbR{\mbox{\bbbold R}}

\def\cA{{\cal A}}
\def\cB{{\cal B}}
\def\cC{{\cal C}}
\def\cD{{\cal D}}
\def\cE{{\cal E}}
\def\cF{{\cal F}}

\def\cH{{\cal H}}
\def\cI{{\cal I}}

\def\cL{{\cal L}}
\def\cM{{\cal M}}
\def\cN{{\cal N}}
\def\cO{{\cal O}}
\def\cP{{\cal P}}
\def\cQ{{\cal Q}}
\def\cR{{\cal R}}
\def\cS{{\cal S}}

\def\cU{{\cal U}}
\def\cV{{\cal V}}

\newfont{\goth}{eufm10 scaled \magstep1}

\def\ge{\mbox{\goth e}}
\def\gf{\mbox{\goth f}}
\def\gg{\mbox{\goth g}}
\def\gh{\mbox{\goth h}}

\def\gl{\mbox{\goth l}}

\def\go{\mbox{\goth o}}

\def\gs{\mbox{\goth s}}

\def\a{\alpha}
\def\b{\beta}
\def\c{\gamma}\def\C{\Gamma}
\def\d{\delta}
\def\ve{\varepsilon}
\def\f{\phi}\def\F{\Phi}

\def\l{\lambda}\def\L{\Lambda}

\def\r{\rho}
\def\S{\Sigma}

\def\th{\theta}\def\Th{\Theta}

\def\be{\begin{equation}}\def\ee{\end{equation}}
\def\bea{\begin{eqnarray}}\def\eea{\end{eqnarray}}
\def\barr{\begin{array}}\def\earr{\end{array}}

\def\o{\omega}\def\O{\Omega}

\def\una{\underline a}\def\unA{\underline A}

\def\unM{\underline M}

\def\xz{\times}

\def\nab{\nabla}

\let\la=\label

\def\nn{\nonumber}
\def\bd{\begin{document}}
\def\ed{\end{document}}
\def\ba{\begin{array}}
\def\ea{\end{array}}
\def\bea{\begin{eqnarray}}
\def\eea{\end{eqnarray}}
\def\ft#1#2{\tfrac{#1}{#2}}
\def\fft#1#2{\frac{#1}{#2}}
\def\sst#1{{\scriptscriptstyle #1}}
\def\oneone{\rlap 1\mkern4mu{\rm l}}

\newcommand{\eq}[1]{(\ref{#1})}
\newcommand{\w}[1]{\\[0.#1cm]}
\def\eqs#1#2{(\ref{#1}-\ref{#2})}
\def\det{{\rm det\,}}
\def\tr{{\rm tr}}
\def\ad{{\rm ad}}

\def\f{\mathfrak{f}}
\def\g{\mathfrak{g}}
\def\so{\mathfrak{so}}
\def\su{\mathfrak{su}}
\def\sp{\mathfrak{sp}}
\def\sl{\mathfrak{sl}}
\def\sq{\mathfrak{sq}}
\def\qed{\hspace{\stretch{1}} $\Box$ \\
\noindent}

\newcommand{\frakk}{\mathfrak{k}}

\newcommand{\al}{\alpha}
\newcommand{\ep}{\varepsilon}
\newcommand{\de}{\delta}
\newcommand{\Ga}{\Gamma}
\newcommand{\Gh}{\hat{\Gamma}}
\newcommand{\Gt}{\tilde{\Gamma}}
\newcommand{\tsigma}{\tilde{\sigma}}
\newcommand{\db}{{\dot{\beta}}}

\newcommand{\dlb}{\ensuremath{[\![}}
\newcommand{\drb}{\ensuremath{]\!]}}

\newtheorem{lemma}[theorem]{Lemma}
\newtheorem{prop}[theorem]{Proposition}
\newtheorem{cor}[theorem]{Corollary}
\def\Pf{\noindent \textbf{Proof. }}

\newcommand{\hoch}[1]{$\, ^{#1}$}
\newcommand{\imperial}{\it\small Theoretical Physics Group, Imperial College London\\ Prince Consort Road, London SW7 2AZ, UK}
\newcommand{\kings}
{\it\small Department of Mathematics, King's College, University of London\\ Strand, London WC2R 2LS, UK}
\newcommand{\uu}
{\it\small Department of Theoretical Physics, Uppsala, Sweden}
\newcommand{\hip}
{\it\small HIP-Helsinki Institute of Physics, P.O. Box 64 FIN-00014
University of Helsinki, Suomi-Finland}
\newcommand{\stock}
{\it\small Department of Theoretical Physics, Stockholm, Sweden}
\newcommand{\golm}
{\it\small AEI, Max Planck Institut f\"ur Gravitationsphysik\\ Am M\"{u}hlenberg 1, D-14476 Potsdam, Germany}
\makeatletter
\renewcommand\theequation{\thesection.\arabic{equation}}
\@addtoreset{equation}{section} \makeatother

\newcommand{\sa}{/ \hspace{-1.2ex}}
\newcommand{\saa}{/ \hspace{-1.4ex}}
\newcommand{\saaa}{\, / \hspace{-1.6ex}}
\newcommand{\Scal}[1]{\Bigl ({#1} \Bigr )}
\newcommand{\scal}[1]{\bigl ({#1} \bigr )}

\newcommand{\CR}{\nonumber \\*}

\newcommand{\trace}{\hbox {tr}~}
\newcommand{\traceS}{\hbox {tr}_{\scriptscriptstyle \mathfrak{S}}~}

\DeclareMathAlphabet{\mathpzc}{OT1}{pzc}{m}{it}
\def\BRST{\,\mathpzc{s}\,}
\def\aBRST{{\scriptstyle (\mathpzc{s})}}
\def\q{{{\scriptscriptstyle (Q)}}}
\def\qs{{\scriptscriptstyle (Q\mathpzc{s})}}
\def\Qsla{{\mathcal{S}_{\q}}}
\def\Slav{{\mathcal{S}_\aBRST}}
\def\epsilonb{{\overline{\epsilon}}}
\def\bulletup{{\scriptstyle \bullet}}

\newcommand{\gra}[2]{{\scriptscriptstyle (#1 , #2 )}}
\newcommand{\ord}[1]{{\scriptscriptstyle (#1)}}

\def\cL{{\cal L}}
\def\cN{\mathcal{N}}
\def\cO{\mathcal{O}}

\def\ie{{\it i.e.}\ }
\def\eg{{\it e.g.}\ }

\newcommand{\sfrac}[2]{{\scriptstyle \frac{#1}{#2}}}
\newcommand{\stfrac}[2]{{\scriptscriptstyle \frac{#1}{#2}}}

 \def\balpha{{\overline{\alpha}}}
 \def\bbeta{{\overline{\beta}}}
 \def\bgamma{{\overline{\gamma}}}
 \def\bdelta{{\overline{\delta}}}
 \def\bepsilon{{\overline{\epsilon}}}
 \def\bvarepsilon{{\overline{\varepsilon}}}
 \def\bzeta{{\overline{\zeta}}}
 \def\bareta{{\overline{\eta}}}
 \def\btheta{{\overline{\theta}}}
 \def\bvartheta{{\overline{\vartheta}}}
 \def\biota{{\overline{\iota}}}
 \def\bkappa{{\overline{\kappa}}}
 \def\blambda{{\overline{\lambda}}}
 \def\bmu{{\overline{\mu}}}
 \def\bnu{{\overline{\nu}}}
 \def\bxi{{\overline{\xi}}}
 \def\bpi{{\overline{\pi}}}
 \def\brho{{\overline{\rho}}}
 \def\bvarrho{{\overline{\varrho}}}
 \def\bsigma{{\overline{\sigma}}}
 \def\bvarsigma{{\overline{\varsigma}}}
 \def\btau{{\overline{\tau}}}
 \def\bphi{{\overline{\phi}}}
 \def\bvarphi{{\overline{\varphi}}}
 \def\bchi{{\overline{\chi}}}
 \def\bpsi{{\overline{\psi}}}
 \def\bomega{{\overline{\omega}}}

\def\thalf{{\textrm{\tiny\textonehalf}}}
\def\tquarter{{\textrm{\tiny\textonequarter}}}
\def\Ko{{\scriptscriptstyle K}}
\def\tKo{\scriptscriptstyle k }
\def\corr{$\clubsuit$}

\newcommand{\auth}{\large 
Paul Howe${}^{a}$\footnote{email: paul.howe@kcl.ac.uk} and Jakob Palmkvist${}^b$\footnote{email: jakobpalmkvist@tamu.edu}}

\begin{document}

\renewcommand{\thefootnote}{\fnsymbol{footnote}}

\null
\begin{flushright}
{\small KCL-MTH-15-01}\\
{\small MI-TH-1505}
\vskip 1.5 cm
\end{flushright}

\begin{center}
{\Large{\bf Forms and algebras in (half-)maximal supergravity theories}}
\vspace{.75cm}

\auth
\end{center}
\vspace{.5cm}

\centerline{${}^a${\it \small Department of Mathematics,
King's College London}}
\centerline{{\it \small The Strand, London WC2R 2LS, UK}}
\vspace{.5cm}
\centerline{${}^b${\it \small Mitchell Institute for Fundamental Physics and Astronomy, Texas A\&M University}}
\centerline{{\it \small College Station, TX 77843, USA }}

\vspace{1cm}

\centerline{{\bf Abstract}}

\noindent
The forms in $D$-dimensional (half-)maximal supergravity theories are discussed for
$3 \leq D \leq 11$.
Superspace methods are used to derive consistent sets of Bianchi identities for all the forms for all degrees, and to show that they are soluble and fully compatible with supersymmetry. The Bianchi identities determine Lie superalgebras that can be extended 
to Borcherds superalgebras of a special type. It is shown that
any Borcherds superalgebra of this type gives the same form spectrum, up to an arbitrary degree, as an
associated Kac-Moody algebra. For maximal supergravity up to $D$-form potentials, this is the very extended Kac-Moody algebra $E_{11}$. 
It is also shown how gauging can be carried out in a simple fashion by deforming the Bianchi identities by means of a new algebraic element related to the embedding tensor. In this case the appropriate extension of the form algebra is a truncated version of the
so-called tensor hierarchy algebra.

\vskip .5cm

\vspace{1cm}

\renewcommand{\thefootnote}{\arabic{footnote}}
\setcounter{footnote}{0}

\pagebreak
\tableofcontents


\section{Introduction}

It has been known for many years that the forms in $D$-dimensional  maximal supergravity theories, when the duals of the physical forms are included, are associated with algebraic structures \cite{Cremmer:1997ct,Cremmer:1998px}. These structures have been interpreted as subalgebras of Borcherds superalgebras \cite{HenryLabordere:2002dk,HenryLabordere:2002xh} and in terms of extended $E$-series algebras
\cite{Julia:1997cy,West:2001as,Damour:2002cu, Riccioni:2007au,Bergshoeff:2007qi,Bergshoeff:2007vb,Riccioni:2007ni,Bergshoeff:2008xv,Riccioni:2009xr}. It has been found that there are also $(D-1)$-form potentials  (also called de-forms, since they are associated with deformations)
and $D$-forms (otherwise known as top forms), both
carrying no physical degrees of freedom, whose existence is implied by these algebraic structures (these were first written down explicitly in $D=10$ \cite{Bergshoeff:2005ac,Bergshoeff:2006qw}). In general, the forms transform under representations $\cR_{\ell}$ of the duality group of the given supergravity theory where the level number $\ell$ coincides with the form-degree of the potentials. In a separate, but related, development, studies of the general structure of gauged supergravity theories \cite{deWit:1981eq,Gunaydin:1984qu,Pernici:1984xx,Hull:1984vg,Hull:1984qz,deWit:1983gs,
Nicolai:2000sc,Nicolai:2001sv,deWit:2002vt,deWit:2003hr}
have revealed that the same sets of forms are needed in that context
(with two exceptions for $D=3$)  and that the gauge transformations of the potentials at level
$\ell$ involve parameters  up to level $(\ell+1)$, 
the whole set of forms giving rise to a tensor hierarchy \cite{deWit:2005hv,deWit:2008ta,deWit:2008gc}. A key feature of this general construction is the use of the embedding tensor that specifies how the gauge group $G_0$ is embedded in the duality group $G$, thereby facilitating a unified description of arbitrary gaugings in any given spacetime dimension. The half-maximal matter-coupled supergravity theories, their forms, algebras and gaugings have also been discussed in the literature \cite{Nicolai:2001ac,deWit:2003ja,Weidner:2006rp,Bergshoeff:2007vb}.

In this article an extended discussion of the above topics is given for all maximal supergravity and half-maximal matter-coupled supergravity theories in dimensions $D\geq 3$. We rederive all of the forms in an elementary way and show that they are consistent with supersymmetry at the full, non-linear level. This is done using a superspace formulation, thus extending the results of \cite{Greitz:2011da,Greitz:2012vp,Greitz:2011vh} to all cases. One advantage of the formalism is that superforms can have arbitrarily high degrees; in particular, this means that the algebraic structures associated with them can be studied covariantly via the Bianchi identities for the field strengths even for potential forms with degree $D$. Indeed, one can even consider over-the-top (OTT) forms that have degrees higher than $D$. In the context of on-shell supergravity theories without any higher-order corrections some $(D+2)$-form field strengths are non-zero; these are necessary for the completion of the gauge hierarchy in a covariant formalism, and for this reason we also classify these. A second advantage of superspace is that one can use cohomological techniques to show that the Bianchi identities for all of the forms can all be solved almost trivially. This gives a very simple way of verifying that the forms are indeed consistent with supersymmetry. 

As noted in the original papers \cite{Cremmer:1997ct,Cremmer:1998px} the Bianchi identities for the forms give rise directly to Lie (super)algebras.  In the superspace context these are naturally infinite-dimensional, except for $D=11$, and in many examples are determined by the level-one forms, together with a level-two consistency condition that can be interpreted as the requirement that the level-two Bianchi identity be soluble, a question that can be settled by cohomology. This puts a constraint on the possible level-two representations in the symmetric tensor product of $\cR_1$ with itself. If we identify level zero (\ie the scalars) with the Lie algebra $\gg$ of the duality group, then we can obtain the full superalgebra of forms by appending an additional odd generator, $e_0$ say,  that generates $\cR_1$ under the action of the raising operators of $\gg$ and which anti-commutes with itself in order that the supersymmetry constraint be satisfied. We can then extend this algebra symmetrically about level zero to obtain a Borcherds superalgebra $\cB$,  where $e_0$ and the corresponding generator $f_0$ at level $-1$ are associated to an odd null root, added to the simple roots of $\gg$.

In maximal supergravity this construction works well for $3\leq D\leq 8$ and we arrive in this way at the Borcherds superalgebras proposed in \cite{HenryLabordere:2002dk,HenryLabordere:2002xh}.
However, ambiguities arise for higher dimensions. This brings us to the second main theme of the paper, namely a re-investigation of the algebraic structures and their interrelationships. In reference \cite{Henneaux:2010ys} it was shown how one could derive Borcherds superalgebras for maximal supergravity theories starting from $E_{11}$, while in \cite{Palmkvist:2011vz}, it was shown how to go in  the other direction. More recently, it was argued in \cite{Kleinschmidt:2013em} that the Borcherds superalgebras given in \cite{HenryLabordere:2002dk,Henneaux:2010ys} for $D\geq8$ do not agree with those obtained by oxidation from lower dimensions. As we shall discuss, it is also the case that the Lie superalgebras determined by the forms do not imply unique Borcherds superalgebras for these cases. We address this problem in a unified construction that includes the form algebras, their Borcherds extensions and extended Kac-Moody algebras. The key observation is that,
in each case, the unique Borcherds superalgebra $\cB$
determined by the Bianchi identities (and, if needed for uniqueness, the oxidation procedure) is
of a special type. Namely, it can be obtained from an associated Kac-Moody algebra $\cA$ by assigning a non-negative
integer, called the V-degree \cite{Henneaux:2010ys,Kleinschmidt:2013em}, to each simple root.
As we shall explain, the V-degrees prescribe completely both how to construct $\cB$ from $\cA$,
and how to extend $\cA$ to another Kac-Moody algebra $\cC$, such that $\cB$ and $\cC$ give identical form spectra, up to an arbitrary level.
If this level is equal to the spacetime dimension of a maximal supergravity theory, then $\cC$ is the ``very extended'' Kac-Moody algebra
$E_{11}$. Our construction is thus a further development of the work of \cite{Henneaux:2010ys,Palmkvist:2011vz}
relating $\cB$ and $E_{11}$ to each other.
It also generalises the result of \cite{Palmkvist:2012nc}, which applies to the generic case described above, where one simple root
has V-degree one (the odd null root of $\cB$), and the others have V-degree zero (the simple roots of $\g$). Our result is much more general since
we can start with not only an arbitrary Kac-Moody algebra $\cA$, but also an arbitrary assignment of V-degrees to its simple roots.
In particular, it is valid for the non-generic cases of
maximal and half-maximal supergravity that appear for $D\geq9$ and $D\geq5$, respectively.\footnote{
Another example is chiral supergravity in $D=6$ coupled to two vector multiplets and two tensor multiplets, recently studied
in \cite{Henneaux:2015gya}.}

The form algebras lead to Borcherds superalgebras by extending them to negative levels in a symmetrical fashion, in the sense that level minus one is the dual of level one, but there is another extension that does not have this property known as the tensor hierarchy algebra (THA)
\cite{Palmkvist:2013vya}. This algebra is instead symmetrical in the sense of spacetime duality, for example level minus one is dual to level $(D-1)$, and has applications to gauged supergravity theories. We shall show that deforming the Bianchi identities by a dimension-one, level-zero field-strength (the embedding tensor), invariant under the gauge group (a subgroup of the duality group), has a natural interpretation in terms of a truncation of the THA. We discuss the consistent deformed Bianchi identities and show how the field strengths  can be solved for in terms of potentials by using the THA \cite{Greitz:2013pua}. We also show that the gauged Bianchi identities can be solved to all orders and thus prove consistency with supersymmetry. 

The organisation of the paper is as follows: in the next two sections we discuss maximal and half-maximal supergravity theories, including all the forms, in a superspace setting. In the maximal case we go directly to the on-shell theories but in the half-maximal case we start from off-shell formulations of supergravity (partially off-shell in the case of $D=10$) and then go on-shell by introducing the forms and scalars. This has the advantage of simplifying the discussion a little. In these sections we also discuss superspace cohomology and demonstrate how it can be used to show that all Bianchi identities are satisfied. Sections 4 and 5 concern the algebras that one obtains from the forms and their interpretation in terms of the Borcherds-Kac-Moody picture sketched above, while in section 6 we discuss gauged supergravity theories and the role of the THA. 
In the first two appendices we list, up to level $\ell=(D+1)$,
all the representations $\cR_\ell$ for maximal supergravity, and all the Bianchi identities for half-maximal supergravity.
There are also appendices on 
Borcherds superalgebras (and the more general contragredient Lie superalgebras) and on extended superspaces for some maximal theories. These superspaces include additional even coordinates corresponding to the level-one representation $\cR_1$. Since all of the forms are generated from this in many cases, formulating the theory in such an extended superspace contains all of the forms implicitly.


\section{Forms, consistent Bianchi identities and cohomology in maximal supergravity}


In this section we describe a simple approach to the extended algebraic structures that arise in maximal supergravity theories based on supersymmetry. In order to make supersymmetry manifest we shall work in superspace. A significant advantage of this approach is that forms in superspace can have arbitrary degrees, because the odd basis forms commute, and this implies that one can work with field-strength forms rather than potentials, even for the top forms for which the field strengths are identically zero in spacetime. Moreover, one can in principal have potential forms that have degrees greater than the dimension of spacetime. We shall refer to these as over-the-top (OTT) forms. Even in supergravity there are examples of OTT potentials with degree $(D+1)$ whose $(D+2)$-form field strengths are non-zero in superspace \cite{Greitz:2012vp,Greitz:2011vh}, but we might expect many more of them to be non-zero when higher-order corrections are taken into account. Such forms fit in naturally with Borcherds superalgebras which are typically infinite-dimensional in the supergravity context.

\subsection{Maximal supergravity}

We begin with maximal supergravity considering only the on-shell Poincar\'e theories. The supergeometries for these theories are straightforward to construct given their field content. We briefly review this to remind the reader how supergravity is presented in superspace. (See, for example, \cite{Howe:1981gz} for $D=4$, $N=8$ supergravity and \cite{Cremmer:1980ru,Brink:1980az} for $D=11$.)
The basic geometrical structure is a choice of odd tangent bundle $T_1$ such that the even tangent bundle $T_0$, considered as a quotient of the tangent bundle $T$, \ie $T_0=T/T_1$, is generated from the odd one by taking Lie brackets of odd vector fields.
This is essentially supersymmetry, for which the translation generator 
is the Lie bracket of two supercharges. In addition, we shall assume that the dimension-zero torsion is the same as it is in flat superspace,
\be
T_{\a\b}{}^c=-i (\C^c)_{\a\b}\ ,
\la{2.1}
\ee
where the indices $\a,\b$ run from 1 to 32 and combine spinor and internal symmetry indices according to the dimension of spacetime, and where small latin indices are vector indices. Both sets of indices are with respect to preferred frame bases for $T_1$ and $T_0^*$. The gamma-matrix in \eq{2.1} is in general a product of an appropriate spacetime gamma-matrix and an invariant tensor for the internal R-symmetry group, so that the diagonal components of the Lie algebra of the structure group should be Lorentzian in the even-even sector and a direct sum of the corresponding spin and R-symmetry Lie algebras in the odd-odd sector. In addition we can choose the dimension one-half torsion $T_{a \b}{}^c$ to vanish. This specifies the dimension one-half Lorentz connection and fixes the splitting of the tangent bundle into even and odd, so that there is no longer any mixed component in the structure algebra. These two conventional constraints do not fix $T_{a \b}{}^c$ to be zero immediately, but the remaining irreducible representation of the spin group that it contains can be shown to vanish by means of the dimension one-half Bianchi identity. Given this, the connection one-form $\O_A{}^B$ and the curvature two-form $R_A{}^B=d\O_A{}^B + \O_A{}^C \O_C{}^B$, both of which take their values in the structure algebra, do not have mixed even-odd components. Let $E^A=(E^a,E^\a)$ denote a preferred basis of one-forms with dual basis $E_A$. The torsion two-form is $T^A=DE^A:=dE^A + E^B\O_B{}^A$. The Bianchi identities are
\be
DT^A=E^B R_B{}^A\ , \qquad DR_A{}^B=0\ .
\la{2.2a}
\ee
Dragon's theorem \cite{Dragon:1978nf}, valid in $D>3$, states that the second set of Bianchi identities is satisfied if the first one is and that the components of the curvature can be expressed in terms of those of the torsion and derivatives thereof.\footnote{In $D=3$ there is a set of scalars in the dimension-one curvature transforming as an R-symmetry four-form that does not appear in the torsion; this has to be specified in terms of the physical fields on-shell \cite{Greitz:2012vp,Greitz:2011vh}.} In order to describe on-shell supergravity we therefore only need to specify the components of the torsion tensor. This can be done straightforwardly given the spectrum of the theory under consideration and dimensional analysis. Since the torsion is gauge-covariant it follows that the non-zero components can only be functions of the field strengths. (Any component field strength can always be considered as the leading component of a superfield in a $\th$-expansion.)

A simple example of this is $D=11$ supergravity.\footnote{In this case the constraint \eq{2.1} is enough to put the theory on-shell \cite{Howe:1997he}.} The fields are the graviton, the gravitino and a three-form gauge field. The corresponding field strengths are the curvature (dimension two) the gravitino field-strength (dimension three-halves) and the four-form field-strength which has dimension one. The superfield starting with this field will be denoted $F_{abcd}$. Since there are no fields at dimension one-half it follows that all torsion components are zero or constant until we get to dimension one. Moreover, the dimension-one torsion $T_{ab}{}^c$ can be set to zero as a conventional constraint corresponding to the dimension-one component of the Lorentz connection. Thus we are left with $T_{a\b}{}^\c$, which must be linear in $F_{abcd}$ and, at dimension three-halves  $T_{ab}{}^\c$, whose leading component can be identified with the gravitino field-strength. (This can be seen from its definition if we make the identification $E_m{}^\a=\psi_m{}^\a$ at $\th=0$.) There are two possible structures in $T_{a\b}{}^\c$ that can contain $F_{abcd}$ and their relative coefficient can be determined by the Bianchi identities. We find
\be
T_{a\b}{}^\c=-\frac{1}{36}(\C_a{}^{bcd})_\b{}^\c F_{abcd} + \frac{1}{8}(\C_a{}^{bcde})_\b{}^\c F_{bcde}\ .
\la{2.3a}
\ee
The remaining Bianchi identities can be used to determine the gravitino field strength and the dimension-two curvature in terms of odd derivatives of $F_{abcd}$, and since the theory is on-shell, one also finds the equations of motion for all of the component fields and the Bianchi identity for $F_{abcd}$. This is a covariant expression, rather than $dF=0$, but we can obtain the latter by constructing a superspace four-form which obeys  this equation. By dimensional analysis the only non-zero components can be at dimension zero and one; the latter is just $F_{abcd}$ while the former is
\be
F_{ab\c\d}=-i(\C_{ab})_{\c\d}\ .
\la{2.4b}
\ee
In addition we can introduce a dual six-form potential with its seven-form field-strength obeying the  Bianchi identity $dF_7=\half F_4^2$. Its non-vanishing components are \cite{Candiello:1993di}
\be
F_{abcde\a\b}=-i(\C_{abcde})_{\a\b} \qquad{\rm\ and}\qquad F_{abcdefg}=\frac{1}{4!} \ve_{abcdefgijkl} F^{ijkl}\ .
\la{2.5a}
\ee
\subsection{Forms}
Similar analyses can be applied to lower-dimensional maximal supergravity theories. They are slightly more complicated due to the presence of dimension-one-half fermions and scalars, although the latter cause few difficulties thanks to duality symmetries. The scalars take their values in the coset space $H\bsh G$ where $G$ is the rigid duality group and $H$ is the local R-symmetry group. Thus the scalars do not appear naked in the torsion and curvature (provided we include $H$ in the structure group). 

As well as  the torsion and curvature tensors, which are forms that also carry superspace indices, there are additional field-strength forms, such as the four- and seven-forms in $D=11$, that do not.
These forms will transform under representations of the duality group when one is present. The components of the field-strength forms can be constructed straightforwardly using dimensional analysis and the Bianchi identities. A potential form $A_{\ell}$ of degree $\ell$ has a field-strength form $F_{\ell+1}$ of degree
$\ell+1$ and there will be a corresponding Bianchi identity $I_{\ell+2}$ of degree $\ell+2$. 
As well as the physical forms and their duals there are additional forms that can be generated from them. The way to do this is to look for all potential forms $A_\ell$ of degree $\ell$ such that the corresponding field-strength forms $F_{\ell +1}$ satisfy consistent Bianchi identities (CBIs) $I_{\ell +2}$ of the form

\be
I_{\ell +2}:=d F_{\ell+1}-\sum_{m+n=\ell} F_{m+1} F_{n+1}\ ,
\la{2.2b}
\ee
where consistency means that the set of Bianchi identities forms a differential ideal,
\be
dI= 0\qquad  {\rm mod}\  I\ .
\la{2.3b}
\ee
Of course, by solving the Bianchi identities we mean setting the $I$s to zero, but this is considerably simplified if one makes use of the consistency conditions together with superspace cohomology, an idea first put forward some years ago \cite{Sohnius:1980iw}.

Clearly the higher-degree forms will transform under representations of the duality group if the physical ones do. We must also require that each CBI should be soluble. It turns out that there is a subset of the physical forms,  which we shall call the generating set, from which all of the forms can be constructed systematically using the above process, and that the full set of CBIs is guaranteed to be satisfied if those for the generating set are. This can be done rather easily using superspace cohomology which we shall now briefly describe in order for the discussion to be self-contained.

\subsection{Cohomology}

Since the tangent bundle splits into even and odd parts the space of $n$-forms splits into spaces of $(p,q)$-forms, $p+q=n$, where a $(p,q)$ form has $p$ even and $q$ odd indices:
\be
\O^{p,q}\ni \o_{p,q}=\frac{1}{p! q!} E^{\b_q}\cdots E^{\b_1} E^{a_p}\cdots E^{a_1} \o_{a_1\cdots a_p\b_1\cdots \b_q}\ .
\la{2.4a}
\ee
The exterior derivative splits into four terms with different bi-degrees:
\be
d=d_0+ d_1 + t_0 + t_1\ ,
\la{2.5b}
\ee
where the bidegrees are $(1,0),(0,1), (-1,2)$ and $(2,-1)$ respectively. The first two, $d_0$ and $d_1$, are essentially even and odd differential operators, while the other two are algebraic operators formed with the dimension-zero and dimension three-halves torsion respectively. In particular,
\be
(t_0 \o_{p,q})_{a_2\cdots a_p \b_1\cdots \b_{q+2}}\propto\  T_{(\b_1\b_2}{}^{a_1}\o_{a_1a_2\cdots a_p\b_3\cdots \b_{q+2})}\ .
\la{2.6}
\ee

The equation $d^2=0$ splits into various parts according to their bi-degrees amongst which one has:
\begin{align}
(t_0)^2&= 0\la{c4}\ ,\w1
t_0 d_1 + d_1 t_0&=0\la{c5}\ ,\w1
d_1^2 +t_0 d_0+ d_0 t_0&=0\ .
\la{2.7}
\end{align}

The first of these enables us to define the cohomology groups $H_t^{p,q}$, the space of $t_0$-closed $(p,q)$-forms modulo the exact ones \cite{Bonora:1986ix}. The other two then allow one to define the spinorial cohomology groups $H_s^{p,q}$. To do this we first introduce the spinorial derivative $d_s$ that maps $H_t^{p,q}$ to $H_t^{p,q+1}$ by $d_s[\o_{p,q}]:=[d_1 \o_{p,q}]$  where the brackets denote cohomology classes in $H_t$. It is easy to check, using \eq{2.7}, that $d_s$ is well-defined, \ie is independent of the choice of representative $\o$, and that it is nilpotent, $d_s^2=0$, so that cohomology groups can be defined \cite{Cederwall:2001dx,Howe:2003cy}. The cohomology groups $H_t^{p,q}$ and $H_s^{p,q}$ are related to spaces of pure spinors (suitably defined) and pure spinor cohomology respectively  \cite{Howe:1991mf,Howe:1991bx,Berkovits:2002zk,Berkovits:2008qw}. We emphasise that  the cohomologies we are interested in here are algebraic, rather than topological, in that the coefficients of the cohomology groups will be covariant fields of the supergravity theory under consideration. 

When the dimension-zero torsion is flat, $H_t$ cohomology is isomorphic to the cohomology of the supersymmetry algebra if $(E^a,E^\a)$ are respectively identified with the translational and supersymmetry ghosts and the BRST operator $Q$ with $t_0$. Supersymmetry cohomology has been discussed in various dimensions in 
\cite{Brandt:2009xv,Brandt:2010fa,Brandt:2010tz,Brandt:2013xpa}
and in the context of $H_t$ in  \cite{Movshev:2010mf,Movshev:2011pr}.

In maximal supergravity the $t_0$-cohomology groups turn out to be trivial for 
$p\geq 1$ when $D\leq 9$, for $p\geq 2$ when $D=10$, and  for $p\geq 3$ when $D=11$. Most of these results can be derived by dimensional reduction from $D=10$ or $D=11$.
The proof that this is so can be found in \cite{Movshev:2011pr,Brandt:2013xpa}, but we can understand it intuitively by thinking about branes. A scalar $p$-brane, \ie one without world-volume gauge fields, has a world-volume action with a kinetic term and a Wess-Zumino term. The latter is the pull-back of a $(p+1)$-form potential in the target superspace to the world volume. In order for the brane action to be invariant under kappa-symmetry, it is necessary that the corresponding $(p+2)$-form field-strength be closed and non-trivial cohomologically. In a flat background, the only non-zero component of this field strength must have dimension zero, \ie it is the component $F_{p,2}$. This can only be a gamma-matrix multiplied by some R-symmetry invariant. For maximal supersymmetry there are just two independent possibilities, $D=11$ or IIB in $D=10$, since the other cases can be obtained by dimensional reduction. In $D=11$ there is a membrane with $F_{2,2}\sim \C_{2,2}$, where $\C_{p,2}$ denotes a symmetric gamma-matrix with $p$ antisymmetric even indices, while in IIB in $D=10$ we have two types of string, reflected by the fact that there are two independent $F_{1,2}$ involving the $16\xz 16$ gamma-matrix $\c_a$ multiplied by two independent two-dimensional euclidean gamma-matrices. These $(1,2)$-forms are $t_0$-closed but not exact. More generally, one can see that $t_0$-cohomology is determined by the $p$-branes in any given dimension for any number of supersymmetries. There is thus a close relationship between this cohomology and the brane scan \cite{Achucarro:1987nc}.

Suppose we have a $t_0$-closed non-trivial gamma matrix $\C_{p,2}$ considered as a $(p,2)$-form and consider a form $\o_{r,s}$ defined by
\be
\o_{a_1\cdots a_r,\a_1\cdots \a_s}=(\C_{a_1\cdots a_r \cdots a_p})_{(\a_1\a_2} \l^{a_{r+1}\cdots a_p}{}_{\a_3\cdots \a_s)}\ ,
\la{2.8a}
\ee
where $\l$ is constructed from the fields and $r\leq p$ and $s\geq2$. Clearly such an $\o$ is $t_0$-closed and is not exact, in general. This illustrates how one can build non-trivial cohomology elements with the aid of the basic brane gamma-matrices. 
Note that, after dimensional reduction, the $t_0 \C_{p,2}=0$ relation will not be of the same form in general.
This explains why there is no $t_0$ cohomology for $p>0$ in $D\leq 9$ for maximal supergravity. (There is cohomology for $p=1$ in $D=10$ IIA because $\C_a\C_{11}$, considered as a (1,2)-form, is $t_0$-closed but not exact.) Moreover, since one needs a gamma-matrix to construct a non-trivial $t_0$-closed form, it is generically the case that $H_t^{p,q}=0$ if $p\geq 1$ and $q<2$. 
 
\subsection{Solving the Bianchi identities}

The above results, when combined with some simple dimensional analysis, greatly simplify the analysis of the CBIs, as we shall now explain. Suppose the dimension of the top (purely even) component of an $n$-form, $\o_{n,0}$, is $k$, then the dimension of $\o_{n-q,q}$ is $k-q/2$. The dimension of the top component of a field-strength form $F_{\ell+1,0}$ is one, and since there are no fields with negative dimensions in supergravity, the only components of a field strength $F_{\ell+1}$ that can be non-zero are $F_{\ell-1,2}(0),\ F_{\ell,1}(1/2)$ and $F_{\ell+1,0}(1)$, where the dimensions are indicated in brackets.
For a CBI $I_{\ell+2}$, the  components that are not identically zero on dimensional grounds are $I_{\ell-2,4}$ up to $I_{\ell+2,0}$, the former having dimension zero and the latter dimension two. The Bianchi identities for the forms with lowest degree must be gauge-trivial, \ie have the form $dF=0$. For $3\leq D\leq 9$, the lowest-degree potentials have $\ell=1$, in $D=10$ the forms start at degree one in IIA and two in IIB while in $D=11$ the forms start at degree three. As we shall see shortly, the only forms with canonical dimensions that can have gauge-trivial Bianchi identities are the generating forms.

Let us consider first the Bianchi identities for the forms of least degree (\ie $\ell=1$) for the cases $3\leq D\leq 9$. The Bianchi identity for the two-form field strength is $I_3=dF_2$ and is clearly consistent as $dI_3=0$. The lowest possible non-zero component of $F_2$ is $F_{0,2}$ which has dimension zero and trivially satisfies $t_0 F_{0,2}=0$. If it is $t_0$-exact it can be set to zero by redefining the top component of the connection $A_{1,0}$. Suppose that this is the case, then $I_{1,2}=t_0 F_{1,1}$, and hence setting $I_{1,2}=0$ implies that $F_{1,1}=0$ by the absence of cohomology, and this in turn implies $F_{2,0}=0$
as well. Indeed, this is a general feature: the possible generating forms all have cohomologically non-trivial dimension-zero components. The dimension-one-half component of the Bianchi identity, $I_{0,3}$, is $d_1 F_{0,2}+ t_0 F_{1,1}=0$. It allows one to solve for $F_{1,1}$ but also places a constraint on $F_{0,2}$, namely $d_s[F_{0,2}]=0$. All of the higher components of the Bianchi identity are then automatically satisfied because of the consistency condition. Given that $I_{0,3}=0$, $d I_3=0$ implies that $t_0 I_{1,2}=0$, and since $H_t^{1,q}=0$ for $D\leq 9$, we have
$I_{1,2}=t_0 J_{2,0}$ for some $J_{2,0}$. So we only need to set $J_{2,0}=0$ to get $I_{1,2}=0$; but the former equation simply allows us to solve for $F_{2,0}$ in terms of derivatives of $F_{0,2}$ and $F_{1,1}$ (as well as the dimension-one torsion which appears via $d_0$). The remaining components of the Bianchi identity are then trivially satisfied. At dimension three-halves, we have $I_{2,1}$, but if the lower-dimensional identities are satisfied then $t_0 I_{2,1}=0\Rightarrow I_{2,1}=0$; similarly, at dimension two we find that $I_{3,0}=0$ if all of the lower-dimensional identities are satisfied.

Given that the Bianchi identity for $F_2$ is satisfied there is one further cohomological obstruction to be overcome at level two. The Bianchi identity has the form $I_4=dF_3- (F_2)^2$, and the consistency condition is  $dI_4=0$ if $I_3=0$. The lowest component of $I_4$ is $I_{0,4}$ and has dimension zero, specifically,
\be 
I_{0,4}=t_0 F_{1,2} - F_{0,2} F_{0,2}\ .
\la{2.8b}
\ee
Now $t_0 I_{0,4}=0$, but $H_t^{0,4}$ need not be trivial. So in order to be able to solve \eq{2.8b} when $I_{0,4}$ is set equal to zero, the second term on the right must be cohomologically trivial to match the first one.

The forms at any level will fall into representations of the duality group which we shall call $\cR_\ell$, so that $\cR_2$ must be in the symmetric product of $\cR_1$ with itself. In general this will give rise to several irreducible representations, so that the constraint we arrive at will be on the possible $\cR_2$ representations for which $I_{0,4}$ is soluble, \ie for which the second term on the right in \eq{2.8b} is cohomologically trivial. This constraint is often called the supersymmetry constraint.

The analysis for the higher components of $I_4$ and all of the higher-degree Bianchi identities is straightforward provided the supersymmetry constraint is fulfilled. As described above one can apply cohomological methods to the consistency conditions on the Bianchi identities and use the fact that there are no $t_0$-cohomological obstructions to show that the remaining identities simply allow us to solve for the non-zero components of the field strength forms. In other words the only non-trivial Bianchi identity components are $I_{0,3}, I_{1,2}$ at level one and $I_{0,4}$ at level two. All of the higher components for all levels are then satisfied by solving for the dimension-zero, one-half and one components of the various field strength forms in terms of those already given, \ie in terms of the supergravity fields.

The analysis for IIB in $D=10$ is similar \cite{Greitz:2011da}. In this case the lowest level is two while $H_t^{p,q}=0$ for $p\geq 2$. When the Bianchi identities for the three-form field-strengths are satisfied then all the higher-degree ones can be satisfied by solving for the components of the higher-degree forms. There are no further constraints or non-trivial consistency conditions on the fields to worry about. A similar situation obtains in $D=11$, where the lowest level is three and $H_t^{p,q}=0$ for $p\geq 3$. In this case there are only two levels, $\ell=3$ and $\ell=6$ and no duality group.

For the IIA case the lowest level is one while the $H_t$ cohomology groups vanish for $p\geq 2$. If the Bianchi identity for $F_2$ (a singlet) is satisfied, the Bianchi identity for the next level should be $I_4= dF_3 - F_2 F_2$, with $dI_4=0$. In this case $F_{0,2}\neq 0$ and is not exact, but it is not difficult to show that $F_{0,2}^2$ is not $t_0$-exact. This means that $I_{0,4}=0$ cannot be satisfied as it stands and the only way out is to allow a constant in front of the $F_2^2$ term which is then chosen to vanish, in other words, the Bianchi identity for $F_3$ should  also be gauge-trivial, $d F_3=0$. This is allowed because $H_t^{1,2}\neq 0$. So in the IIA case there are two generating forms at levels one and two. Given that the Bianchi identities for both of these are satisfied one can easily show that those for all of the forms generated from them are as well and that their components are specified by them \cite{Bergshoeff:2010mv}. 

To conclude this section we give two examples for which the superspace method can be used straightforwardly to find the representations of the forms including some of the OTT ones.

\subsection{Examples}

In this subsection we explicitly work out the form representations for two examples, IIB supergravity in $D=10$ and $D=3$, $N=16$ supergravity. In the first case the duality group  is $SL(2,\bbR)$, which is rather easy to work with, so that the problem is tractable by hand. This example allows us to demonstrate explicit agreement with the Borcherds algebra predictions for the form representations and their multiplicities even beyond the spacetime limit. In the second case the duality group is $E_8$, and the 
dimensionalities of the representations
rapidly become rather large with increasing form degree.
On the other hand,
we only need to go up to level four to accommodate all the forms with non-vanishing field strengths in supergravity.

\subsubsection*{$D=10$ IIB}

For IIB the generating forms are two-forms in the doublet representation of $SL(2,\bbR)$; the corresponding field strengths are a doublet $F_3^{\cM}$. The superspace formulation of the theory was given in \cite{Howe:1983sra}, the dual physical forms were added in \cite{Cederwall:1996ri,Dall'Agata:1998va} and the components of the full set of field strengths up to degree eleven were written down in \cite{Bergshoeff:2007ma}; the CBIs are
\begin{align}
dF_3^\cM&=0\ ,\nn\w1
dF_5&=\ve_{\cM\cN} F_3^\cM F_3^\cN \ ,\nn\w1
dF_7^\cM&=F_3^\cM F_5\ ,\nn\w1
dF_9^{\cM\cN}&= F_3^{(\cM} F_7^{\cN)}\ ,\nn\w1
dF_{11}^{\cM\cN\cP}&= F_3^{(\cM} F_9^{\cN\cP)}\ ,\nn\w1
dF_{11}^{\cM}&= \ve_{\cN\cP} F_3^{\cN} F_9^{\cP\cM} + \frac{3}{4}F_5 F_7^\cM\ .
\label{2.10}
\end{align}
It is  rather simple to see that these equations are indeed consistent. It is also easy to explicitly work out the first few OTT forms \cite{Greitz:2011da}. For $\ell=12$ we find three possible representations,
\begin{align}
dF_{13}^{\cM\cN\cP\cQ}&= F_3^{(\cM} F_{11}^{\cN\cP\cQ)} \ ,\nn\w1
dF_{13}^{\cM\cN}&=\ve_{\cP\cQ} F_3^\cP F_{11}^{\cQ\cM\cN} + \frac{8}{15}F_3^{(\cM} F_{11}^{\cN)} + \frac{2}{5}F_5 F_9^{\cM\cN}\ , \nn\w1
dF_{13}&=\ve_{\cM\cN} F_3^\cM F_{11}^\cN +\frac{3}{8} \ve_{\cM\cN} F_7^\cM F_7^\cN\ ,
\label{2.11}
\end{align}
and only a little work has to be done to fix the coefficients by consistency. At $\ell=14$ we find that there are again three possible representations, but that this time there are degeneracies,
\begin{align}
dF_{15}^{\cM\cN\cP\cQ\cR}&= F_3^{(\cM} F_{13}^{\cN\cP\cQ\cR)}\ ,\nn \w1
dF_{15}^{\cM\cN\cP}&=a\ve_{\cQ\cR} F_3^\cQ F_{13}^{\cR\cM\cN\cP} +bF_3^{(\cM} F_{13}^{\cN\cP)} + cF_5 F_{11}^{\cM\cN\cP} +  d F_7^{(\cM} F_9^{\cN\cP)}\ ,\nn\w1
dF_{15}^\cM&=e\ve_{\cN\cP} F_3^\cN F_{13}^{\cP\cM}+f F_3^\cM F_{13} + g F_5 F_{11}^\cM + h\ve_{\cN\cP} F_7^\cN F_9^{\cP\cM}\ .
\label{2.12}
\end{align}

At first sight it looks as if there can be four fifteen-form field strengths in both the quadruplet and doublet representations, but the consistency conditions imply that there are only two of each.  Note that, although the thirteen-form  field strengths are identically zero in supergravity, not all of the forms in the bilinear terms on the right-hand sides of \eq{2.11} are because $(F\wedge F)_{10,4}$ has dimension zero,
so that it is not entirely trivial that these Bianchi identities are satisfied. In fact, this is the case, by the cohomological argument that we have discussed above.

The above results are summarised in the table below, where the numbers in brackets denote multiplicities. The representations and degeneracies agree with those computed for the positive roots  of a Borcherds superalgebra known as the Slansky algebra. This algebra appeared in the physics literature some time ago in a different context \cite{Slansky:1991dx}, and was discussed in the supergravity context in \cite{HenryLabordere:2002dk,Henneaux:2010ys}. 
\begin{center}
\begin{tabular*}{0.75\textwidth}{@{\extracolsep{\fill}} |c|c|l|}
\hline
Level $k$ & Form degree $\ell=2k$& $\gs\gl(2)$ representation(s) \\
\hline\hline
1 & 2& {\bf 2}\\
\hline
2&4& {\bf 1}\\
\hline
3&6&{\bf 2}\\
\hline
4&8&{\bf 3}\\
\hline
5&10&{\bf 4}+{\bf 2}\\
\hline
6&12& {\bf 5}+{\bf 3}+{\bf 1}\\
\hline
 7& 14& {\bf 6}+{\bf 4}(2)+{\bf 2}(2)\\
\hline
\end{tabular*}
\end{center}


\subsubsection*{$D=3$}


The forms for $D=3$ maximal supergravity were discussed in a superspace setting in \cite{Greitz:2011vh}, up to level four, but some representations that did not appear in gauged supergravity were omitted. We include them here for completeness and also because this is another case that can be computed by elementary means using the CBIs. At level one we have the generating forms in the adjoint ({\bf 248}) of $\ge_8$ and at level two there are three representations in the two-fold symmetric product of the adjoint,
${\bf 1}+{\bf 3875}+{\bf 27000}$. The last of these is excluded by the supersymmetry constraint because the Bianchi identity
\be
I_4=d F_3 -(F_2 F_2)
\label{2.13}
\ee
has no solution for the $(0,4)$ component if the right-hand side is in the {\bf 27000} representation. To make this clearer we write out $I_{0,4}$
explicitly using $\a=1,2$ for spinor indices and $i=1,\ldots, 16$ for internal $SO(16)$ vector indices ($\a\rightarrow \a i$). We have $t_0 F_{1,2}=F_{0,2} F_{0,2}$ or
\be
-i\d_{ij} (\c^a)_{\a\b} F_{a \c k\d l}=F_{\a i\b j} F_{\c k\d l}\ ,
\label{2.14}
\ee
where total symmetrisation over the four spinor-index pairs is understood. Now since $F_{0,2}$ and $F_{1,2}$ can only contain scalar fields we must have
\be
F_{\a i\b j}\sim\ve_{\a\b} F_{ij}\qquad {\rm and} \qquad F_{a\a i\b j}\sim (\c_a)_{\a\b} G_{ij}
\label{2.15}
\ee
where $F_{ij}$ is antisymmetric and $G_{ij}$ is symmetric. Of course, these objects also carry $\ge_8$ indices which can be projected onto $\gs\go(16)$ representations, only the tensorial ones being relevant in this case. Now the {\bf 27000} representation contains an $\gs\go(16)$ representation that has the symmetries of the Weyl tensor. This object does not drop out of the right-hand side of \eq{2.14} when one symmetrises over the joint indices, but clearly cannot be contained on the left. Thus we conclude that this representation is not allowed. On the other hand, the {\bf 3875}
contains the {\bf 1820} representation of $\gs\go(16)$. This is totally skew on four vector indices. But since the right-hand side of \eq{2.14} must be totally symmetric on  the four joint $\a i$-type  indices it follows that the two-component spinor indices must  be totally antisymmetric. Hence  this representation drops out of the Bianchi identity implying that the {\bf 3875} representation is indeed allowed at level two.

The possible representations at level three will be contained in the tensor product
\begin{align}
{\bf248}\otimes ({\bf1}+{\bf3875})={\bf248} + {\bf3875} + {\bf147250}\ .
\end{align}
The ${\bf248}$ was left out of the discussion given in \cite{Greitz:2011vh}  because the gauged models were restricted to those with the embedding tensor in the ${\bf3875}$, but it is allowed as has been pointed out elsewhere \cite{deWit:2008ta}. The Bianchi identity is
\be
dF_4^\cM=a F_3^{\langle\cM\cN\rangle} F_{2\cN} + b F_3 F_2^\cM\ ,
\label{2.16}
\ee
where $\langle\cM\cN\rangle$ denotes the symmetrised product of two ${\bf248}$ representations projected onto the ${\bf3875}$, and $a,b$ are constants. Since $dF_3\sim F_2 F_2$, applying $d$ to \eq{2.16} yields the symmetrised product of three ${\bf248}$s projected onto the ${\bf248}$ from both terms on the right-hand side. As the {\bf248} appears only once in this triple product it follows that there is a unique choice of constants (up to an overall scale that can be absorbed into $F_4^\cM$) such that \eq{2.16} is consistent. The non-zero components of all of the forms up to level four, except for this one, were given in \cite{Greitz:2011vh}. For $F_4^\cM$ the possible non-zero components are $F_{2,2}$ and $F_{3,1}$, since $F_{4,0}$ is identically zero. We have
\be
F_{2,2}^\cM\rightarrow F^\cM_{ab\c k\d l}\sim (\c_{ab})_{\c\d} F^\cM_{kl}\ ,
\ee
where $F_{kl}^\cM$ is symmetric on $kl$. But the {\bf248} branches into ${\bf120}+{\bf128}$  in $\gs\go(16)$, so $F_{kl}^\cM=0$. On the other hand, at dimension one-half we have
\be
F_{3,1}^\cM\rightarrow F^\cM_{abc\d l}\sim \ve_{abc} (\S_l)^{IJ'}\L_{\d J'} \cV_I{}^\cM\ ,
\ee
where $I,I'$ are $\gs\go(16)$ Weyl spinor indices, both running over 128 values, $\L_{\a I'}$ is the physical spinor field, and $\cV_I{}^\cM$ is the matrix of scalar fields projected onto the spinor representation. (The other component is $\cV_{ij}{}^\cM$ in the {\bf120} of $\gs\go(16)$.)

At level four the possible representations are in the tensor product of the {\bf248} with the level-three representations, or in the antisymmetric product of two level-two representations, corresponding to the Bianchi identities $dF_5= F_2 F_4 + F_3 F_3$. After a little group theory, and taking the consistency conditions into account, we find that the allowed representations at level four are
\begin{align}
{\bf248}(2)+ {\bf3875}+ {\bf30380}(2)+{\bf147250}+ {\bf779247}+{\bf6696000}\ ,
\end{align}
where the numbers in brackets denote the degeneracies. 

We shall not give explicit details of the forms in the other maximal theories here, \ie for $3<D<10$, but the representations (up to level $D+1$) are tabulated in appendix \ref{maxreps-app}. However, since for these cases $H_t^{p,q}=0$ unless $p=0$, it follows that solving the level-one and -two identities, (in fact, it is only necessary to solve $I_{0,3}, I_{1,2}$ and $I_{0,4}$),  means that all the higher ones are automatically guaranteed to be consistent and soluble by the general arguments we have given.\footnote{The level-two forms in $D=4$,
which transform under the ${\bf 133}$ of $\ge_7$, were given explicitly in \cite{Bandos:2015ila}.}

\section{Half-maximal supergravity theories}

\subsection{Supergravity and vector multiplets} \label{svm-subsection}

In this section we study half-maximal supergravity coupled to vector supermultiplets in dimensions three to ten. This topic was studied in components in \cite{Bergshoeff:2007vb}, but here we give a slightly different approach based on the superspace formalism. The models we discuss are $N=1$,
$D=10$ supergravity coupled to $n$ vector multiplets and the lower-dimensional ones
obtained from this by dimensional reduction.\footnote{We can take $n\geq -k$ in $D=10-k$ dimensions, $D\geq 4$ and $n\geq -8$ in $D=3$. From an algebraic point of view the special cases are those for which the duality algebra $\gs\go(k,n+k)$ is split. We shall discuss this further in section 4.} In $D=10-k$ dimensions, $k\leq 5$, the physical bosonic component fields consist of the graviton, the dilaton, a two-form gauge field, $n+2 k$ vectors (of which $k$ belong to the supergravity multiplet) and $k(n+k)$ scalars. The duality group is $\bbR^+ \xz SO(k,n+k)$ and the scalars belong to the coset with isometry group $SO(k)\xz SO(n+k)$, where the former is the local R-symmetry group factor of the superspace structure group. In $D=4$ dimensions the two-form can be dualised to another scalar so that there is an extra $SL(2,\bbR)$ factor in the duality group (but on the other hand no $\mathbb{R}^+$ factor), with this scalar and the dilaton in the coset $U(1)\bsh SL(2,\bbR)$. In $D=3$ dimensions the vectors can be dualised to scalars so that we have $8(n+8)$ of them in the coset $(SO(8)\xz SO(n+8))\bsh SO(8,n+8)$. The special case of $D=6b$, where the vector multiplets are replaced by tensors, is not derivable from $D=10$ and needs a separate treatment
(see section \ref{6b-section}  below).\footnote{Following \cite{Bergshoeff:2007vb} we use the notation $6a$ and $6b$ for the supergravity theories with (1,1) and (2,0) supersymmetry in six dimensions.}

One way of thinking about the on-shell theory is to start from an off-shell supergravity multiplet with 128 + 128 components \cite{Nilsson:1985si,Howe:1986ed}. This multiplet is dual to a supercurrent multiplet which is conformal in $D=4$ \cite{Bergshoeff:1980is} but not in higher dimensions \cite{Howe:1981nz,Bergshoeff:1981um}. It is, however,  local, except in $D=10$ \cite{Bergshoeff:1982av,Howe:1982mt}. In $D=10$ the supergravity multiplet consists of the graviton, a six-form gauge field (dual to the two-form potential) and the gravitino, with constraints on the curvature scalar and the spin-one-half part of the gravitino field-strength (and therefore not fully off-shell, but in the $D=10$ case only). For $5\leq D\leq 9$ the off-shell bosons consist of the graviton, a set of non-abelian $SO(k)$ gauge fields and  a $(D-4)$-form potential (dual to the supergravity two-form), together with some auxiliary scalars, $S_{IJK}=S_{[IJK]}$, totally antisymmetric on the $SO(k)$ vector indices, and a set of antisymmetric tensors $N_{ab I}=N_{[ab]I}$ at dimension one, as well as dimension-two scalar fields $C_{IJ}$ which are symmetric and traceless. The fermions are the gravitino and a set of $(k-1)$ sixteen-component dimension-three-halves spinors. This multiplet does not contain the dilaton, the physical two-form or the dilatino which are therefore introduced via the form sector of the theory. Note that the absence of the dilaton and the dilatino implies that the dimension-zero torsion is flat and that the dimension one-half torsion is zero, a situation that can be maintained when we go on-shell. In $D=4$, since the gravitino is superconformal, there are also four dimension-one-half spinors \cite{Howe:1980sy}. In $D=3$ it is also possible to use an off-shell superconformal multiplet in the supergravity sector \cite{Howe:1995zm,Kuzenko:2011xg,Cederwall:2011pu,Greitz:2012vp}.

The strategy is to go on-shell by introducing the scalars, vectors and higher-degree form fields and by imposing suitable constraints on the components of the field-strength forms\footnote{There is an extensive literature on on-shell $N=1$, $D=10$ supergravity in superspace. See, for example, \cite{Nilsson:1981bn,Bonora:1986ix,Bellucci:1988ff} and for more recent reviews \cite{Lechner:2008uz,Howe:2008vb}.}. The dilaton appears in the dimension-zero component of the three-form field-strength ($F_{1,2}$) dual to the $(D-3)$-form field-strength in the off-shell multiplet. The dimension-zero component of the latter is an appropriate gamma-matrix but 
contains no dilatonic factor in this approach; this corresponds to choosing a brane-frame rather than the string or Einstein frames\footnote{For example, in $D=10$ the six-form potential couples to a five-brane whose seven-form field-strength has $F_{5,2}=-i\C_{5,2}$, whereas the string frame would have $F_{1,2}=-i \C_{1,2}$.}. The non-dilatonic scalars (for $D>4$) are taken to belong to the coset $(SO(k)\xz SO(n+k))\bsh SO(k,n+k)$. Let $\cV_{\cA}{}^{\cM}=(\cV_I{}^{\cM},V_{I'}{}^{\cM})$ denote the matrix of scalar fields in the vector representation of $SO(k,n+k)$, where $I$ and $I'$ are respectively $SO(k)$ and $SO(k+n)$ vector indices, and $\cM$ is a vector index for $SO(k,n+k)$. As usual, we define
\be
d\cV \cV^{-1}=P + Q\ ,
\la{4.1}
\ee
where $Q$ is a composite connection taking its values in $\gs\go(k)\oplus\gs\go(n+k)$ and $P$ takes its values in the quotient algebra. The Maurer-Cartan equation gives
\be
R+ DP=-P\wedge P\ ,
\la{4.2}
\ee
where $R$ is the $\gs\go(k)\oplus\gs\go(n+k)$ curvature, $D$ is the $\gs\go(k)\oplus\gs\go(n+k)$ covariant derivative with connection $Q$, and where $\gs\go(k)$ is identified with the R-symmetry part of the superspace structure algebra. The right-hand side takes its values in $\gs\go(k)\oplus\gs\go(n+k)$, so $DP=0$. In indices we write $P^{I J'}$. There is a constraint on $P$ at dimension one-half that reads
\be
P_{\a JK'}=(\S_J  \r)_{\a K'}\ ,
\la{4.3}
\ee
where $\S_J$ is a suitable spin matrix, $\a$ is a sixteen-component spinor index and where we have not explicitly exhibited the $SO(k)$ spinor indices. The spinors $\r$ belong to the vector multiplets. It follows from this and from \eq{4.2} that the dimension-one $SO(k)$ curvature components are determined as bilinears in these fermions. In particular, this allows us to solve for the dimension-one auxiliary fields in terms of the physical ones. This is turn will determine the higher-dimensional auxiliaries on-shell. 

\subsection{Forms}

The basic field-strength forms are those for the vector multiplets, $F_2^\cM$, and the $(D-3)$-form, $F_{D-3}$, that appears in the off-shell supergravity multiplet.  Their duals are  the three-form field strength, $F_3$,  of the supergravity two-form potential and the duals of the vector fields, $F_{D-2}^\cM$. This set of forms generates all the rest, although we know that there must be forms, $F_{D-1}$ and $F_{D-1}^{\cM\cN}$, dual to the scalars  in the
$1+\Yboxdim{5pt}{\yng(1,1)}$ representations of the duality group. 

The basic Bianchi identities are, for $D>4$:
\begin{align}
dF_2^\cM&=0\ , 
&dF_3&=F_2\cdot F_2\ , \nn\w1
dF_{D-3}&=0\ , 
&dF_{D-2}^\cM&= F_{D-3} F_2^\cM \ .
\end{align}
We then find that the $(D-1)$-form field strengths Bianchis (\ie level $(D-2)$) are, as expected,
\begin{align}
dF_{D-1}&=F_{D-2}\cdot F_2-F_{D-3} F_3\ ,\nn\w1
dF_{D-1}^{\cM\cN}&=2F_{D-2}^{[\cM} F_2^{\cN]}\ .
\end{align}
The details of the higher-level forms vary slightly with dimension, but they are all fixed by the basic Bianchi identities  
(3.4) above. At level $(D-1)$ we have
\begin{align}
dF_D^\cM&=F_{D-1}^{\cM\cN}F_{2\cN}-F_{D-2}^\cM F_3 + F_{D-1} F_2^\cM\ ,\nn\w1
dF_D^{\cM\cN\cP}&=3F_{D-1}^{[\cM\cN}F_2^{\cP]}\ ,
\end{align}
at level $D$,
\begin{align}
dF_{D+1}&=F_D^\cM F_{2\cM}-F_{D-1} F_3\ ,\nn\w1
dF_{D+1}^{\cM\cN}&=F_D^{\cM\cN\cP} F_{2\cP} + 2F_D^{[\cM} F_2^{\cN]}-F_{D-1}^{\cM\cN} F_3\ ,\nn\w1
dF_{D+1}^{\cM\cN\cP\cQ}&=4 F_D^{[\cM\cN\cP} F_2^{\cQ]}\ ,
\end{align}
and at level $(D+1)$,
\begin{align}
dF_{D+2}^{\cM}&=F_{D+1}^{\cM\cN} F_{2\cN}+F_{D+1} F_2^\cM -F_{D}^\cM F_3\ ,\nn\w1
dF_{D+2}^{\cM\cN\cP}&=F_{D+1}^{\cM\cN\cP\cQ} F_{2\cQ} +3F_{D+1}^{[\cM\cN} F_{2}^{\cP]} -F_{D}^{\cM\cN\cP} F_3\ ,\nn\w1
dF_{D+2}^{\cM\cN\cP\cQ\cR}&=5 F_D^{[\cM\cN\cP\cQ} F_2^{\cR]}\ .
\end{align}
This last set of forms is over the top, but not all of them are necessarily zero in supergravity. The dimension-zero component of a $(D+2)$-form is $F_{D,2}$ and has to be of the form of a spacetime epsilon tensor multiplied by a symmetric $16\xz 16$ matrix that carries no spacetime vector indices. The R-symmetry factor in this must match the index of the form. As a simple example, consider $F_{11}^\cM$ in $D=9$. The non-zero component is $F^\cM_{9,2}\sim\ve_{9,0} \C_{0,2} \cV_1{}^\cM$, where $\C_{0,2}$ is symmetric. 

The forms in $D=7,6a,5$ consist of the above sets together with some additional ones which we now describe. In $D=6a,7$ there are extra forms at level $(D-1)$ given by  Bianchi identities $dF_D^\cM=F_{D-3}F_{D-2}^\cM$  and $dF_D=F_{D-3} F_{D-3}$ respectively. These correspond to deformations of type $p=2$ and $p=3$ in the nomenclature of \cite{Bergshoeff:2007vb}. For example, in $D=7$ we have $F_{D-3}=F_4$ and it is clear that $F_4 F_4$ is closed. But it is also exact as one can see by using cohomology. This means that the super seven-form $L_7=F_7-A_3 F_4$, where $F_4=dA_3$, is closed and so determines, via the ectoplasm formalism \cite{Gates:1997kr,Gates:1997ag},  a superinvariant given as the integral over spacetime of the top component of $L_7$ in a coordinate basis. This has dimension one and so gives rise to a massive deformation if we include a mass parameter in $L_7$. In $D=6a$ the new six-form leads to a possible deformation of type $p=2$ where a two-form acquires mass. Both types of term affect the higher-degree forms but do not change the cohomological argument given above for solubility of the complete set of Bianchi identities. In fact, terms of this type also exist in higher dimension but they are over the top;
they occur at level $(D+1)$ in $D=8,9$ and level $(D+3)$ in $D=10$.

In $D=5$, $F_{D-2}$ is another two-form, so that in this case all of the forms are generated from the level-one forms. The same is true in $D=4$ and $D=3$. In the former case vectors are in the representation $(2,\Yboxdim{5pt}\yng(1))$ of the duality group while in $D=3$ the scalars and vectors are dual to each other. This implies that the level-one forms can be taken to be in the adjoint representation of $SO(8,8+n)$, \ie $\Yboxdim{5pt}\yng(1,1)$.

We give all of the forms and their Bianchi identities, for $D>3$ up to level $(D+1)$, and for $D=3$ up to level $D$, in Appendix \ref{bianchi-app}.

\subsection{Half-maximal cohomology}
In order to analyse the Bianchi identities we shall need the cohomology for half-maximal superspaces which we now briefly review following \cite{Berkovits:2008qw}.

In $N=1$, $D=10$ superspace, with the dimension-zero torsion taking the same form as in flat superspace, the $H_t^{p,q}$ groups are more complicated than in the maximal case due to the fact that there is both a string and a five-brane. There are therefore non-trivial groups for $p\leq 5$. The ones associated with the five-brane are given by
\be
H_t^{p,q}=H_t^{q-2}(\L^{5-p}T_0)\ ,\qquad p\in\{1,2,3,4,5\},\ q\geq 2\ ,
\la{4.4}
\ee
while the ones associated with the string are $H_t^{1,1}$, which can be identified with sections of the odd tangent bundle, and $H_t^{1,2}$ which is given by functions. In addition, $H_t^{0,q}:=H_t^q$ is just the usual space of $q$-th rank pure spinors.  The notation $H_t^{q-2}(\L^{5-p}T_0)$ denotes the space of $(0,q-2)$-forms taking their values in $\L^{5-p}T_0$, where $T_0$ is the even tangent bundle, modulo two equivalences. The first is just the action of $t_0$ with respect to the $(q-2)$ spinor indices, while the second is given by an operation that we describe below. These spaces can be presented in terms of irreducible representations of the Lorentz group \cite{Movshev:2010mf}.

Informally, these results can be understood as follows. For the string, the associated gamma-matrix is $\C_{1,2}$, the symmetric $16\xz 16$ matrix with one vector index considered as a $(1,2)$-form. Clearly this can only lead to cohomology with $p=1$. The space $H_t^{1,1}$ consists of elements of the form
\be
\o_{a\b}=(\C_a)_{\a\b} \l^\b\ .
\la{4.5}
\ee
Clearly such a $(1,1)$-form is $t_0$-closed, using the identity $t_0 \C_{1,2}=0$, because the latter can be written with explicit symmetrisation over only three of the spinor indices, but it is not exact. A non-trivial element of $H_t^{1,2}$ just has the form $\C_{1,2} f$ where $f$ is any function, as one can easily check. It might be thought that one could define non-trivial groups in a similar fashion for $q>2$, but this is not the case because $t_0 \C_{1,2}=0$. For example, a three-form $\o_{1,3}=\C_{1,2}\l_{0,1}$ is certainly closed but can be rewritten as $t_0 \r_{2,1}$ where $\r_{ab\c}\sim (\C_{ab}\l)_\c$. 

The five-brane is associated with the symmetric five-index matrix $\C_{5,2}$ which can be considered as a $(5,2)$-form. A $t_0$-closed $(p,q)$-form, for $1\leq p\leq 5$, can be written $\o_{p,q}=\C_{5,2} \l^{5-p,0}_{0,q-2}$ where the notation indicates that the upper indices on $\l$ are even and are to be contracted with indices on the gamma-matrix, while the lower indices are purely odd. If we  change $\l$ by $t_0\r^{5-p,0}_{1,q-4}$, \ie ignoring the upper indices on $\l$, then $\o_{p,q}$ will change by a $t_0$ term. We can also change $\l$ by a term of the form $\r^{[a_1\ldots a_{5-p-1} |\c|}_{(\a_1\ldots \a_{q-1} }(\C^{a_{5-p}]})_{\a_q)\c}$ which will also change $\o_{p,q}$ by a $t_0$ factor (see \cite{Berkovits:2008qw} for more details).

For $D<10$ there are no strings with sixteen supersymmetries, so the cohomology can be obtained by dimensional reduction from the five-brane sector. In $5\leq D\leq 10$ dimensions the non-zero cohomology groups are $H_t^{p,q}$ for $p\leq  (D-5)$, and $q>1$ if $p>0$. For $D\leq 5$ we only have $H_t^{0,q}$ which is similar to the maximal case for $D\leq 9$.

\subsection{Solving the Bianchi identities}
In this subsection we shall show that all of the consistent Bianchi identities can be solved with the aid of cohomology. This means that the complete sets of forms for all half-maximal theories are compatible with supersymmetry. Since zero is the lowest dimension for which a Bianchi identity component can be non-zero in supergravity, the highest level at which there is a non-trivial Bianchi component is
$\ell=D-3$, for $D\geq 5$;
it is $I_{D-5,4}$. Beyond this level, the Bianchi identities, provided they are consistent, will automatically be soluble as we saw in the maximal case.

To illustrate this we consider the case $D=10$. The non-trivial Bianchi identity components are 
\bea
\ell=1&&\ I_3: \qquad{\phantom {I_{0,4}}}\qquad I_{0,3}\qquad I_{1,2}        \nn\w1
\ell=2&&\ I_4: \qquad I_{0,4}\qquad I_{1,3}\qquad I_{2,2} \nn\w1
\ell=3&&\ I_5: \qquad I_{1,4}\qquad I_{2,3}\qquad I_{3,2} \nn\w1
\ell=4&&\ I_6: \qquad I_{2,4}\qquad I_{3,3}\qquad I_{4,2} \nn\w1
\ell=5&&\ I_7: \qquad I_{3,4}\qquad I_{4,3}\qquad I_{5,2} \nn\w1
\ell=6&&\ I_8: \qquad I_{4,4}\qquad I_{5,3}\qquad  \nn\w1
\ell=7&&\ I_9: \qquad I_{5,4}\qquad\qquad\qquad\qquad ,
\la{4.11}
\eea
where the columns correspond to dimension zero, one-half and one respectively.

At level one the Bianchi identities function in the same way as they do for SYM. $F_{0,2}^\cM$ has to be zero as it does not contain a Lorentz scalar; $I_{0,3}$ requires $F_{1,1}^\cM$ to be $t_0$-closed, but not exact (otherwise it could be redefined away), which is where the cohomology group $H_t^{1,1}$ fits in, and $I_{1,2}=d_1 F_{1,1} + t_0 F_{2,0}$, the same as in the flat case. This therefore determines $F_{2,0}$ in terms of the odd derivative of the spin-one-half field, $\r$, in $F_{1,1}$ and constrains the other components in the spinorial derivative of $\r$ to vanish.

The level-two Bianchi identity involves the field-strength for the supergravity two-form, $F_3$, which we recall is not in the (partially) off-shell supergravity multiplet. It equates $dF_3$ with $F_2\cdot F_2$, but the latter only contributes starting at $I_{2,2}$ because there are no scalars in $D=10$ SYM. The dimension-zero component $I_{0,4}=t_0 F_{1,2}$, so that $F_{1,2}=-i\C_{1,2} S+\C_{5,2} X^4$, where $S$ is a function of the dilaton, explicitly, $S=\exp(-\frac{2\phi}{3})$, and $X$ is a four-form which must be set to zero because the only dimension-zero fields are scalars. Again, the remaining $S$-term is cohomologically non-trivial as it should be. At dimension one-half we then have $-i\C_{1,2} d_1 S + t_0 F_{2,1}=0$, but this is clearly satisfied because the first term is indeed $t_0$-exact as we remarked previously. So $F_{ab\c}\sim (\c_{ab} \l)_\c$ where $\l=D_\a S$ is the dilatino. Given that $I_{0,4}=I_{1,3}=0$ we have
\begin{align}
t_0 I_{2,2}=0\Rightarrow I_{2,2}=t_0 J_{3,0} + \C_{5,2} K^3{}_{,0}
\end{align}
%
so that in principle there are two even three-form components. These give linear relations between $F_{3,0}$, the bilinear in the SYM fermions (coming from $F_{1,1}\cdot F_{1,1}$) and a three-index field $G_{abc}$ in the dimension-one torsion. This is the dual of the seven-form field strength in the off-shell supergravity multiplet. Clearly these two relations cannot be independent, but this is not obviously the case; it is necessary to carry out the actual computation. The presence of the dimension-one torsion in this equation arises from the term $d_0 F_{1,2}$, since $d_0$ acting on a fermion gives $D_a \l_\b+ T_{a\b}{}^\c \l_\c$. 

Now we know that there are no fields at levels 3, 4 or 5 in $D=10$, but it is nevertheless interesting to understand this from a cohomological point of view. It is easy to see that the only consistent Bianchi identities are gauge-trivial, \ie $dF_{\ell+1}=0,\ \ell=3,4,5$. Considering the dimension-zero Bianchis we have $t_0 F_{\ell-1,2}=0$, which has the non-trivial solution $F_{\ell-1,2}= \C_{5,2} X^{6-\ell}$. However, there are no dimension-zero fields in the theory that are not scalars, so that all of these components must vanish. The dimension one-half and one Bianchi identities then imply that the remaining components of these field-strengths are also zero.

The level-six Bianchi identity is $dF_7=0$. As $F_7$ is contained in the off-shell supergravity multiplet the non-trivial solution to $I_{4,4}=0$ has to be $F_{5,2}=-i\C_{5,2}$, since the dilaton is not present off-shell.  $I_{5,3}=0$ is $t_0 F_{6,1}=0$, which implies $F_{6,1}=0$, while at dimension one we find that $F_{7,0}$ is the spacetime dual of $G_{abc}$ (because $d_0 F_{5,2}$ includes the dimension-one torsion). Finally, the level-seven identity $I_{5,4}$ is trivial because $F_{0,2}=0$. This completes the analysis since the remaining, cohomologically trivial, Bianchi identity components simply determine the remaining unknown components of the various field-strengths in terms of the physical fields and their duals. In particular, this analysis proves that the forms are compatible with supersymmetry and that it is not necessary to check the supersymmetry algebra on the component fields.

The analysis goes through in exactly the same fashion for $D<10$, although there are fewer non-trivial Bianchis to consider because of the 
non-trivial cohomology limit $\ell=D-3$, for $D>4$. For $D=3,4,5$ the only non-trivial Bianchi identity components are $I_{0,3}$ and $I_{0,4}$, as in the case of maximal supergravity for $D<10$.

In the half-maximal case the fact that $F_3$ is a singlet is, of course, due to supersymmetry, but one can argue that this must be the case by solubility of the Bianchi identities for $D<9$. To see this, let us consider the dimension-zero component of the putative Bianchi identity $dF_3^{\cM\cN}=F_2^\cM F_2^\cN$ where we allow for the possibility of a level-two term in the symmetric, traceless representation. It reads:

\be
t_0 F_{1,2}^{\cM\cN}=F_{0,2}^\cM F_{0,2}^\cN\ .
\label{4.13}
\ee

Now $F_{0,2}$ and $F_{1,2}$ must be scalar fields multiplied by appropriate gamma-matrices. If we write the $D=10$ gamma-matrices as $\C^{\una}=(\C^a,\C^I)$, where $a=0,\ldots, (D-1)$, and $I=D,\ldots, (D+k-1)$, in $D=(10-k)$ dimensions, it is clear that we must have
\begin{align}
F_{0,2}^\cM &=(\C^I)_{0,2} \cV_I{}^\cM\ , \nn\w1
F_{1,2}^{\cM\cN} &= \C_{1,2}  f^{\cM\cN}
\end{align}
up to constants, where $\C_{1,2}$ denotes the $D$-dimensional gamma-matrix considered as a $(1,2)$-form, $\cV_I{}^\cM$ denotes part of the scalar field matrix and $f^{\cM\cN}$ is a function of the scalar fields that is to be determined. The relation  $t_0 \C_{1,2}=0$ in $D=10$  translates to $t_0 \C_{1,2} + (\C^I)_{0,2} (\C_I)_{0,2}=0$ in $D=(10-k)$ dimensions. This implies that \eq{4.13} can only be solved for the singlet representation if $D<9$.

\subsection{$D=6b$ supergravity} \label{6b-section}

The $D=6b$ theory has $(2,0)$ supersymmetry as opposed to the $(1,1)$ supersymmetry of the $6a$ case discussed above. The vector multiplets are replaced by tensor multiplets, so to get the same number of degrees of freedom we shall need $(n+4)$ of these. The physical fields of the full theory consist of the graviton, $5(n+5)$ scalars, $(n+10)$ two-forms with anti-self-dual field-strengths, together with the gravitino and $(n+5)$ sixteen-component spin-one-half fields. Note that there is no dilaton or dilatino. The duality group is $SO(5,n+5)$ and the scalars belong to the coset of this group defined by the isotropy group $SO(5)\xz SO(n+5)$. In a covariant formalism in which the isotropy group is local the superspace torsion is flat for dimension less than one.

The free $6b$ tensor multiplet is conformal and so gives rise to a superconformal current multiplet \cite{Howe:1983fr}. We can therefore start from an off-shell conformal supergravity multiplet in this case. Its components are the graviton, the $(2,0)$ gravitino, the $SO(5)$ gauge fields, a set of five dimension-one self-dual three-forms (not field strengths), five sixteen-component dimension three-halves auxiliary fermions and fourteen dimension-two scalars. We can go on-shell from this starting point by introducing the physical scalars as in \eq{4.1} and \eq{4.2}, and the field-strength forms for the tensor gauge fields. The $SO(5)$ gauge fields are then determined as composite, while the other auxiliaries can also be found in terms of the physical fields. For example, the self-dual dimension-one three-forms must vanish on-shell. Note that since the conformal theory is formulated in conventional superspace there will be other fields in the conformal supergeometry that can be gauged away by higher-dimensional components of the scale superfield parameter. For example, at dimension one, we should expect to find fields corresponding to the $\th^2$ components of a scalar superfield, \ie a set of 1+5 anti-self-dual tensors and 10 vectors. These fields are given as fermion bi-linears on-shell.

The dimension-one-half component of the one-form $P^{IJ'}$ is constrained in a similar fashion to \eq{4.3}; we have
\be
D_{\a i}\cV_I{}^\cM (\cV^{-1})_{\cM J'}=P_{\a i I J'}=(\c_I)_i{}^j \r_{\a j J'}\ .
\la{4.11.1}
\ee

The forms in this theory are generated by the tensors at level two. The field-strengths are three-forms $F_3^{\cM}$ obeying the Bianchi identity
\be
dF_3^\cM=0\ .
\la{4.12}
\ee
From this one can easily see that the remaining consistent Bianchi identities are at levels four and six, although there are also OTT forms starting at level eight. Since the latter are zero in supergravity we shall not consider them further here. The Bianchi identities are
\begin{align}
dF_5^{\cM\cN}&= F_3^\cM F_3^\cN \ , \nn\w1
dF_7^{\cM\cN,\cP}&=F_5^{\cM\cN} F_3^\cP\ .
\end{align}
Consistency requires that the totally antisymmetric part of $F_7^{\cM\cN,\cP}$ must be zero leaving two irreducible representations, mixed symmetry and vector.

To show that these Bianchi identities can be satisfied on-shell we can again make use of cohomological methods. For $D=6b$ the $H_t^{p,q}$ groups are empty for $p>1$ but can be non-zero for $p=0,1$. The $p=1$ case corresponds to the existence of membranes. In fact there are five of these arising from the fact that $\C^I_{1,2}$, $I=1,\ldots 5$, is $t_0$-closed but not exact. Here
\be 
(\C^I_a)_{\a\b}\rightarrow (\C^I_a)_{\a i\b j}=(\c^I)_{ij} (\c_a)_{\a\b} 
\la{4.14a}
\ee
in a two-step notation, with $\a, \b=1,\ldots, 4$ being six-dimensional chiral spinor indices and $i,j=1,\ldots, 4$ being $Sp(2)$ indices. Since there is no non-trivial cohomology for $p>1$ it follows that the only components of the Bianchi identities that need to be checked are $I_{0,4}$ and $I_{1,3}$ at level two. The former is
\be
t_0 F^\cM_{1,2}=0
\la{4.14b}
\ee
since $F^\cM_{0,3}=0$ (it has negative dimension). This must have a cohomologically non-trivial solution because otherwise $F^\cM_{1,2}$ could be set to zero by a field redefinition of the potential $A^\cM_{2,0}$. So the solution must be
\be
F^\cM_{1,2}=-i (\C^I)_{1,2} \cV_I{}^\cM\ ,
\la{4.15}
\ee
where the scalar matrix $\cV_\cA=(\cV_I{}^\cM, \cV_{I'}{}^\cM)$ as in the other cases. Given that $I_{0,4}=0$ and that $dI_4=0$ it follows that $t_0 I_{1,3}=0$. But this is not satisfied automatically due to cohomology and so $I_{1,3}=0$ has to be examined directly. It reads
\be
(\C^I)_{1,2} d_1 \cV_I{}^\cM + t_0 F^\cM_{2,1}=0\ ,
\la{4.16}
\ee
where we have made use of \eq{4.15} and the fact that the dimension one-half torsion is trivial. Disregarding for the moment the duality vector index, we see that the first term contains two representations of $Sp(2)$, whereas the second term only has one, the four-dimensional spin representation. So the gamma-traceless $SO(5)$ vector-spinor representation in $d_1 \cV_I{}^\cM$ has to be set to zero, 
\be
D_{\a i} \cV_I{}^\cM= (\c_I)_{ij} \r_{\a}^{j\cM}\ .
\la{4.17}
\ee
This is similar to the usual constraint for the tensor multiplet, the difference being that there are only $(n+5)$ sixteen-component dimension one-half spinors, so that if we contract \eq{4.17} with $(\cV^{-1})_\cM{}^J$  we get zero (\ie the spinors are $\r_{\a I'}$) in accordance with \eq{4.11.1}. The higher-dimensional components of $I_4$ and all of the components of the higher-level Bianchi identities can now be solved by specifying the non-zero components of the field-strength forms with no  further constraints arising.


\section{Algebras from forms}


The set of consistent Bianchi identities written in the form
\be
d F_{\ell+1}=\sum_{m+n=\ell} F_{m+1} F_{n+1}
\la{2.9a}
\ee
gives rise directly to an algebraic structure, namely a
co-algebra $\gf^\ast$ dual to a Lie superalgebra $\gf$. This is a $\mathbb{Z}_2$-graded vector space together
with a co-product, a linear map $d:\gf^\ast\rightarrow \gf^\ast\wedge\gf^\ast$ ($\mathbb{Z}_2$-graded antisymmetry) that extends to a
$\mathbb{Z}_2$-graded derivation of the exterior algebra of $\gf^\ast$ satisfying the nilpotency condition $d^2=0$ (equivalent to the Jacobi identity for 
$\gf$).

In the supergravity context the vector 
space $\gf^\ast$ is spanned by all the field-strength forms, and thus also has a $\mathbb{Z}_+$-grading consistent with the $\mathbb{Z}_2$-grading.
For the dual Lie superalgebra $\gf$ this means
\be
\gf=\bigoplus_{\ell\in \mathbb{Z},\, \ell\geq 1} \gf_{\ell}=\gf_{(0)} \oplus \gf_{(1)}
\la{2.9b}
\ee
where the level $\ell$ is the degree of the corresponding potential form
and the even and odd parts of $\gf$ correspond to even and odd $\ell$.

In this section we shall try to identify the Lie superalgebras $\gf$ dual to the co-algebras determined by the CBIs as
subalgebras of Borcherds superalgebras.
We shall also consider  $\gg_{\gf}$, the semi-direct sum of the duality algebra $\gg$ and $\gf$,
with the adjoint action of $\gg$ on $\gf$ given by the representations $\cR_\ell$ that the forms at level $\ell$ transform
under.

As a special case of a Kac-Moody algebra $\gg$ is
generated by $3r$ Chevalley generators $e_i,f_i$ and $h_i$ ($i=1,2,\ldots,r$)
modulo the Chevalley relations
\begin{align} \label{chevrelfinite}
[h_i,e_j]&=a_{ij}e_j\ , & 
[h_i,f_j]&=-a_{ij}f_j\ , & [e_i,f_j]&=\delta_{ij}h_j  
\end{align}
and the Serre relations
\begin{align} \label{serrelfinite}
({\rm ad}\,e_i)^{1-2a_{ij}/a_{ii}}(e_j)=({\rm ad}\,f_i)^{1-2a_{ij}/a_{ii}}(f_j)=0\ ,
\end{align}
where $a_{ij}$ is the (symmetrised) Cartan matrix of $\gg$.
When extending $\gg$ to $\gg_{\gf}$
we add one more Chevalley generator for each irreducible representation of generating forms, and demand it to be a 
lowest-weight vector of that representation, with respect to the adjoint action of $\gg$. 
In the case of only one irreducible representation of generating forms, we denote the corresponding generator by $e_0$.
Acting with the $h_i$ and $f_i$ generators on $e_0$ we then get
\begin{align} \label{additionalchevrel}
[f_i,e_0]&=0\ , & [h_i,e_0]&=-p_ie_0\ ,
\end{align}
where $p_i$ are the Dynkin labels of the representation, and we recognize (\ref{additionalchevrel}) as some of the 
additional relations generalising (\ref{chevrelfinite}) to
\begin{align} \label{chevrelfinitegen}
[h_I,e_J]&=B_{IJ}e_J\ , & 
[h_I,f_J]&=-B_{IJ}f_J\ , & [e_I,f_J]&=\delta_{IJ}h_J  
\end{align}
($I=0,1,\ldots,r$), where $B_{ij}=a_{ij}$ and $B_{0i}=B_{i0}=-p_i$. The remaining entry $B_{00}$ of the matrix $B_{IJ}$ can then be chosen 
such that it satisfies the conditions for 
a Cartan matrix of a Borcherds superalgebra $\cB$ (see appendix \ref{borcherdsapp}), and (\ref{chevrelfinitegen}) are the associated Chevalley relations.
In the case of more than one irreducible representation of generating forms, 
each of them corresponds to an additional diagonal entry in the Cartan matrix, but each pair of them also corresponds to two (equal)
additional off-diagonal
entries that have to be determined.

In the construction of the Borcherds superalgebra we must also include a Chevalley generator $f_0$ for each $e_0$ (and a Cartan element
$h_0=[e_0,f_0]$), extending
$\gg_{\gf}$ to negative levels symmetrically around level zero (so that $\cR_{-1}$ is the representation conjugate to $\cR_1$).
However, in section \ref{gauging-section} we will consider a a 
different extension, leading to a tensor hierarchy algebra \cite{Palmkvist:2013vya}, which is in some respects better suited for applications to gauged supergravity.
 
Each Chevalley generator $e_I$ of $\cB$ is a root vector corresponding to a simple root $\beta_I$, and defines a 
$\mathbb{Z}$-grading of $\cB$. This is a decomposition into a direct sum of subspaces
labelled by integer levels $k_I$ such that $e_I$ and $f_I$ are at level 1 and $-1$, respectively, and all the other Chevalley generators at level zero.
From all these different $\mathbb{Z}$-gradings of $\cB$, one for each simple root $\beta_I$,
we can obtain a single one,
for which the levels are given by
$\ell=\sum_I v_I k_I$, where $v_I$ is an integer assigned to each simple root $\beta_I$. Following \cite{Henneaux:2010ys,Kleinschmidt:2013em} we call $v_I$ the V-degree of $\beta_I$.
When we extend $\gg$ to $\cB$ we let the simple roots $\beta_i$ of $\gg$ have V-degree zero, while the additional ones, corresponding to irreducible representations of generating forms, have positive V-degrees given by the form degrees of the corresponding potential forms.

As shown in \cite{HenryLabordere:2002dk,Henneaux:2010ys} it is always possible 
to choose the additional entries in the Cartan matrix, and the V-degrees of the additional simple roots, such that
$\mathfrak{f}_\ell=\cB_\ell$ for all $\ell>0$, and thus $\mathfrak{f}$ is the subalgebra of $\cB$ corresponding to positive levels.
While, as described above, the V-degrees and the additional off-diagonal entries of type $B_{0i}=B_{i0}$ are directly given by the degrees of the generating forms and the Dynkin labels of the corresponding irreducible representations,
it is less trivial to choose the additional diagonal entries, and the off-diagonal entries corresponding to pairs of additional simple roots, such that the additional Serre relations precisely correspond to the supersymmetry constraint on the representations $\cR_\ell$.
In fact, the choice is not always unique, but there is always a distinguished choice from the point of view of
oxidation, in the sense that the Borcherds superalgebra
relevant for the $D$-dimensional theory can be embedded into
the one relevant for the $(D-1)$-dimensional theory
\cite{Kleinschmidt:2013em}.

The results of the considerations described here are presented in Table \ref{maxtable} (for maximal supergravity) and Table \ref{halftable} (for half-maximal supergravity with $|n|\leq 1$).
Instead of displaying the Cartan matrices themselves we have given the corresponding V-diagrams,
which contain the same information in a much more compact way, and at the same time the information about the V-degrees. 
A V-diagram of a Borcherds superalgebra (or,
more generally, a contragredient Lie superalgebra)
$\cB$ is a Dynkin diagram of a Kac-Moody algebra $\cA$ of the same rank as $\cB$,
where node $I$ is labelled by the corresponding V-degree $v_I$ if $v_I \neq 0$.
From the Dynkin diagram one first obtains the Cartan matrix $A_{IJ}$ of $\cA$
and then, taking the V-degrees into account, the V-diagram gives the Cartan matrix $B_{IJ}$ of $\cB$ by
\begin{align} \label{fromatob0}
B_{IJ}=A_{IJ}-w(v_I,v_J)\ ,
\end{align}
where $w$ is a symmetric map $w:\mathbb{Z} \times \mathbb{Z} \to \mathbb{Z}$ defined by $w(a,b)=a(b+1)$ for
$0 \leq a \leq b$ and $w(-a,b)=w(a,-b)=-w(a,b)$.
A more general discussion of V-degrees and V-diagrams can be found in appendix \ref{borcherdsapp}.
In the following two subsections we will instead give explicit examples of $\cA$ and $\cB$
in some cases.

Before going into details about the 
Borcherds superalgebras relevant for (half-)maximal supergravity in different dimensions
we mention that, besides the subsequent embeddings of them
into each other, each of them can also be extended to a 
Borcherds superalgebra $\cD$ which is the same for different $D$ \cite{Kleinschmidt:2013em}.
For maximal supergravity it has the V-diagram
\begin{align} \label{e11-extension}
\begin{picture}(160,40)(-5,-10)
\put(0,0){\circle{5}}
\put(2.5,0){\line(1,0){10}}
\put(30,2.5){\line(0,1){10}}
\put(30,15){\circle{5}}
\put(15,0){\circle{5}}
\put(30,0){\circle{5}}
\put(17.5,0){\line(1,0){10}}
\put(32.5,0){\line(1,0){10}}
\put(47.5,0){\line(1,0){10}}
\put(62.5,0){\line(1,0){10}}
\put(77.5,0){\line(1,0){10}}
\put(92.5,0){\line(1,0){10}}
\put(107.5,0){\line(1,0){10}}
\put(122.5,0){\line(1,0){10}}
\put(137.5,0){\line(1,0){10}}
\put(45,0){\circle{5}}
\put(60,0){\circle{5}}
\put(75,0){\circle{5}}
\put(90,0){\circle{5}}
\put(105,0){\circle{5}}
\put(120,0){\circle{5}}
\put(135,0){\circle{5}}
\put(150,0){\circle{5}}
\put(148,-12){\scriptsize $1$}
\end{picture}
\end{align}
and can be obtained from the Kac-Moody algebra $E_{11}$ by adding an odd null root to the simple roots.
Like any other Borcherds superalgebra $\cD$ can also be considered as a contragredient Lie superalgebra,
for which the conditions on the Cartan matrix are less restrictive (see appendix \ref{borcherdsapp}).
As a consequence, 
it does not have a unique Cartan matrix, but different ones that can be obtained from each other by so-called generalised
Weyl transformations.
The different Cartan matrices are naturally associated to different dimensions $D$
so that the Cartan matrix of the $\cB$ subalgebra can be obtained by just removing rows and columns.
As can be seen in Table \ref{maxtable} for the maximal case
this amounts to removing a chain of nodes from the V-diagram, where the last one corresponds to a simple root with
V-degree $-1$. Continuing the rightmost column of Table~\ref{maxtable} by generalised
Weyl transformations, one would end up with the ``distinguished''
V-diagram (\ref{e11-extension}) of $\cD$ corresponding to $D=0$.

\newcommand{\elevenv}{
\begin{picture}(100,40)(-5,-10)
\put(28,3){\scriptsize $3$}
\put(30,15){\circle{5}}
\end{picture}}

\newcommand{\maxtenav}{
\begin{picture}(100,40)(-5,-10)
\put(0,0){\circle{5}}
\put(-2,-12){\scriptsize $1$}
\put(28,3){\scriptsize $2$}
\put(30,15){\circle{5}}
\end{picture}}

\newcommand{\maxtenbv}{
\begin{picture}(100,40)(-5,-10)
\put(2.5,0){\line(1,0){10}}
\put(15,0){\circle{5}}
\put(0,0){\circle{5}}
\put(13,-12){\scriptsize $2$}
\end{picture}}

\newcommand{\maxninev}{
\begin{picture}(100,40)(-5,-10)
\put(28,3){\scriptsize $1$}
\put(0,0){\circle{5}}
\put(15,0){\circle{5}}
\put(30,15){\circle{5}}
\put(2.5,0){\line(1,0){10}}
\put(13,-12){\scriptsize $1$}
\end{picture}}

\newcommand{\maxeightv}{
\begin{picture}(100,40)(-5,-10)
\put(0,0){\circle{5}}
\put(2.5,0){\line(1,0){10}}
\put(17.5,0){\line(1,0){10}}
\put(30,2.5){\line(0,1){10}}
\put(15,0){\circle{5}}
\put(30,0){\circle{5}}
\put(30,15){\circle{5}}
\put(28,-12){\scriptsize $1$}
\end{picture}}

\newcommand{\maxsevenv}{
\begin{picture}(115,40)(-5,-10)
\put(0,0){\circle{5}}
\put(2.5,0){\line(1,0){10}}
\put(30,2.5){\line(0,1){10}}
\put(30,15){\circle{5}}
\put(15,0){\circle{5}}
\put(30,0){\circle{5}}
\put(17.5,0){\line(1,0){10}}
\put(32.5,0){\line(1,0){10}}
\put(45,0){\circle{5}}
\put(43,-12){\scriptsize $1$}
\end{picture}}

\newcommand{\maxsixv}{
\begin{picture}(115,40)(-5,-10)
\put(0,0){\circle{5}}
\put(2.5,0){\line(1,0){10}}
\put(30,2.5){\line(0,1){10}}
\put(30,15){\circle{5}}
\put(15,0){\circle{5}}
\put(30,0){\circle{5}}
\put(17.5,0){\line(1,0){10}}
\put(32.5,0){\line(1,0){10}}
\put(47.5,0){\line(1,0){10}}
\put(45,0){\circle{5}}
\put(60,0){\circle{5}}
\put(58,-12){\scriptsize $1$}
\end{picture}}

\newcommand{\maxfivev}{
\begin{picture}(115,40)(-5,-10)
\put(0,0){\circle{5}}
\put(2.5,0){\line(1,0){10}}
\put(30,2.5){\line(0,1){10}}
\put(30,15){\circle{5}}
\put(15,0){\circle{5}}
\put(30,0){\circle{5}}
\put(17.5,0){\line(1,0){10}}
\put(32.5,0){\line(1,0){10}}
\put(47.5,0){\line(1,0){10}}
\put(62.5,0){\line(1,0){10}}
\put(45,0){\circle{5}}
\put(60,0){\circle{5}}
\put(75,0){\circle{5}}
\put(73,-12){\scriptsize $1$}
\end{picture}}

\newcommand{\maxfourv}{
\begin{picture}(115,40)(-5,-10)
\put(0,0){\circle{5}}
\put(2.5,0){\line(1,0){10}}
\put(30,2.5){\line(0,1){10}}
\put(30,15){\circle{5}}
\put(15,0){\circle{5}}
\put(30,0){\circle{5}}
\put(17.5,0){\line(1,0){10}}
\put(32.5,0){\line(1,0){10}}
\put(47.5,0){\line(1,0){10}}
\put(62.5,0){\line(1,0){10}}
\put(77.5,0){\line(1,0){10}}
\put(45,0){\circle{5}}
\put(60,0){\circle{5}}
\put(75,0){\circle{5}}
\put(90,0){\circle{5}}
\put(88,-12){\scriptsize $1$}
\end{picture}}

\newcommand{\maxthreev}{
\begin{picture}(115,40)(-5,-10)
\put(0,0){\circle{5}}
\put(2.5,0){\line(1,0){10}}
\put(30,2.5){\line(0,1){10}}
\put(30,15){\circle{5}}
\put(15,0){\circle{5}}
\put(30,0){\circle{5}}
\put(17.5,0){\line(1,0){10}}
\put(32.5,0){\line(1,0){10}}
\put(47.5,0){\line(1,0){10}}
\put(62.5,0){\line(1,0){10}}
\put(77.5,0){\line(1,0){10}}
\put(92.5,0){\line(1,0){10}}
\put(45,0){\circle{5}}
\put(60,0){\circle{5}}
\put(75,0){\circle{5}}
\put(90,0){\circle{5}}
\put(105,0){\circle{5}}
\put(103,-12){\scriptsize $1$}
\end{picture}}


\newcommand{\elevenvc}{
\begin{picture}(100,40)(-5,-10)
\put(0,0){\circle{5}}
\put(2.5,0){\line(1,0){10}}
\put(-4,-12){\scriptsize $-1$}
\put(35,13){\scriptsize $3$}
\put(15,0){\circle{5}}
\put(30,0){\circle{5}}
\put(47.5,0){\line(1,0){10}}
\put(17.5,0){\line(1,0){10}}
\put(32.5,0){\line(1,0){10}}
\put(62.5,0){\line(1,0){10}}
\put(2.7,1.2){\line(2,1){25}}
\put(77.5,0){\line(1,0){10}}
\put(92.5,0){\line(1,0){10}}
\put(107.5,0){\line(1,0){10}}
\put(122.5,0){\line(1,0){10}}
\put(137.5,0){\line(1,0){10}}
\put(45,0){\circle{5}}
\put(60,0){\circle{5}}
\put(30,15){\circle{5}}
\put(75,0){\circle{5}}
\put(90,0){\circle{5}}
\put(105,0){\circle{5}}
\put(120,0){\circle{5}}
\put(135,0){\circle{5}}
\put(150,0){\circle{5}}
\end{picture}}

\newcommand{\maxtenavc}{
\begin{picture}(100,40)(-5,-10)
\put(0,0){\circle{5}}
\put(2.5,0){\line(1,0){10}}
\put(-2,-12){\scriptsize $1$}
\put(11,-12){\scriptsize $-1$}
\put(35,13){\scriptsize $2$}
\put(15,0){\circle{5}}
\put(30,0){\circle{5}}
\put(47.5,0){\line(1,0){10}}
\put(17.5,0){\line(1,0){10}}
\put(32.5,0){\line(1,0){10}}
\put(62.5,0){\line(1,0){10}}
\put(17,2){\line(1,1){11}}
\put(77.5,0){\line(1,0){10}}
\put(92.5,0){\line(1,0){10}}
\put(107.5,0){\line(1,0){10}}
\put(122.5,0){\line(1,0){10}}
\put(137.5,0){\line(1,0){10}}
\put(45,0){\circle{5}}
\put(60,0){\circle{5}}
\put(30,15){\circle{5}}
\put(75,0){\circle{5}}
\put(90,0){\circle{5}}
\put(105,0){\circle{5}}
\put(120,0){\circle{5}}
\put(135,0){\circle{5}}
\put(150,0){\circle{5}}
\end{picture}}

\newcommand{\maxtenbvc}{
\begin{picture}(100,40)(-5,-10)
\put(0,0){\circle{5}}
\put(2.5,0){\line(1,0){10}}
\put(13,-12){\scriptsize $2$}
\put(35,13){\scriptsize $-1$}
\put(15,0){\circle{5}}
\put(30,0){\circle{5}}
\put(47.5,0){\line(1,0){10}}
\put(32.5,0){\line(1,0){10}}
\put(62.5,0){\line(1,0){10}}
\put(30,2.5){\line(0,1){10}}
\put(17,2){\line(1,1){11}}
\put(77.5,0){\line(1,0){10}}
\put(92.5,0){\line(1,0){10}}
\put(107.5,0){\line(1,0){10}}
\put(122.5,0){\line(1,0){10}}
\put(137.5,0){\line(1,0){10}}
\put(45,0){\circle{5}}
\put(60,0){\circle{5}}
\put(30,15){\circle{5}}
\put(75,0){\circle{5}}
\put(90,0){\circle{5}}
\put(105,0){\circle{5}}
\put(120,0){\circle{5}}
\put(135,0){\circle{5}}
\put(150,0){\circle{5}}
\end{picture}}

\newcommand{\maxninevc}{
\begin{picture}(100,40)(-5,-10)
\put(0,0){\circle{5}}
\put(2.5,0){\line(1,0){10}}
\put(15,0){\circle{5}}
\put(30,0){\circle{5}}
\put(47.5,0){\line(1,0){10}}
\put(17.5,0){\line(1,0){10}}
\put(32.5,0){\line(1,0){10}}
\put(62.5,0){\line(1,0){10}}
\put(30,2.5){\line(0,1){10}}
\put(77.5,0){\line(1,0){10}}
\put(92.5,0){\line(1,0){10}}
\put(107.5,0){\line(1,0){10}}
\put(122.5,0){\line(1,0){10}}
\put(137.5,0){\line(1,0){10}}
\put(45,0){\circle{5}}
\put(60,0){\circle{5}}
\put(30,15){\circle{5}}
\put(75,0){\circle{5}}
\put(90,0){\circle{5}}
\put(105,0){\circle{5}}
\put(120,0){\circle{5}}
\put(135,0){\circle{5}}
\put(150,0){\circle{5}}
\put(13,-12){\scriptsize $1$}
\put(26,-12){\scriptsize $-1$}
\put(35,13){\scriptsize $1$}
\end{picture}}

\newcommand{\maxeightvc}{
\begin{picture}(100,40)(-5,-10)
\put(0,0){\circle{5}}
\put(2.5,0){\line(1,0){10}}
\put(30,2.5){\line(0,1){10}}
\put(30,15){\circle{5}}
\put(15,0){\circle{5}}
\put(30,0){\circle{5}}
\put(47.5,0){\line(1,0){10}}
\put(17.5,0){\line(1,0){10}}
\put(32.5,0){\line(1,0){10}}
\put(62.5,0){\line(1,0){10}}
\put(77.5,0){\line(1,0){10}}
\put(92.5,0){\line(1,0){10}}
\put(107.5,0){\line(1,0){10}}
\put(122.5,0){\line(1,0){10}}
\put(137.5,0){\line(1,0){10}}
\put(45,0){\circle{5}}
\put(60,0){\circle{5}}
\put(75,0){\circle{5}}
\put(90,0){\circle{5}}
\put(105,0){\circle{5}}
\put(120,0){\circle{5}}
\put(135,0){\circle{5}}
\put(150,0){\circle{5}}
\put(28,-12){\scriptsize $1$}
\put(41,-12){\scriptsize $-1$}
\end{picture}}

\newcommand{\maxsevenvc}{
\begin{picture}(100,40)(-5,-10)
\put(0,0){\circle{5}}
\put(2.5,0){\line(1,0){10}}
\put(30,2.5){\line(0,1){10}}
\put(30,15){\circle{5}}
\put(15,0){\circle{5}}
\put(30,0){\circle{5}}
\put(47.5,0){\line(1,0){10}}
\put(17.5,0){\line(1,0){10}}
\put(32.5,0){\line(1,0){10}}
\put(62.5,0){\line(1,0){10}}
\put(77.5,0){\line(1,0){10}}
\put(92.5,0){\line(1,0){10}}
\put(107.5,0){\line(1,0){10}}
\put(122.5,0){\line(1,0){10}}
\put(137.5,0){\line(1,0){10}}
\put(45,0){\circle{5}}
\put(60,0){\circle{5}}
\put(75,0){\circle{5}}
\put(90,0){\circle{5}}
\put(105,0){\circle{5}}
\put(120,0){\circle{5}}
\put(135,0){\circle{5}}
\put(150,0){\circle{5}}
\put(43,-12){\scriptsize $1$}
\put(56,-12){\scriptsize $-1$}
\end{picture}}

\newcommand{\maxsixvc}{
\begin{picture}(100,40)(-5,-10)
\put(0,0){\circle{5}}
\put(2.5,0){\line(1,0){10}}
\put(30,2.5){\line(0,1){10}}
\put(30,15){\circle{5}}
\put(15,0){\circle{5}}
\put(30,0){\circle{5}}
\put(47.5,0){\line(1,0){10}}
\put(17.5,0){\line(1,0){10}}
\put(32.5,0){\line(1,0){10}}
\put(62.5,0){\line(1,0){10}}
\put(77.5,0){\line(1,0){10}}
\put(92.5,0){\line(1,0){10}}
\put(107.5,0){\line(1,0){10}}
\put(122.5,0){\line(1,0){10}}
\put(137.5,0){\line(1,0){10}}
\put(45,0){\circle{5}}
\put(60,0){\circle{5}}
\put(75,0){\circle{5}}
\put(90,0){\circle{5}}
\put(105,0){\circle{5}}
\put(120,0){\circle{5}}
\put(135,0){\circle{5}}
\put(150,0){\circle{5}}
\put(58,-12){\scriptsize $1$}
\put(71,-12){\scriptsize $-1$}
\end{picture}}

\newcommand{\maxfivevc}{
\begin{picture}(100,40)(-5,-10)
\put(0,0){\circle{5}}
\put(2.5,0){\line(1,0){10}}
\put(30,2.5){\line(0,1){10}}
\put(30,15){\circle{5}}
\put(15,0){\circle{5}}
\put(30,0){\circle{5}}
\put(17.5,0){\line(1,0){10}}
\put(32.5,0){\line(1,0){10}}
\put(47.5,0){\line(1,0){10}}
\put(62.5,0){\line(1,0){10}}
\put(77.5,0){\line(1,0){10}}
\put(92.5,0){\line(1,0){10}}
\put(107.5,0){\line(1,0){10}}
\put(122.5,0){\line(1,0){10}}
\put(137.5,0){\line(1,0){10}}
\put(45,0){\circle{5}}
\put(60,0){\circle{5}}
\put(75,0){\circle{5}}
\put(90,0){\circle{5}}
\put(105,0){\circle{5}}
\put(120,0){\circle{5}}
\put(135,0){\circle{5}}
\put(150,0){\circle{5}}
\put(73,-12){\scriptsize $1$}
\put(86,-12){\scriptsize $-1$}
\end{picture}}

\newcommand{\maxfourvc}{
\begin{picture}(100,40)(-5,-10)
\put(0,0){\circle{5}}
\put(2.5,0){\line(1,0){10}}
\put(30,2.5){\line(0,1){10}}
\put(30,15){\circle{5}}
\put(15,0){\circle{5}}
\put(30,0){\circle{5}}
\put(17.5,0){\line(1,0){10}}
\put(32.5,0){\line(1,0){10}}
\put(47.5,0){\line(1,0){10}}
\put(62.5,0){\line(1,0){10}}
\put(77.5,0){\line(1,0){10}}
\put(92.5,0){\line(1,0){10}}
\put(107.5,0){\line(1,0){10}}
\put(122.5,0){\line(1,0){10}}
\put(137.5,0){\line(1,0){10}}
\put(45,0){\circle{5}}
\put(60,0){\circle{5}}
\put(75,0){\circle{5}}
\put(90,0){\circle{5}}
\put(105,0){\circle{5}}
\put(120,0){\circle{5}}
\put(135,0){\circle{5}}
\put(150,0){\circle{5}}
\put(88,-12){\scriptsize $1$}
\put(101,-12){\scriptsize $-1$}
\end{picture}}

\newcommand{\maxthreevc}{
\begin{picture}(160,40)(-5,-10)
\put(0,0){\circle{5}}
\put(2.5,0){\line(1,0){10}}
\put(30,2.5){\line(0,1){10}}
\put(30,15){\circle{5}}
\put(15,0){\circle{5}}
\put(30,0){\circle{5}}
\put(17.5,0){\line(1,0){10}}
\put(32.5,0){\line(1,0){10}}
\put(47.5,0){\line(1,0){10}}
\put(62.5,0){\line(1,0){10}}
\put(77.5,0){\line(1,0){10}}
\put(92.5,0){\line(1,0){10}}
\put(107.5,0){\line(1,0){10}}
\put(122.5,0){\line(1,0){10}}
\put(137.5,0){\line(1,0){10}}
\put(45,0){\circle{5}}
\put(60,0){\circle{5}}
\put(75,0){\circle{5}}
\put(90,0){\circle{5}}
\put(105,0){\circle{5}}
\put(120,0){\circle{5}}
\put(135,0){\circle{5}}
\put(150,0){\circle{5}}
\put(103,-12){\scriptsize $1$}
\put(116,-12){\scriptsize $-1$}
\end{picture}}

\newcommand{\eeleven}{
\begin{picture}(145,35)(-5,-10)
\put(0,0){\circle{5}}
\put(2.5,0){\line(1,0){10}}
\put(30,2.5){\line(0,1){10}}
\put(30,15){\circle{5}}
\put(15,0){\circle{5}}
\put(30,0){\circle{5}}
\put(17.5,0){\line(1,0){10}}
\put(32.5,0){\line(1,0){10}}
\put(47.5,0){\line(1,0){10}}
\put(62.5,0){\line(1,0){10}}
\put(77.5,0){\line(1,0){10}}
\put(92.5,0){\line(1,0){10}}
\put(107.5,0){\line(1,0){10}}
\put(122.5,0){\line(1,0){10}}
\put(45,0){\circle{5}}
\put(60,0){\circle{5}}
\put(75,0){\circle{5}}
\put(90,0){\circle{5}}
\put(105,0){\circle{5}}
\put(120,0){\circle{5}}
\put(135,0){\circle{5}}
\end{picture}}

\newcommand{\tio}{
\begin{picture}(0,0)(-2.7,-14)
\put(0,0){10}
\end{picture}}

\setlength{\arraycolsep}{9.95pt}
\begin{table} 
{\renewcommand{\arraystretch}{1.5}
\begin{align*}
\begin{array}{|c|l|l|}
\hline
D & \text{V-diagram of $\mathcal{B}$}& \text{V-diagram of $\mathcal{D}$}\\
\hline 
11&\elevenv &\elevenvc\\ \hline
\tio\text{IIA}&\maxtenav &\maxtenavc\\ \hline
\tio\text{IIB}
&\maxtenbv &\maxtenbvc\\ \hline
9&\maxninev &\maxninevc\\ \hline
8& \maxeightv &\maxeightvc\\ \hline
7&\maxsevenv &\maxsevenvc\\ \hline
6&\maxsixv &\maxsixvc\\ \hline
5&\maxfivev &\maxfivevc\\ \hline
4&\maxfourv &\maxfourvc\\ \hline
3& \maxthreev &\maxthreevc\\ \hline
\end{array}
\end{align*}
}
\caption{\it V-diagrams of the Borcherds superalgebras $\cB$ and $\cD$ relevant for
maximal supergravity in $D$ dimensions. The Cartan matrices of the Borcherds superalgebras can be obtained from those of the
corresponding Kac-Moody algebras by (\ref{fromatob0}). Note that the Borcherds superalgebra $\cD$ does not depend on $D$, but there are different V-diagrams
of the same algebra $\cD$ corresponding to the various cases. On the other hand, the Borcherds superalgebras $\cB$
are different subalgebras of $\cD$, depending on $D$. (The rightmost column, continued to $D=0$, contains the same information as Figure 3 in \cite{Kleinschmidt:2013em}, but
there Dynkin diagrams were used instead of V-diagrams, with coloured nodes, multiple lines and some entries in the Cartan matrix written out explicitly to
avoid sign ambiguities.)
}\label{maxtable}
\end{table}

\newpage

\subsection{Maximal supergravity}

For maximal supergravity in $D$ dimensions with $3 \leq D \leq 7$ the generating forms are at level one and transform under a single
irreducible representation $\cR_1$ of the
simple duality group $E_{11-D}$.
Thus we add a simple root of V-degree one with a corresponding Chevalley generator
$e_0$, and since the V-degree is an odd integer, $e_0$ is an odd element in the resulting Borcherds superalgebra $\cB$.
The Cartan matrix $B_{IJ}$ of $\cB$ will have the form
\be
B_{IJ}= \left( \begin{array}{cc} B_{00} & B_{0i}\cr B_{i0} & B_{ij} \end{array} \right)
\label{3.1}
\ee
discussed above, where the Dynkin labels of $\cR_1$ (with the opposite sign) constitute the row $B_{0i}$ and the column $B_{i0}$, and $B_{ij}=a_{ij}$ is
the (symmetrised) Cartan matrix of $\gg$.
It remains to determine $B_{00}$. At  level two it is easy to show that $[e_0,e_0]$ is the lowest weight state for an irreducible representation which is not allowed by the supersymmetry constraint, \ie for which $(F_{0,2})^2$ is not $t_0$-exact. We therefore have the constraint $[e_0,e_0]=0$, which in turn, considered
as a Serre relation, leads to $B_{00}=0$. The V-diagram of $\cB$ is the Dynkin diagram of $\ge_{11-D+1}$, with the node added  
to the Dynkin diagram of $\ge_{11-D}$ labelled by the
V-degree $v_0=0$.

The $D=8$ case is similar to the ones we have just discussed, the only difference being that the duality group, $SL(2,\bbR) \xz SL(3,\bbR)$, is not simple. The generating forms are still at level one and transform under the product of the doublet and triplet representations of the two simple subgroups. By similar reasoning to that used in the above cases we find that the Cartan matrix of $\cB$ in this case is
\be
B_{IJ}= \left( \begin{array}{rrrr} 0&-1& -1&0\cr
-1&2&0&0\cr
-1&0&2&-1\cr
0&0&-1&2\end{array} \right)
\ee
where the $2\xz 2$ matrix in the bottom right corner is the Cartan matrix of $\gs\gl(3)$. The V-diagram of $\cB$ is the Dynkin diagram of 
$\gs\gl(5)$
corresponding to the Cartan matrix
\be
A_{IJ}= \left( \begin{array}{rrrr} 2&-1& -1&0\cr
-1&2&0&0\cr
-1&0&2&-1\cr
0&0&-1&2\end{array} \right)
\ee
with the node corresponding to the 
first row and column
labelled by the V-degree $v_0=1$.

The situation is different for $D\geq9$. For $D=9$ we have two irreducible representations of generating forms, both at level one, which are the doublet and singlet representations of $\gs\gl(2)$. This suggests that we should introduce two odd algebra generators, $e_0, e_{0'}$, that transform under the doublet and singlet representations of $\gs\gl(2)$, together with a corresponding negative pair, $f_0, f_{0'}$. Denoting the generators of $\gs\gl(2)$ by $\{e_1,f_1,h_1\}$, the corresponding
V-degrees are $v_0=v_{0'}=1$ and $v_1=0$. The duality algebra is the direct sum of $\gs\gl(2)$ and a one-dimensional Lie algebra
spanned by a linear combination of $h_0$ and $h_{0'}$. This appearance of a singlet at level zero is thus a consequence of
the fact that there are two irreducible representations of generating forms, as we will see also in the half-maximal case below.
The supersymmetry constraint at level two implies that
both $[e_0,e_0]$ and $[e_{0'},e_{0'}]$ must vanish, whereas $[e_0,e_{0'}]\neq0$.
As a consequence we conclude that both $B_{00}$ and $B_{0'0'}$ must be zero, but there is no constraint on the off-diagonal entries $B_{00'}$ and $B_{0'0}$
more than that they should be equal to each other and negative. As explained in \cite{Kleinschmidt:2013em} the distinguished choice from the point of view of oxidation is $B_{00'}=B_{0'0}=-2$, which gives the Cartan matrix 
\be
B_{IJ}= \left( \begin{array}{rrr} 0& -2 & 0\cr
-2 & 0 & -1\cr 0& -1 & 2 \end{array} \right)
\ee
where $I=(0',0,1)$.
This can be written as $B_{IJ}=A_{IJ}-w(v_I,v_J)$, where
\be
A_{IJ}= \left( \begin{array}{rrr} 2& 0 & 0\cr
0 & 2 & -1\cr 0& -1 & 2 \end{array} \right)\ ,
\ee
and thus the V-diagram of $\cB$ is the Dynkin diagram
corresponding to this Cartan matrix $A_{IJ}$,
with the first two nodes labelled by the V-degrees $v_0=v_{0'}=1$.

In IIB supergravity the generating forms are at level two in the doublet of $\gs\gl(2)$. This means that we need two sets of algebraic generators, $\{e_I, f_I, h_I\}$, $I=(0,1)$, where the generator $e_0$ is associated with the level-two potentials and the index 1 with the $\gs\gl(2)$ subalgebra. 
Thus the corresponding V-degrees are $v_0=2$ and $v_1=0$.
The algebraic elements for the level-two forms are $e_0$ and $[e_1,e_0]$, and we must have $(\ad\, e_1)^2 (e_0)=0$ so that we indeed have a doublet at this level. Clearly we also have $[h_0,e_1]=-e_1$ since $e_0$ is taken to be a lowest weight vector, and the Cartan matrix therefore has the form
\be
B_{IJ}= \left( \begin{array}{rr} B_{00}&-1\cr
-1&2 \end{array} \right).
\ee
The representations that the forms transform under are the same whether one chooses $B_{00}$ to be zero or any negative number.
In \cite{HenryLabordere:2002dk,Henneaux:2010ys} it was assumed that $B_{00}=0$ which gives the Cartan matrix for the Slansky algebra discussed previously.
In \cite{Greitz:2011da}  it was argued that $B_{00}=0$ was desirable from a superspace point of view because in this case $\{e_0,f_0,h_0\}$ generate a Heisenberg algebra, whereas if one were to take $B_{00}$ to be negative, there would be a second $\gs\gl(2)$ subalgebra under which the whole algebra would split into infinite-dimensional representations.
It was argued in \cite{Greitz:2011da}  that this would be unnatural because these representations would necessarily involve different form degrees, but it is not clear that this is a necessary restriction because the second $\gs\gl(2)$ is not a symmetry of the theory by itself.
In \cite{Kleinschmidt:2013em} it was argued that $B_{00}$ should be chosen to be $-4$ as this is required by oxidation,
and with this distinguished choice we can write
$B_{IJ}=A_{IJ}-w(v_I,v_J)$, where $A_{IJ}$ is the Cartan matrix of $\gs\gl(3)$.
The Dynkin diagram of $\sl(3)$, with the node corresponding to $e_0$ labelled by the V-degree $v_0=2$, is the V-diagram of $\cB$.

In IIA supergravity there is no duality group, but the generating forms, which are at level one and two, give rise to a Borcherds superalgebra $\cB$ by themselves. 
As in $D=9$ and IIB supergravity, $\cB$ is not uniquely determined by the spectrum of forms,
but there is again a distinguished choice given by the V-diagram
shown in {Table~\ref{maxtable}}. Finally, in $D=11$, the three-form potential gives rise to the finite-dimensional Lie superalgebra
$\mathfrak{osp}(1|2)$. The single entry $B_{00}$ in the Cartan matrix can be any nonzero integer, but since the corresponding V-degree is $v_0=3$, it is natural to choose $B_{00}=2-w(v_0,v_0)=-10$.

\newcommand{\streck}{
\begin{picture}(85,40)(-5,-10)
\put(33,5){$-$}
\end{picture}}

\newcommand{\halften}{
\begin{picture}(100,40)(-5,-10)
\thicklines
\put(0,0){\circle*{5}}
\put(2.5,0){\line(1,0){10}}
\put(15,0){\circle*{5}}
\put(-2,-12){\scriptsize $2$}
\put(13,-12){\scriptsize $6$}
\end{picture}}

\newcommand{\halfnine}{
\begin{picture}(100,40)(-5,-10)
\thicklines
\put(0,0){\circle*{5}}
\put(2.5,0){\line(1,0){10}}
\put(15,2.5){\line(0,1){10}}
\put(15,15){\circle*{5}}
\put(15,0){\circle*{5}}
\put(1,1){\line(1,1){12}}
\put(-2,-12){\scriptsize $1$}
\put(13,-12){\scriptsize $5$}
\put(20,13){\scriptsize $1$}
\end{picture}}

\newcommand{\halfeight}{
\begin{picture}(100,40)(-5,-10)
\thicklines
\put(0,0){\circle{5}}
\put(2.5,0){\line(1,0){10}}
\put(15,2.5){\line(0,1){10}}
\put(15,15){\circle{5}}
\put(15,0){\circle*{5}}
\put(30,0){\circle*{5}}
\put(17.5,0){\line(1,0){10}}
\put(13,-12){\scriptsize $1$}
\put(28,-12){\scriptsize $4$}
\end{picture}}

\newcommand{\halfseven}{
\begin{picture}(100,40)(-5,-10)
\thicklines
\put(0,0){\circle{5}}
\put(2.5,0){\line(1,0){10}}
\put(15,2.5){\line(0,1){10}}
\put(15,15){\circle{5}}
\put(15,0){\circle{5}}
\put(17.5,0){\line(1,0){10}}
\put(32.5,0){\line(1,0){10}}
\put(30,0){\circle*{5}}
\put(45,0){\circle*{5}}
\put(28,-12){\scriptsize $1$}
\put(43,-12){\scriptsize $3$}
\end{picture}}

\newcommand{\halfsixa}{
\begin{picture}(100,40)(-5,-10)
\thicklines
\put(0,0){\circle{5}}
\put(2.5,0){\line(1,0){10}}
\put(15,2.5){\line(0,1){10}}
\put(15,15){\circle{5}}
\put(15,0){\circle{5}}
\put(30,0){\circle{5}}
\put(47.5,0){\line(1,0){10}}
\put(17.5,0){\line(1,0){10}}
\put(32.5,0){\line(1,0){10}}
\put(30,0){\circle{5}}
\put(45,0){\circle*{5}}
\put(60,0){\circle*{5}}
\put(43,-12){\scriptsize $1$}
\put(58,-12){\scriptsize $2$}
\end{picture}}

\newcommand{\halfsixb}{
\begin{picture}(100,40)(-5,-10)
\thicklines
\put(0,0){\circle{5}}
\put(2.5,0){\line(1,0){10}}
\put(15,2.5){\line(0,1){10}}
\put(15,15){\circle{5}}
\put(15,0){\circle{5}}
\put(30,0){\circle{5}}
\put(17.5,0){\line(1,0){10}}
\put(32.5,0){\line(1,0){10}}
\put(47.5,0){\line(1,0){10}}
\put(62.5,0){\line(1,0){10}}
\put(30,0){\circle{5}}
\put(45,0){\circle{5}}
\put(60,0){\circle{5}}
\put(75,0){\circle*{5}}
\put(73,-12){\scriptsize $2$}
\end{picture}}

\newcommand{\halffive}{
\begin{picture}(100,40)(-5,-10)
\thicklines
\put(0,0){\circle{5}}
\put(2.5,0){\line(1,0){10}}
\put(15,2.5){\line(0,1){10}}
\put(15,15){\circle{5}}
\put(15,0){\circle{5}}
\put(30,0){\circle{5}}
\put(17.5,0){\line(1,0){10}}
\put(32.5,0){\line(1,0){10}}
\put(47.5,0){\line(1,0){10}}
\put(61,1){\line(1,1){12}}
\put(30,0){\circle{5}}
\put(45,0){\circle{5}}
\put(60,0){\circle*{5}}
\put(75,15){\circle*{5}}
\put(58,-12){\scriptsize $1$}
\put(80,13){\scriptsize $1$}
\end{picture}}

\newcommand{\halffour}{
\begin{picture}(100,40)(-5,-10)
\thicklines
\put(0,0){\circle{5}}
\put(2.5,0){\line(1,0){10}}
\put(15,2.5){\line(0,1){10}}
\put(15,15){\circle{5}}
\put(15,0){\circle{5}}
\put(30,0){\circle{5}}
\put(17.5,0){\line(1,0){10}}
\put(32.5,0){\line(1,0){10}}
\put(47.5,0){\line(1,0){10}}
\put(62.5,0){\line(1,0){10}}
\put(75,2.5){\line(0,1){10}}
\put(30,0){\circle{5}}
\put(45,0){\circle{5}}
\put(60,0){\circle{5}}
\put(75,15){\circle{5}}
\put(75,0){\circle*{5}}
\put(73,-12){\scriptsize $1$}
\end{picture}}

\newcommand{\halfthree}{
\begin{picture}(100,40)(-5,-10)
\thicklines
\put(0,0){\circle{5}}
\put(2.5,0){\line(1,0){10}}
\put(15,2.5){\line(0,1){10}}
\put(15,15){\circle{5}}
\put(15,0){\circle{5}}
\put(30,0){\circle{5}}
\put(17.5,0){\line(1,0){10}}
\put(32.5,0){\line(1,0){10}}
\put(47.5,0){\line(1,0){10}}
\put(62.5,0){\line(1,0){10}}
\put(75,2.5){\line(0,1){10}}
\put(77.5,0){\line(1,0){10}}
\put(30,0){\circle{5}}
\put(45,0){\circle{5}}
\put(60,0){\circle{5}}
\put(75,15){\circle{5}}
\put(75,0){\circle{5}}
\put(90,0){\circle*{5}}
\put(88,-12){\scriptsize $1$}
\end{picture}}


\newcommand{\halftenv}{
\begin{picture}(100,40)(-5,-10)
\put(15,15){\circle{5}}
\put(13,3){\scriptsize $2$}
\put(28,3){\scriptsize $6$}
\put(30,15){\circle{5}}
\end{picture}}

\newcommand{\halfninev}{
\begin{picture}(100,40)(-5,-10)
\put(13,3){\scriptsize $1$}
\put(28,3){\scriptsize $5$}
\put(0,0){\circle{5}}
\put(15,15){\circle{5}}
\put(30,15){\circle{5}}
\put(-2,-12){\scriptsize $1$}
\end{picture}}

\newcommand{\halfeightv}{
\begin{picture}(100,40)(-5,-10)
\put(0,0){\circle{5}}
\put(2.5,0){\line(1,0){10}}
\put(15,2.5){\line(0,1){10}}
\put(15,15){\circle{5}}
\put(15,0){\circle{5}}
\put(30,15){\circle{5}}
\put(13,-12){\scriptsize $1$}
\put(28,3){\scriptsize $4$}
\end{picture}}

\newcommand{\halfsevenv}{
\begin{picture}(100,40)(-5,-10)
\put(0,0){\circle{5}}
\put(2.5,0){\line(1,0){10}}
\put(15,2.5){\line(0,1){10}}
\put(15,15){\circle{5}}
\put(15,0){\circle{5}}
\put(17.5,0){\line(1,0){10}}
\put(30,0){\circle{5}}
\put(45,15){\circle{5}}
\put(28,-12){\scriptsize $1$}
\put(43,3){\scriptsize $3$}
\end{picture}}

\newcommand{\halfsixav}{
\begin{picture}(100,40)(-5,-10)
\put(0,0){\circle{5}}
\put(2.5,0){\line(1,0){10}}
\put(15,2.5){\line(0,1){10}}
\put(15,15){\circle{5}}
\put(15,0){\circle{5}}
\put(30,0){\circle{5}}
\put(17.5,0){\line(1,0){10}}
\put(32.5,0){\line(1,0){10}}
\put(45,0){\circle{5}}
\put(60,15){\circle{5}}
\put(43,-12){\scriptsize $1$}
\put(58,3){\scriptsize $2$}
\end{picture}}

\newcommand{\halfsixbv}{
\begin{picture}(100,40)(-5,-10)
\put(0,0){\circle{5}}
\put(2.5,0){\line(1,0){10}}
\put(15,2.5){\line(0,1){10}}
\put(15,15){\circle{5}}
\put(15,0){\circle{5}}
\put(30,0){\circle{5}}
\put(17.5,0){\line(1,0){10}}
\put(32.5,0){\line(1,0){10}}
\put(47.5,0){\line(1,0){10}}
\put(45,0){\circle{5}}
\put(60,0){\circle{5}}
\put(58,-12){\scriptsize $2$}
\end{picture}}

\newcommand{\halffivev}{
\begin{picture}(100,40)(-5,-10)
\put(0,0){\circle{5}}
\put(2.5,0){\line(1,0){10}}
\put(15,2.5){\line(0,1){10}}
\put(15,15){\circle{5}}
\put(15,0){\circle{5}}
\put(30,0){\circle{5}}
\put(17.5,0){\line(1,0){10}}
\put(32.5,0){\line(1,0){10}}
\put(47.5,0){\line(1,0){10}}
\put(45,0){\circle{5}}
\put(60,0){\circle{5}}
\put(75,15){\circle{5}}
\put(58,-12){\scriptsize $1$}
\put(73,3){\scriptsize $1$}
\end{picture}}

\newcommand{\halffourv}{
\begin{picture}(100,40)(-5,-10)
\put(0,0){\circle{5}}
\put(2.5,0){\line(1,0){10}}
\put(15,2.5){\line(0,1){10}}
\put(15,15){\circle{5}}
\put(15,0){\circle{5}}
\put(30,0){\circle{5}}
\put(17.5,0){\line(1,0){10}}
\put(32.5,0){\line(1,0){10}}
\put(47.5,0){\line(1,0){10}}
\put(62.5,0){\line(1,0){10}}
\put(75,2.5){\line(0,1){10}}
\put(45,0){\circle{5}}
\put(60,0){\circle{5}}
\put(75,15){\circle{5}}
\put(75,0){\circle{5}}
\put(73,-12){\scriptsize $1$}
\end{picture}}

\newcommand{\halfthreev}{
\begin{picture}(100,40)(-5,-10)
\put(0,0){\circle{5}}
\put(2.5,0){\line(1,0){10}}
\put(15,2.5){\line(0,1){10}}
\put(15,15){\circle{5}}
\put(15,0){\circle{5}}
\put(30,0){\circle{5}}
\put(17.5,0){\line(1,0){10}}
\put(32.5,0){\line(1,0){10}}
\put(47.5,0){\line(1,0){10}}
\put(62.5,0){\line(1,0){10}}
\put(75,2.5){\line(0,1){10}}
\put(77.5,0){\line(1,0){10}}
\put(45,0){\circle{5}}
\put(60,0){\circle{5}}
\put(75,15){\circle{5}}
\put(75,0){\circle{5}}
\put(90,0){\circle{5}}
\put(88,-12){\scriptsize $1$}
\end{picture}}


\newcommand{\halftenvm}{
\begin{picture}(85,40)(10,-10)
\put(15,15){\circle*{5}}
\put(13,3){\scriptsize $2$}
\put(28,3){\scriptsize $6$}
\put(30,15){\circle{5}}
\end{picture}}

\newcommand{\halfninevm}{
\begin{picture}(85,40)(10,-10)
\put(13,3){\scriptsize $1$}
\put(28,3){\scriptsize $5$}
\put(15,15){\circle*{5}}
\put(30,15){\circle{5}}
\end{picture}}

\newcommand{\halfeightvm}{
\begin{picture}(85,40)(10,-10)
\put(14.3,2.5){\line(0,1){10}}
\put(15.7,2.5){\line(0,1){10}}
\put(10.75,10.5){$\times$}
\put(15,15){\circle*{5}}
\put(15,0){\circle{5}}
\put(30,15){\circle{5}}
\put(13,-12){\scriptsize $1$}
\put(28,3){\scriptsize $4$}
\end{picture}}

\newcommand{\halfsevenvm}{
\begin{picture}(85,40)(10,-10)
\put(14.3,2.5){\line(0,1){10}}
\put(15.7,2.5){\line(0,1){10}}
\put(10.75,10.5){$\times$}
\put(15,15){\circle*{5}}
\put(15,0){\circle{5}}
\put(17.5,0){\line(1,0){10}}
\put(30,0){\circle{5}}
\put(45,15){\circle{5}}
\put(28,-12){\scriptsize $1$}
\put(43,3){\scriptsize $3$}
\end{picture}}

\newcommand{\halfsixavm}{
\begin{picture}(85,40)(10,-10)
\put(14.3,2.5){\line(0,1){10}}
\put(15.7,2.5){\line(0,1){10}}
\put(10.75,10.5){$\times$}
\put(15,15){\circle*{5}}
\put(15,0){\circle{5}}
\put(30,0){\circle{5}}
\put(17.5,0){\line(1,0){10}}
\put(32.5,0){\line(1,0){10}}
\put(45,0){\circle{5}}
\put(60,15){\circle{5}}
\put(43,-12){\scriptsize $1$}
\put(58,3){\scriptsize $2$}
\end{picture}}

\newcommand{\halfsixbvm}{
\begin{picture}(85,40)(10,-10)
\put(14.3,2.5){\line(0,1){10}}
\put(15.7,2.5){\line(0,1){10}}
\put(10.75,10.5){$\times$}
\put(15,15){\circle*{5}}
\put(15,0){\circle{5}}
\put(30,0){\circle{5}}
\put(17.5,0){\line(1,0){10}}
\put(32.5,0){\line(1,0){10}}
\put(47.5,0){\line(1,0){10}}
\put(45,0){\circle{5}}
\put(60,0){\circle{5}}
\put(58,-12){\scriptsize $2$}
\end{picture}}

\newcommand{\halffivevm}{
\begin{picture}(85,40)(10,-10)
\put(14.3,2.5){\line(0,1){10}}
\put(15.7,2.5){\line(0,1){10}}
\put(10.75,10.5){$\times$}
\put(15,15){\circle*{5}}
\put(15,0){\circle{5}}
\put(30,0){\circle{5}}
\put(17.5,0){\line(1,0){10}}
\put(32.5,0){\line(1,0){10}}
\put(47.5,0){\line(1,0){10}}
\put(45,0){\circle{5}}
\put(60,0){\circle{5}}
\put(75,15){\circle{5}}
\put(58,-12){\scriptsize $1$}
\put(73,3){\scriptsize $1$}
\end{picture}}

\newcommand{\halffourvm}{
\begin{picture}(85,40)(10,-10)
\put(14.3,2.5){\line(0,1){10}}
\put(15.7,2.5){\line(0,1){10}}
\put(10.75,10.5){$\times$}
\put(15,15){\circle*{5}}
\put(15,0){\circle{5}}
\put(30,0){\circle{5}}
\put(17.5,0){\line(1,0){10}}
\put(32.5,0){\line(1,0){10}}
\put(47.5,0){\line(1,0){10}}
\put(62.5,0){\line(1,0){10}}
\put(75,2.5){\line(0,1){10}}
\put(45,0){\circle{5}}
\put(60,0){\circle{5}}
\put(75,15){\circle{5}}
\put(75,0){\circle{5}}
\put(73,-12){\scriptsize $1$}
\end{picture}}

\newcommand{\halfthreevm}{
\begin{picture}(85,40)(10,-10)
\put(14.3,2.5){\line(0,1){10}}
\put(15.7,2.5){\line(0,1){10}}
\put(10.75,10.5){$\times$}
\put(15,15){\circle*{5}}
\put(15,0){\circle{5}}
\put(30,0){\circle{5}}
\put(17.5,0){\line(1,0){10}}
\put(32.5,0){\line(1,0){10}}
\put(47.5,0){\line(1,0){10}}
\put(62.5,0){\line(1,0){10}}
\put(75,2.5){\line(0,1){10}}
\put(77.5,0){\line(1,0){10}}
\put(45,0){\circle{5}}
\put(60,0){\circle{5}}
\put(75,15){\circle{5}}
\put(75,0){\circle{5}}
\put(90,0){\circle{5}}
\put(88,-12){\scriptsize $1$}
\end{picture}}


\newcommand{\halftenvp}{
\begin{picture}(100,40)(-5,-10)
\put(0,15){\circle*{5}}
\put(-2,3){\scriptsize $1$}
\put(13,3){\scriptsize $6$}
\put(15,15){\circle{5}}
\end{picture}}

\newcommand{\halfninevp}{
\begin{picture}(100,40)(-5,-10)
\put(-0.7,2.5){\line(0,1){10}}
\put(0.7,2.5){\line(0,1){10}}
\put(-4.25,10.5){$\times$}
\put(13,3){\scriptsize $5$}
\put(0,0){\circle{5}}
\put(0,15){\circle*{5}}
\put(15,15){\circle{5}}
\put(-2,-12){\scriptsize $1$}
\end{picture}}

\newcommand{\halfeightvp}{
\begin{picture}(100,40)(-5,-10)
\put(0,0){\circle{5}}
\put(2.5,0){\line(1,0){10}}
\put(-0.7,2.5){\line(0,1){10}}
\put(0.7,2.5){\line(0,1){10}}
\put(-4.25,10.5){$\times$}
\put(0,15){\circle*{5}}
\put(15,0){\circle{5}}
\put(30,15){\circle{5}}
\put(13,-12){\scriptsize $1$}
\put(28,3){\scriptsize $4$}
\end{picture}}

\newcommand{\halfsevenvp}{
\begin{picture}(100,40)(-5,-10)
\put(0,0){\circle{5}}
\put(2.5,0){\line(1,0){10}}
\put(-0.7,2.5){\line(0,1){10}}
\put(0.7,2.5){\line(0,1){10}}
\put(-4.25,10.5){$\times$}
\put(0,15){\circle*{5}}
\put(15,0){\circle{5}}
\put(17.5,0){\line(1,0){10}}
\put(30,0){\circle{5}}
\put(45,15){\circle{5}}
\put(28,-12){\scriptsize $1$}
\put(43,3){\scriptsize $3$}
\end{picture}}

\newcommand{\halfsixavp}{
\begin{picture}(100,40)(-5,-10)
\put(0,0){\circle{5}}
\put(2.5,0){\line(1,0){10}}
\put(-0.7,2.5){\line(0,1){10}}
\put(0.7,2.5){\line(0,1){10}}
\put(-4.25,10.5){$\times$}
\put(0,15){\circle*{5}}
\put(15,0){\circle{5}}
\put(30,0){\circle{5}}
\put(17.5,0){\line(1,0){10}}
\put(32.5,0){\line(1,0){10}}
\put(45,0){\circle{5}}
\put(60,15){\circle{5}}
\put(43,-12){\scriptsize $1$}
\put(58,3){\scriptsize $2$}
\end{picture}}

\newcommand{\halfsixbvp}{
\begin{picture}(100,40)(-5,-10)
\put(0,0){\circle{5}}
\put(2.5,0){\line(1,0){10}}
\put(-0.7,2.5){\line(0,1){10}}
\put(0.7,2.5){\line(0,1){10}}
\put(-4.25,10.5){$\times$}
\put(0,15){\circle*{5}}
\put(15,0){\circle{5}}
\put(30,0){\circle{5}}
\put(17.5,0){\line(1,0){10}}
\put(32.5,0){\line(1,0){10}}
\put(47.5,0){\line(1,0){10}}
\put(45,0){\circle{5}}
\put(60,0){\circle{5}}
\put(58,-12){\scriptsize $2$}
\end{picture}}

\newcommand{\halffivevp}{
\begin{picture}(100,40)(-5,-10)
\put(0,0){\circle{5}}
\put(2.5,0){\line(1,0){10}}
\put(-0.7,2.5){\line(0,1){10}}
\put(0.7,2.5){\line(0,1){10}}
\put(-4.25,10.5){$\times$}
\put(0,15){\circle*{5}}
\put(15,0){\circle{5}}
\put(30,0){\circle{5}}
\put(17.5,0){\line(1,0){10}}
\put(32.5,0){\line(1,0){10}}
\put(47.5,0){\line(1,0){10}}
\put(45,0){\circle{5}}
\put(60,0){\circle{5}}
\put(75,15){\circle{5}}
\put(58,-12){\scriptsize $1$}
\put(73,3){\scriptsize $1$}
\end{picture}}

\newcommand{\halffourvp}{
\begin{picture}(100,40)(-5,-10)
\put(0,0){\circle{5}}
\put(2.5,0){\line(1,0){10}}
\put(-0.7,2.5){\line(0,1){10}}
\put(0.7,2.5){\line(0,1){10}}
\put(-4.25,10.5){$\times$}
\put(0,15){\circle*{5}}
\put(15,0){\circle{5}}
\put(30,0){\circle{5}}
\put(17.5,0){\line(1,0){10}}
\put(32.5,0){\line(1,0){10}}
\put(47.5,0){\line(1,0){10}}
\put(62.5,0){\line(1,0){10}}
\put(75,2.5){\line(0,1){10}}
\put(45,0){\circle{5}}
\put(60,0){\circle{5}}
\put(75,15){\circle{5}}
\put(75,0){\circle{5}}
\put(73,-12){\scriptsize $1$}
\end{picture}}

\newcommand{\halfthreevp}{
\begin{picture}(100,40)(-5,-10)
\put(0,0){\circle{5}}
\put(2.5,0){\line(1,0){10}}
\put(-0.7,2.5){\line(0,1){10}}
\put(0.7,2.5){\line(0,1){10}}
\put(-4.25,10.5){$\times$}
\put(0,15){\circle*{5}}
\put(15,0){\circle{5}}
\put(30,0){\circle{5}}
\put(17.5,0){\line(1,0){10}}
\put(32.5,0){\line(1,0){10}}
\put(47.5,0){\line(1,0){10}}
\put(62.5,0){\line(1,0){10}}
\put(75,2.5){\line(0,1){10}}
\put(77.5,0){\line(1,0){10}}
\put(45,0){\circle{5}}
\put(60,0){\circle{5}}
\put(75,15){\circle{5}}
\put(75,0){\circle{5}}
\put(90,0){\circle{5}}
\put(88,-12){\scriptsize $1$}
\end{picture}}


\newcommand{\halftenvc}{
\begin{picture}(100,40)(-5,-10)
\put(0,0){\circle{5}}
\put(2.5,0){\line(1,0){10}}
\put(15,15){\circle{5}}
\put(20,13){\scriptsize $2$}
\put(-2,-12){\scriptsize $-1$}
\put(35,13){\scriptsize $6$}
\put(15,0){\circle{5}}
\put(30,0){\circle{5}}
\put(47.5,0){\line(1,0){10}}
\put(17.5,0){\line(1,0){10}}
\put(32.5,0){\line(1,0){10}}
\put(62.5,0){\line(1,0){10}}
\put(2,2){\line(1,1){11}}
\put(2.7,1.2){\line(2,1){25}}
\put(77.5,0){\line(1,0){10}}
\put(92.5,0){\line(1,0){10}}
\put(107.5,0){\line(1,0){10}}
\put(122.5,0){\line(1,0){10}}
\put(45,0){\circle{5}}
\put(60,0){\circle{5}}
\put(30,15){\circle{5}}
\put(75,0){\circle{5}}
\put(90,0){\circle{5}}
\put(105,0){\circle{5}}
\put(120,0){\circle{5}}
\put(135,0){\circle{5}}
\end{picture}}

\newcommand{\halfninevc}{
\begin{picture}(100,40)(-5,-10)
\put(0,0){\circle{5}}
\put(2.5,0){\line(1,0){10}}
\put(15,2.5){\line(0,1){10}}
\put(15,15){\circle{5}}
\put(15,0){\circle{5}}
\put(30,0){\circle{5}}
\put(47.5,0){\line(1,0){10}}
\put(17.5,0){\line(1,0){10}}
\put(32.5,0){\line(1,0){10}}
\put(62.5,0){\line(1,0){10}}
\put(17,2){\line(1,1){11}}
\put(77.5,0){\line(1,0){10}}
\put(92.5,0){\line(1,0){10}}
\put(107.5,0){\line(1,0){10}}
\put(122.5,0){\line(1,0){10}}
\put(45,0){\circle{5}}
\put(60,0){\circle{5}}
\put(30,15){\circle{5}}
\put(75,0){\circle{5}}
\put(90,0){\circle{5}}
\put(105,0){\circle{5}}
\put(120,0){\circle{5}}
\put(135,0){\circle{5}}
\put(-2,-12){\scriptsize $1$}
\put(11,-12){\scriptsize $-1$}
\put(20,13){\scriptsize $1$}
\put(35,13){\scriptsize $5$}
\end{picture}}

\newcommand{\halfeightvc}{
\begin{picture}(100,40)(-5,-10)
\put(0,0){\circle{5}}
\put(2.5,0){\line(1,0){10}}
\put(15,2.5){\line(0,1){10}}
\put(15,15){\circle{5}}
\put(15,0){\circle{5}}
\put(30,0){\circle{5}}
\put(47.5,0){\line(1,0){10}}
\put(17.5,0){\line(1,0){10}}
\put(32.5,0){\line(1,0){10}}
\put(62.5,0){\line(1,0){10}}
\put(30,2.5){\line(0,1){10}}
\put(77.5,0){\line(1,0){10}}
\put(92.5,0){\line(1,0){10}}
\put(107.5,0){\line(1,0){10}}
\put(122.5,0){\line(1,0){10}}
\put(45,0){\circle{5}}
\put(60,0){\circle{5}}
\put(30,15){\circle{5}}
\put(75,0){\circle{5}}
\put(90,0){\circle{5}}
\put(105,0){\circle{5}}
\put(120,0){\circle{5}}
\put(135,0){\circle{5}}
\put(13,-12){\scriptsize $1$}
\put(26,-12){\scriptsize $-1$}
\put(35,13){\scriptsize $4$}
\end{picture}}

\newcommand{\halfsevenvc}{
\begin{picture}(100,40)(-5,-10)
\put(0,0){\circle{5}}
\put(2.5,0){\line(1,0){10}}
\put(15,2.5){\line(0,1){10}}
\put(15,15){\circle{5}}
\put(15,0){\circle{5}}
\put(30,0){\circle{5}}
\put(47.5,0){\line(1,0){10}}
\put(17.5,0){\line(1,0){10}}
\put(32.5,0){\line(1,0){10}}
\put(62.5,0){\line(1,0){10}}
\put(45,2.5){\line(0,1){10}}
\put(77.5,0){\line(1,0){10}}
\put(92.5,0){\line(1,0){10}}
\put(107.5,0){\line(1,0){10}}
\put(122.5,0){\line(1,0){10}}
\put(45,0){\circle{5}}
\put(60,0){\circle{5}}
\put(45,15){\circle{5}}
\put(75,0){\circle{5}}
\put(90,0){\circle{5}}
\put(105,0){\circle{5}}
\put(120,0){\circle{5}}
\put(135,0){\circle{5}}
\put(28,-12){\scriptsize $1$}
\put(41,-12){\scriptsize $-1$}
\put(50,13){\scriptsize $3$}
\end{picture}}

\newcommand{\halfsixavc}{
\begin{picture}(100,40)(-5,-10)
\put(0,0){\circle{5}}
\put(2.5,0){\line(1,0){10}}
\put(15,2.5){\line(0,1){10}}
\put(15,15){\circle{5}}
\put(15,0){\circle{5}}
\put(30,0){\circle{5}}
\put(47.5,0){\line(1,0){10}}
\put(17.5,0){\line(1,0){10}}
\put(32.5,0){\line(1,0){10}}
\put(62.5,0){\line(1,0){10}}
\put(60,2.5){\line(0,1){10}}
\put(77.5,0){\line(1,0){10}}
\put(92.5,0){\line(1,0){10}}
\put(107.5,0){\line(1,0){10}}
\put(122.5,0){\line(1,0){10}}
\put(45,0){\circle{5}}
\put(60,0){\circle{5}}
\put(60,15){\circle{5}}
\put(75,0){\circle{5}}
\put(90,0){\circle{5}}
\put(105,0){\circle{5}}
\put(120,0){\circle{5}}
\put(135,0){\circle{5}}
\put(43,-12){\scriptsize $1$}
\put(56,-12){\scriptsize $-1$}
\put(65,13){\scriptsize $2$}
\end{picture}}

\newcommand{\halfsixbvc}{
\begin{picture}(100,40)(-5,-10)
\put(0,0){\circle{5}}
\put(2.5,0){\line(1,0){10}}
\put(15,2.5){\line(0,1){10}}
\put(15,15){\circle{5}}
\put(15,0){\circle{5}}
\put(30,0){\circle{5}}
\put(17.5,0){\line(1,0){10}}
\put(32.5,0){\line(1,0){10}}
\put(47.5,0){\line(1,0){10}}
\put(62,2){\line(1,1){11}}
\put(75,2.5){\line(0,1){10}}
\put(77.5,0){\line(1,0){10}}
\put(92.5,0){\line(1,0){10}}
\put(107.5,0){\line(1,0){10}}
\put(122.5,0){\line(1,0){10}}
\put(45,0){\circle{5}}
\put(60,0){\circle{5}}
\put(75,15){\circle{5}}
\put(75,0){\circle{5}}
\put(90,0){\circle{5}}
\put(105,0){\circle{5}}
\put(120,0){\circle{5}}
\put(135,0){\circle{5}}
\put(58,-12){\scriptsize $2$}
\put(80,13){\scriptsize $-1$}
\end{picture}}

\newcommand{\halffivevc}{
\begin{picture}(100,40)(-5,-10)
\put(0,0){\circle{5}}
\put(2.5,0){\line(1,0){10}}
\put(15,2.5){\line(0,1){10}}
\put(15,15){\circle{5}}
\put(15,0){\circle{5}}
\put(30,0){\circle{5}}
\put(17.5,0){\line(1,0){10}}
\put(32.5,0){\line(1,0){10}}
\put(47.5,0){\line(1,0){10}}
\put(62.5,0){\line(1,0){10}}
\put(75,2.5){\line(0,1){10}}
\put(77.5,0){\line(1,0){10}}
\put(92.5,0){\line(1,0){10}}
\put(107.5,0){\line(1,0){10}}
\put(122.5,0){\line(1,0){10}}
\put(45,0){\circle{5}}
\put(60,0){\circle{5}}
\put(75,15){\circle{5}}
\put(75,0){\circle{5}}
\put(90,0){\circle{5}}
\put(105,0){\circle{5}}
\put(120,0){\circle{5}}
\put(135,0){\circle{5}}
\put(58,-12){\scriptsize $1$}
\put(80,13){\scriptsize $1$}
\put(71,-12){\scriptsize $-1$}
\end{picture}}

\newcommand{\halffourvc}{
\begin{picture}(100,40)(-5,-10)
\put(0,0){\circle{5}}
\put(2.5,0){\line(1,0){10}}
\put(15,2.5){\line(0,1){10}}
\put(15,15){\circle{5}}
\put(15,0){\circle{5}}
\put(30,0){\circle{5}}
\put(17.5,0){\line(1,0){10}}
\put(32.5,0){\line(1,0){10}}
\put(47.5,0){\line(1,0){10}}
\put(62.5,0){\line(1,0){10}}
\put(75,2.5){\line(0,1){10}}
\put(77.5,0){\line(1,0){10}}
\put(92.5,0){\line(1,0){10}}
\put(107.5,0){\line(1,0){10}}
\put(122.5,0){\line(1,0){10}}
\put(45,0){\circle{5}}
\put(60,0){\circle{5}}
\put(75,15){\circle{5}}
\put(75,0){\circle{5}}
\put(90,0){\circle{5}}
\put(105,0){\circle{5}}
\put(120,0){\circle{5}}
\put(135,0){\circle{5}}
\put(73,-12){\scriptsize $1$}
\put(86,-12){\scriptsize $-1$}
\end{picture}}

\newcommand{\halfthreevc}{
\begin{picture}(145,40)(-5,-10)
\put(0,0){\circle{5}}
\put(2.5,0){\line(1,0){10}}
\put(15,2.5){\line(0,1){10}}
\put(15,15){\circle{5}}
\put(15,0){\circle{5}}
\put(30,0){\circle{5}}
\put(17.5,0){\line(1,0){10}}
\put(32.5,0){\line(1,0){10}}
\put(47.5,0){\line(1,0){10}}
\put(62.5,0){\line(1,0){10}}
\put(75,2.5){\line(0,1){10}}
\put(77.5,0){\line(1,0){10}}
\put(92.5,0){\line(1,0){10}}
\put(107.5,0){\line(1,0){10}}
\put(122.5,0){\line(1,0){10}}
\put(45,0){\circle{5}}
\put(60,0){\circle{5}}
\put(75,15){\circle{5}}
\put(75,0){\circle{5}}
\put(90,0){\circle{5}}
\put(105,0){\circle{5}}
\put(120,0){\circle{5}}
\put(135,0){\circle{5}}
\put(88,-12){\scriptsize $1$}
\put(101,-12){\scriptsize $-1$}
\end{picture}}

\newcommand{\deightppp}{
\begin{picture}(145,40)(-5,-10)
\put(0,0){\circle{5}}
\put(2.5,0){\line(1,0){10}}
\put(15,2.5){\line(0,1){10}}
\put(15,15){\circle{5}}
\put(15,0){\circle{5}}
\put(30,0){\circle{5}}
\put(17.5,0){\line(1,0){10}}
\put(32.5,0){\line(1,0){10}}
\put(47.5,0){\line(1,0){10}}
\put(62.5,0){\line(1,0){10}}
\put(75,2.5){\line(0,1){10}}
\put(77.5,0){\line(1,0){10}}
\put(92.5,0){\line(1,0){10}}
\put(107.5,0){\line(1,0){10}}
\put(45,0){\circle{5}}
\put(60,0){\circle{5}}
\put(75,15){\circle{5}}
\put(75,0){\circle{5}}
\put(90,0){\circle{5}}
\put(105,0){\circle{5}}
\put(120,0){\circle{5}}
\end{picture}}


\setlength{\arraycolsep}{9.95pt}
\begin{table}
{\renewcommand{\arraystretch}{1.5}
\begin{align*}
\begin{array}{|c|l|l|l|}
\hline
\multirow{2}{*}{$D$} & \multicolumn{3}{|c|}{\!\!\!\!\!\!\!\!\text{V-diagram of $\mathcal{B}$ for $|n|\leq1$}}\\
\cline{2-4}
&  
n=-1
&  
n=0
& 
n=1
\\
\hline 
10&\streck
&\halftenv&\halftenvp\\ \hline
9&\halfninevm &\halfninev&\halfninevp\\ \hline
8& \halfeightvm &\halfeightv&\halfeightvp\\ \hline
7&\halfsevenvm &\halfsevenv&\halfsevenvp\\ \hline
6a&\halfsixavm &\halfsixav&\halfsixavp\\ \hline
6b& \halfsixbvm &\halfsixbv&\halfsixbvp\\ \hline
5&\halffivevm &\halffivev&\halffivevp\\ \hline
4&\halffourvm &\halffourv&\halffourvp\\ \hline
3& \halfthreevm &\halfthreev&\halfthreevp\\ \hline
\end{array}
\end{align*}
}
\caption{\it V-diagrams of the Borcherds superalgebras $\cB$ relevant for
half-maximal supergravity in $D$ dimensions with $|n|\leq1$.
The Cartan matrices of the Borcherds superalgebras can be obtained from those of the
corresponding Kac-Moody algebras by (\ref{fromatob0}). The black nodes 
represent short roots of the corresponding Kac-Moody algebras, see appendix \ref{borcherdsapp}. The table should be compared
to Table 2 in \cite{Bergshoeff:2007vb} (but note that black nodes are used differently there).}\label{halftable}
\end{table}

\subsection{Half-maximal supergravity}

We can apply a similar analysis to the half-maximal theories, and derive
a Borcherds superalgebra $\cB$ for any $n$ and any $D$
(such that $k$ and $n+k$ below are non-negative).
However, there is an important difference compared to the maximal case, as we shall now explain.

We recall from section \ref{svm-subsection} that
the duality algebra $\mathfrak{g}$ for half-maximal supergravity
is either $\mathfrak{so}(k,n+k)$ for $k=11-D$ (if $D=3$ or $D=6b$) or the direct sum of $\mathfrak{so}(k,n+k)$ for
$k=10-D$ and a subalgebra which is either $\mathfrak{sl}(2)$ (if $D=4$) or one-dimensional (otherwise).
The subalgebra $\mathfrak{so}(k,n+k)$ is a real form of the complex Lie algebra
$B_r$
(if $n$ is odd, $r=k+\tfrac{n-1}2$)
or
$D_r$
(if $n$ is even, $r=k+\tfrac{n}2$). If $|n|\leq 1$, then this is the split real form, and only in these cases
$\mathfrak{so}(k,n+k)$ is directly given by its Cartan matrix $a_{ij}$ as the real Lie algebra generated by $e_i,f_i,h_i$
modulo the Chevalley-Serre relations (\ref{chevrelfinite})--(\ref{serrelfinite}). In the other cases $\mathfrak{so}(k,n+k)$
is instead spanned over the real numbers by complex linear combinations of the basis elements in the complex Lie algebra
$B_r$ or $D_r$ (generated in the same way as the split real form, but over the complex numbers).
This of course also applies to the Borcherds superalgebra
$\cB$, obtained by adding simple roots to those of $\g$, and its Cartan matrix $B_{IJ}$. However, the Chevalley generators
associated to the additional simple roots will always be genuine basis elements of $\cB$,
so once $\mathfrak{so}(k,n+k)$ is identified as a real subalgebra of $B_r$ or $D_r$, the procedure is the same as for maximal supergravity (where the duality algebras are always split real forms). In Table \ref{halftable} we 
display the result in terms of V-diagrams, for simplicity only for the split cases $|n|\leq1$.
It is (hopefully) evident from the table how it could be extended to include any other possible value of $n$, and also how
each V-diagram can be extended to a V-diagram of a Borcherds superalgebra $\cD$, like in Table \ref{maxtable}. 
Below we also give some Cartan matrices for $n=0$ explicitly.

For $D=3$, the situation is very similar to $3\leq D\leq 7$ in the maximal case. There is just one generator $e_0$ to be added to those of the duality algebra $\gg$. It can be taken to be a lowest-weight vector for the adjoint representation of $\gg$, and the supersymmetry constraint implies that $[e_0,e_0]=0$. Thus $B_{IJ}$ has the form of \eq{3.1} with $B_{00}=0$ and $B_{0i}=-p_i$ where $p_i$ are the Dynkin labels of the adjoint representation of $\gg$.

For $D=4$ the situation resembles that of $D=8$ maximal supergravity since the duality group is a product. The Cartan matrix, for the case $n=0$,  is 
\be
B_{IJ}= \left( \begin{array}{rrrrrrrr} 
0&-1& -1&0&0&0&0&0\cr
-1&2&0&0&0&0&0&0\cr
-1&0&2&-1&0&0&0&0\cr
0&0&-1&2&-1&0&0&0\cr
0&0&0&-1&2&-1&0&0\cr
0&0&0&0&-1&2&-1&-1\cr
0&0&0&0&0&-1&2&0\cr
0&0&0&0&0&-1&0&2\cr
\end{array} \right)\ .
\ee

For $D\geq 5$ the half-maximal case is similar to $D=9$ in the maximal case, in that there is an extra odd generator $e_0$ in the vector representation of the duality group and a second additional generator $e_{0'}$ at level $\ell=(6-k)$, as reflected in the V-diagram in table 2. In this case
$B_{00}=0$ but both $B_{00'}=B_{0'0}$ and $B_{0'0'}$
are nonzero (although the nodes 0 and $0'$ are not connected to each other in the V-diagram)\black. As an example let us consider $D=7$, again with $n=0$. The duality algebra in this case is $\bbR \oplus\mathfrak{so}(3,3)$
and the Cartan matrix $A_{IJ}$ can be determined from the V-diagram to be
\be
A_{IJ}=\left(\begin{array}{rrrrr}
2&0&0&0&0\cr
0&2&-1&0&0\cr
0&-1&2&-1&-1\cr
0&0&-1&2&0\cr
0&0&-1&0&2\cr
\end{array}\right)\ ,
\ee
where the first row corresponds to the level-three root vector $e_{0'}$ and the second to the level-one one, $e_0$. The Cartan matrix for the Borcherds superalgebra $B_{IJ}=A_{IJ}-w(v_{I},v_{J})$ is therefore
\be
B_{IJ}=\left(\begin{array}{rrrrr}
-10&-4&0&0&0\cr
-4&0&-1&0&0\cr
0&-1&2&-1&-1\cr
0&0&-1&2&0\cr
0&0&-1&0&2\cr
\end{array}\right)\ .
\ee
%

\section{The Borcherds-Kac-Moody correspondence}

In the preceding section we saw that the Lie algebra $\gg$ of the duality group in $D$-dimensional (half-)maximal supergravity can be extended
by adding simple roots with positive V-degrees to the simple roots $\beta_i$ of $\gg$ (which have V-degree zero).
This corresponds to adding rows and columns to the Cartan matrix $a_{ij}$ of $\gg$.
We did this in two ways, leading to the Cartan matrix $A_{IJ}$ of a Kac-Moody algebra $\cA$, and to the Cartan matrix $B_{IJ}$
of a Borcherds superalgebra $\cB$. The two Cartan matrices are related to each other by the relation (\ref{fromatob0}).

In this section we will consider a further extension of $\cA$, a Kac-Moody algebra $\cC$, obtained by adding an $A_d=\gs\gl(d+1)$ algebra,
whose Dynkin diagram form a ``gravity line'' of $d$ nodes, where $d$ is an arbitrary positive integer.
For any simple root $\alpha_I$ of $\cA$ with V-degree $v_I \geq 0$,
the corresponding node is then connected with a single line to the $v_I$-th node of the Dynkin diagram of $\gs\gl(d+1)$
(counted from one of the ends) if $v_I \geq 1$, and disconnected from it if $v_I=0$ or $v_I>d$.
In the same way as for $\cB$ we get a $\mathbb{Z}$-grading
of $\cC$, with the overall level given by $\ell=\sum_I v_I k_I$, where $k_I$ is the level with respect to $\alpha_I$.
However, now we have at each level $\ell$
not only a representation of $\gg$, but each representation of $\gg$ also comes together with a representation of $\gs\gl(d+1)$.
In particular the antisymmetric product of $\ell$ fundamental representations of $\gs\gl(d+1)$ appears at level $\ell$,
for $1 \leq \ell \leq d+1$.

It has been known that in the cases of (half-)maximal supergravity in $D$ dimensions with split duality algebras $\gg$
the representation of $\g$ that comes together with
this antisymmetric product of $\ell$ fundamental $\gs\gl(d+1)$ representations precisely coincides
with the representation $\cR_\ell$ of $\g$ at level $\ell$ in $\cB$, up to level $\ell=d+1$.
Thus the form spectrum up to $D$-form potentials can be derived from the Kac-Moody algebra $\cC$ with $d=D-1$,
which is $E_8{}^{+++}=E_{11}$ for maximal supergravity, and $B_7{}^{+++}$, $D_8{}^{+++}$, $B_8{}^{+++}$
for half-maximal supergravity with $n=-1,0,1$, respectively \cite{West:2001as,Riccioni:2007au,Bergshoeff:2007qi,Bergshoeff:2007vb}.\footnote{
Note that although we set $d=D-1$, the Kac-Moody algebra $\cC$ is the same for any $D$, since the Borcherds superalgebras $\cB$ that we start with
also depend on $D$.}
However, if we are also interested in the forms at level $(D+1)$, then $E_{11}$ is not enough; we need to take $d=D$ and go to
$E_{12}$. This continues to infinity, so in order to include {\it all} the representations $\cR_\ell$ one would need to consider Kac-Moody algebras of infinite rank, although they are all still contained in the (finite rank) Borcherds superalgebra $\cB$. An advantage of this correspondence is that
the representations $\cR_\ell$ can be computed recursively using the denominator formula for Borcherds superalgebras, which efficiently can be rewritten in terms of partition functions \cite{Martin}.

The aim of this section is to show that 
the above correspondence holds also in the most general case with an arbitrary Kac-Moody algebra $\g$,
extended to a Borcherds superalgebra $\cB$ and to Kac-Moody algebras $\cA$ and $\cC$ in the way described above, by adding simple roots with positive V-degrees and
an $\sl(d+1)$ algebra.
The case where there is only one additional simple root, with V-degree $v_0=v$, is illustrated in Figure \ref{gAB}.
If furthermore $v_0=1$ we have the special case already proven in \cite{Palmkvist:2012nc}.

The idea is to consider both $\mathcal{B}$ and $\mathcal{C}$ as subalgebras of a contragredient Lie superalgebra $\mathcal{D}$ (already mentioned in the preceding section). In the case of only one simple root $\beta_0$ of $\cB$ with positive V-degree $v_0=v\geq 1$, the V-diagram of $\cD$ is shown in
Figure \ref{gAB}. There it is also assumed that $(\alpha_0,\alpha_0)=2$, since the nodes corresponding to $\alpha_0$ and $\beta_0$ are white, but they can also be
black or correspond to any other value of the diagonal entry $A_{00}=(\alpha_0,\alpha_0)$ in the Cartan matrix of $\cA$
(see appendix \ref{borcherdsapp} for our conventions for colouring the nodes). 
In the case of more than one simple root of $\cB$ with positive V-degree, each of the corresponding nodes is connected with a single line to 
the same node in the V-diagram of $\cD$, corresponding to the simple root $\delta_0$ with V-degree $-1$.
In all cases relevant for maximal supergravity in $D$ dimensions, the V-diagrams of $\cD$ are given explicitly in the right column of Table \ref{maxtable}.
In fact, $\mathcal{D}$ is the same for different $D$, but described by different V-diagrams.

It is clear that $\mathcal{B}$ is a subalgebra of $\mathcal{D}$, since the V-diagram of $\mathcal{B}$ is obtained by removing nodes from that
of $\mathcal{D}$.
The embedding of $\mathcal{C}$ into $\mathcal{D}$ is less obvious, but can be understood by setting
\begin{align} \label{alpha0}
\alpha_0 = \beta_0 + v \delta_0 + (v-1)\delta_1 + (v-2) \delta_2+\cdots+ 2 \delta_{v-2}+\delta_{v-1}\ ,
\end{align}
and $\gamma_\iota=\delta_\iota$ for $\iota=1,2,\ldots,d$ (where $\delta_0,\delta_1,\ldots,\delta_{d}$ are the additional simple roots of $\cD$ according to Figure \ref{gAB}).
We then get
\begin{align} \label{gammagamma}
(\alpha_0,\alpha_0)=(\beta_0,\beta_0)+w(v,v)
\end{align}
and $(\alpha_0,\alpha_\iota)=-\delta_{\iota v}$
as we should, and $\alpha_0$
also satisfies the same 
inner product relations with the roots of $\g$ as $\beta_0$.
In the case of more than one simple root of $\cB$ with positive V-degree, we can let $\beta_{0'}$ be another one, with V-degree $v'=v_{0'}\geq1$
and a corresponding simple root $\alpha_{0'}$ of $\cC$, and we find that (\ref{gammagamma}) can be generalised to include this case, 
\begin{align}
(\alpha_0,\alpha_{0'})=(\beta_0,\beta_{0'})+w(v,v')
\end{align}
for the inner product of $\alpha_0$ and $\alpha_{0'}$.
It remains to show that the linear combination (\ref{alpha0}) 
of simple roots of $\mathcal{D}$ indeed itself is a root of $\mathcal{D}$. This will be done below.

It is convenient to extend the $A_{d}=\sl(d+1)$ subalgebra of $\cD$ to $\mathfrak{gl}(d+1)$,
including the Cartan element corresponding to the simple root $\delta_0$, with the basis elements
$K^a{}_b$ ($a,b=0,1,\ldots,d$) and the commutation relations
\begin{align}
[K^a{}_b,K^c{}_d]&=\de_b{}^c K^a{}_d - \de_d{}^a K^c{}_b\ .
\end{align}

\begin{figure}[h!]
\begin{picture}(450,105)(30,20)
\put(100,-7){$\gamma_1$}
\put(139.5,-7){$\gamma_2$}
\put(199,-7){$\gamma_{v-1}$}
\put(244,-7){$\gamma_v$}
\put(279,-7){$\gamma_{v+1}$}
\put(344,-7){$\gamma_{d-1}$}
\put(389,-7){$\gamma_{d}$}
\put(260,45){$\alpha_0$}  
\thicklines
\multiput(210,10)(40,0){3}{\circle{10}}
\multiput(215,10)(40,0){2}{\line(1,0){30}}
\put(145,10){\circle{10}}
\put(105,10){\circle{10}}
\put(110,10){\line(1,0){30}}
\put(150,10){\line(1,0){10}}
\put(195,10){\line(1,0){10}}
\multiput(165,10)(10,0){3}{\line(1,0){5}}
\multiput(310,10)(10,0){3}{\line(1,0){5}}
\put(355,10){\circle{10}}
\put(360,10){\line(1,0){30}}
\put(295,10){\line(1,0){10}}
\put(340,10){\line(1,0){10}}
\put(395,10){\circle{10}}
\put(250,50){\circle{10}} \put(250,15){\line(0,1){30}}
\put(254,53){\line(1,1){10}}               
\put(246,53){\line(-1,1){10}}              
\put(243,62){$\cdots$}                     
\thinlines
\put(250,59){\line(1,0){60}}
\put(250,59){\line(-1,0){60}}
\put(190,59){\line(0,1){46}}
\put(310,59){\line(0,1){15}}
\put(307,80.5){$\g$}
\put(310,90){\line(0,1){15}}
\put(250,105){\line(1,0){60}}
\put(250,105){\line(-1,0){60}}
\put(85,-15){\line(1,0){330}}
\put(85,135){\line(1,0){330}}
\put(415,-15){\line(0,1){89}}
\put(412,78){$\mathcal{C}$}
\put(415,90){\line(0,1){45}}
\put(85,-15){\line(0,1){150}}
\put(150,30){\line(1,0){200}}
\put(150,120){\line(1,0){200}}
\put(350,30){\line(0,1){44}}
\put(350,90){\line(0,1){30}}
\put(345,78){$\mathcal{A}$}
\put(150,30){\line(0,1){90}}
\end{picture} 

\begin{picture}(450,310)(30,-50)
\put(100,-7){$\delta_1$}
\put(139.5,-7){$\delta_2$}
\put(199,-7){$\delta_{v-1}$}
\put(244,-7){$\delta_v$}
\put(279,-7){$\delta_{v+1}$}
\put(344,-7){$\delta_{d-1}$}
\put(389,-7){$\delta_{d}$}
\put(260,45){$\delta_0$}
\thicklines
\multiput(210,10)(40,0){3}{\circle{10}}
\multiput(215,10)(40,0){2}{\line(1,0){30}}
\put(145,10){\circle{10}}
\put(105,10){\circle{10}}
\put(110,10){\line(1,0){30}}
\put(150,10){\line(1,0){10}}
\put(195,10){\line(1,0){10}}
\multiput(165,10)(10,0){3}{\line(1,0){5}}
\multiput(310,10)(10,0){3}{\line(1,0){5}}
\put(355,10){\circle{10}}
\put(360,10){\line(1,0){30}}
\put(295,10){\line(1,0){10}}
\put(340,10){\line(1,0){10}}
\put(395,10){\circle{10}}
\put(250,50){\circle{10}}
\put(109,13){\line(4,1){136.5}}
\put(250,90){\circle{10}} \put(250,55){\line(0,1){30}}
\put(236,90){\scriptsize $v$}
\put(230,50){\scriptsize $-1$}
\put(254,93){\line(1,1){10}}                
\put(246,93){\line(-1,1){10}}               
\put(243,102){$\cdots$}                     
\thinlines
\put(250,99){\line(1,0){60}}
\put(250,99){\line(-1,0){60}}
\put(190,99){\line(0,1){46}}
\put(310,99){\line(0,1){15}}
\put(307,120.5){$\g$}
\put(310,130){\line(0,1){15}}
\put(250,145){\line(1,0){60}}
\put(250,145){\line(-1,0){60}}
\put(85,-15){\line(1,0){330}}
\put(85,175){\line(1,0){330}}
\put(415,-15){\line(0,1){129}}
\put(410,118){$\mathcal{D}$}
\put(415,130){\line(0,1){45}}
\put(85,-15){\line(0,1){190}}
\put(150,70){\line(1,0){200}}
\put(150,160){\line(1,0){200}}
\put(350,70){\line(0,1){44}}
\put(350,130){\line(0,1){30}}
\put(346,118){$\mathcal{B}$}
\put(150,70){\line(0,1){90}}
\put(260,85){$\beta_0$}
\end{picture} 
\caption{\it Illustration of how the Dynkin diagrams of the
Kac-Moody algebras $\g$, $\cA$ and $\cC$, and the V-diagrams of the Borcherds superalgebras $\cB$ and $\cD$
are related to each other in the general case.
The node corresponding to $\alpha_0$ or $\beta_0$ can be connected 
to any number of nodes in the Dynkin diagram of $\g$ (which is itself not visible in the figure)
and the corresponding off-diagonal entries in the Cartan matrix can take any negative values
(symmetrically).
The V-degree of $\delta_0$ is $-1$ 
and the V-degree of $\beta_0$  
is $v$,
as written next to the nodes, and all other
simple roots have V-degree zero. Thus $\delta_{0}$ is an odd null root,
whereas $\beta_0$ is odd if and only if $v$ is an odd integer, and all the other simple roots are always even.
It also follows from the V-degrees that the length squared of the simple root $\beta_0$ is
$(\beta_0,\beta_0)=(\alpha_0,\alpha_0)-w(v,v)=(\alpha_0,\alpha_0)-v(v+1)$,
and its scalar product with $\delta_0$ is $(\beta_0,\delta_0)=(-1)-w(-1,v)=
v$. In the figure $\alpha_0$ and $\beta_0$ are represented by a white node, which means that $(\alpha_0,\alpha_0)=2$, but
$\alpha_0$ can also have a different length squared. In particular the node can be black, which means that $(\alpha_0,\alpha_0)=1$.}
\label{gAB}
\end{figure}

\noindent
Then we have $h_\iota=K^\iota{}_\iota - K^{\iota-1}{}_{\iota-1}$ for the Cartan elements of the $\sl(d+1)$ subalgebra, 
and if $a \neq b$, then $K^a{}_b$
is a root vector corresponding to the root 
$\delta_{b+1}+\delta_{b+2}+\cdots+\delta_{a}$ (if $b<a$), or 
$-\delta_{a+1}-\delta_{a+2}-\cdots-\delta_{b}$
(if $a<b$).
Furthermore, we let $E^a$ and $F_a$ be root vectors in $\mathcal{D}$ corresponding to the root
$\delta_0+\delta_1+\cdots+\delta_{a}$ and its negative, respectively, such that
\begin{align}
[K^a{}_b,F_c]&=-\de_c{}^a F_b\ , & [K^a{}_b,E^c]&=\de_b{}^c E^a\ , & [E^a,F_b]&=K^a{}_b-\de^a{}_b K\ ,
\end{align}
where
\begin{align}
K=K^a{}_a=-\frac1{d}\big((d+1)\,h_0+d\,h_1+\cdots+2\,h_{d-1}+h_{d}\big)\ .
\end{align}

From now on, we simplify the discussion by restricting to the case of only one simple root $\beta_0$ of $\cB$ with positive V-degree $v_0=v\geq 1$, illustrated 
in Figure \ref{gAB}, but it is straightforward to extend it to the general case. Thus the formula $\ell=\sum_I v_I k_I$ simplifies to $\ell=vk$, and the two different $\mathbb{Z}$-gradings of $\cB$ that we consider differ only by the factor $v$. For the Chevalley generator $e_0$ corresponding to 
$\beta_0$ we thus have $k=1$ and $\ell=v$, and for all other Chevalley generators $e_i$, corresponding to all other simple roots, we have $k=\ell=0$.

Let $e_\cM$ be a basis of the level $k=1$ subspace of $\cB$. The subspace at a general level $k$ is then spanned by elements
\begin{align}
e_{\cM_k\cdots\cM_1} = [e_{\cM_k},[e_{\cM_{k-1}},\ldots, [e_{\cM_{2}},e_{\cM_{1}}]\cdots]]\ .
\end{align}
When we extend $\cB$ to $\cD$ these elements are eigenvectors under the action of the $\sl(d+1)$ subalgebra,
\begin{align}
[K^a{}_b,e_{\cM_k \cdots \cM_1}]&=- \frac{\ell}{d} \de_b{}^a e_{\cM_k \cdots \cM_1} &&\Rightarrow &
[K,e_{\cM_k \cdots \cM_1}]&=-\frac{\ell}{d}(d+1) e_{\cM_k \cdots \cM_1}\ ,
\end{align}
while acting with $F_a$ gives zero, and acting with $E^a$ gives new elements, which we denote by
\begin{align} \label{tensor}
E^{\,a_1 \cdots a_m}{}_{\cM_k \cdots \cM_1}=[E^{a_1},[E^{a_2},\ldots, [E^{a_m},e_{\cM_k \cdots \cM_1}]\cdots]]\ .
\end{align}
Since the elements $E^a$ anticommute with each other, $[E^a,E^b]=0$,
the expression $E^{\,a_1 \cdots a_m}{}_{\cM_k \cdots \cM_1}$ is totally antisymmetric in the
upper indices. Thus it certainly vanishes for $m \geq d+2$, but the following lemma tells us that this
in fact happens already for $m \geq \ell+1$.


\noindent
{\bf Lemma.}
{\em For any element $X=e_{\cM_k\cdots\cM_1}X^{\cM_1\cdots\cM_k}$ at level $k$ in $\cB$, set
\begin{align}
X^{a_1\cdots a_m}=[E^{a_1},[E^{a_2},\ldots, [E^{a_m},X]\cdots]]
\end{align}
in $\cC$,
where the indices
$a_1,a_2,\ldots,a_m$ take $m$ distinct values among $0,1,\ldots,d$. For $1 \leq m \leq \ell$ we have
$X^{\,a_1 \cdots a_m}=0$
if and only if
$X=0$,
whereas for $m \geq \ell+1$ we always have
$X^{\,a_1 \cdots a_m}=0$}.

\Pf
This is most easily shown by a calculation which is not $\mathfrak{sl}(d+1)$ covariant (and thus
the repeated index $a_1$ should not be summed over),
\begin{align} \label{noncovcal}
[F_{a_1}, X^{\,a_1 \cdots a_m}]&=
[[F_{a_1}, E^{\,a_1}], X^{\,a_2 \cdots a_m}]
=[K^{a_1}{}_{a_1}, X^{\,a_2 \cdots a_m}]
-[K,X^{\,a_2 \cdots a_m}]\nn\\
&=\Big(-\frac{\ell}{d}\,
-(m-1)+\frac{\ell}{d}(d+1)\Big)
X^{\,a_2 \cdots a_m}{}
=(\ell+1-m)X^{\,a_2 \cdots a_m}.
\end{align}
The lemma can now be proven by induction.
\qed

Since in particular 
$E^{\,a_1\cdots a_v}{}_\cM$ is nonzero, 
the linear combination (\ref{alpha0}) of simple roots is indeed a root
of $\cD$ and, identifying this root with $\alpha_0$, we can indeed consider $\cC$ as a subalgebra of $\cD$.
The level $k$ subspace of this subalgebra, with respect to $\alpha_0$,
is spanned by elements
\begin{align} \label{5.12}
[E^{\,a_1 \cdots a_v}{}_{\cM_k},[E^{\,a_{v+1} \cdots a_{2v}}{}_{\cM_{k-1}},\cdots,
[E^{\,a_{\ell-2v+1} \cdots a_{\ell-v}}{}_{\cM_{2}},E^{\,a_{\ell-v+1} \cdots a_\ell}{}_{\cM_1}]\cdots]],
\end{align}
and the restricted 
subspace corresponding to the antisymmetric product of $p$ fundamental representations is spanned by elements obtained from
(\ref{5.12}) by antisymmetrising the upper indices,
\begin{align} \label{antisym}
[E^{\,[a_1 \cdots a_v}{}_{\cM_k},[E^{\,a_{v+1} \cdots a_{2v}}{}_{\cM_{k-1}},\cdots,
[E^{\,a_{\ell-2v+1} \cdots a_{\ell-v}}{}_{\cM_{2}},E^{\,a_{\ell-v+1} \cdots a_\ell]}{}_{\cM_1}]\cdots]].
\end{align}
By repeated use of the lemma and the Jacobi identity, it can be shown that
(\ref{tensor}) for $m=\ell$ is proportional to (\ref{antisym}), and thus that
$X^{a_1\cdots\a_\ell}$ is proportional to
\begin{align} \label{antisym2}
[E^{\,[a_1 \cdots a_v}{}_{\cM_k},[E^{\,a_{v+1} \cdots a_{2v}}{}_{\cM_{k-1}},\cdots,
[E^{\,a_{\ell-2v+1} \cdots a_{\ell-v}}{}_{\cM_{2}},E^{\,a_{\ell-v+1} \cdots a_\ell]}{}_{\cM_1}]\cdots]]X^{\cM_1\cdots\cM_k}.
\end{align}
It then follows from the lemma
that $X$ is zero if and only if (\ref{antisym2}) 
is zero, and we conclude that the lower indices in
$e_{\cM_k \cdots \cM_1}$ and (\ref{antisym}) determine the same representation of $\g$.

\section{Gauging} \label{gauging-section}

\subsection{Deformed Bianchi identities}

In this section we consider the gauged version of the Bianchi identities, following the discussion given in \cite{Greitz:2013pua}, but generalised to all (half-)maximal cases. We shall focus on the standard case where the gauge algebra is a subalgebra of the duality algebra $\gg$ (including a one-dimensional algebra corresponding to shifts of the dilaton where appropriate). We shall not discuss the gauging of the constant scaling (or trombone) symmetry of the equations of motion, for which we refer to the literature, see, for example \cite{LeDiffon:2008sh,Riccioni:2010xx,FernandezMelgarejo:2012ne,Prins:2013yra}. We shall also discuss the solubility and supersymmetry of the gauged tensor hierarchies and briefly mention the need to amend the superspace curvature. Gauging will involve a different extension of the form algebra to the ones we have discussed above. To begin, we first reformulate the ungauged system of Bianchi identities in terms of $\gg_{\gf}$-valued forms. We define
\begin{align}
A_{\ell}&=e_{\cM_{\ell}\cdots \cM_1} A_{\ell}^{\cM_1\cdots \cM_{\ell}}\nn\ ,\w1
F_{\ell+1}&=e_{\cM_{\ell}\cdots \cM_1} F_{\ell+1}^{\cM_1\cdots \cM_{\ell}}\ .
\la{g1}
\end{align}
We shall use the convention that odd elements in $\gg_{\gf}$ anti-commute with odd forms, so that the potentials are all even objects while the field-strengths are all odd. We denote the sums of all of these forms by $A$ and $F$, with the sums starting from $\ell=1$. The consistent Bianchi identities can then be rewritten as\footnote{The sign change with respect to (2.6) is for later convenience.}
\be
dF=-\half[F,F]\ .
\la{g2}
\ee
Consistency follows immediately by applying $d$ to both sides and using \eq{g2} again followed by the Jacobi identity. 

At level $(D-1)$ the field-strength form is a $D$-form and so has a dual that is a 0-form. In the ungauged case this object is constant, and, as it has dimension one, parametrises the possible massive deformations of the theory. Its ``virtual potential'' would be at level minus one, so this suggests incorporating such a level into the algebraic structure with $\cR_{-1}=\bar\cR_{D-1}$, by spacetime duality. Note that this is different to the Borcherds superalgebra extension of the form algebra which is symmetrical about level zero
(in the sense that $\cR_{-1}=\bar\cR_{1}$). In the cases where the forms are generated from the level-one forms, we know that $\cR_{D-1}$ is contained in (but is not identical to) the product $\cR_1\otimes adj$, because $\cR_{D-2}\sim adj$.
Here, $adj$ denotes the adjoint representation. This holds for the maximal theories in $3\leq D\leq 9$ and half-maximal theories in $D\leq 5$. It also holds for half-maximal theories in $D=8,9,10$ because the second generating form, at level $(D-4)$, does not give rise a new form at level $(D-1)$. For $D=6a,7$, there are additional forms at level $(D-1)$  involving the level $(D-4)$-form as we saw previously in section 3.2, but these actually give rise to massive deformations of type $p=2,3$ \cite{Bergshoeff:2007vb}, as we discussed there. There are no gaugings in $D=10,11$ maximal or $D=6b$ half-maximal theories, so the net upshot is that all gaugings are associated with elements in $\cR_{D-1}$ that are contained in $\cR_1\otimes adj$.\footnote{There are two massive deformations of $D=10$ IIA supergravity; the Romans theory \cite{Romans:1985tz}, and the generalised Schwarz-Scherk reduction \cite{Howe:1997qt,Lavrinenko:1997qa}, see also \cite{Riccioni:2010xx}, for which there is no component Lagrangian.}

Letting $\mathpzc{m}$, etc, denote indices running from 1 to dim $\gg$, we can write the dimension-zero field-strength as $\Th_\cM{}^{\mathpzc{m}}$, indicating that it is to be identified with the embedding tensor \cite{deWit:1981eq,deWit:2002vt,deWit:2003hr}. We can also introduce a basis for level minus-one, $\phi_{\mathpzc{m}}{}^\cM$, which will be in the dual representation. We can then form the single level minus-one element $\Th:=\Th_\cM{}^{\mathpzc{m}}\,\phi_{\mathpzc{m}}{}^\cM$.  We have the commutation relations
\begin{align}
[\phi_{\mathpzc{m}}{}^\cM,e_\cN]&=\delta_\cN{}^{\lceil\cM} t_{\mathpzc{m}\rfloor} \nn\ ,\w1
[t_{\mathpzc{m}},\phi_{\mathpzc{n}}{}^{\cM}]&=f_{\mathpzc{m}\lfloor\mathpzc{n}}{}^{\mathpzc{p}} \phi_{\mathpzc{p}}{}^{\cM\rceil}  - t_{\mathpzc{m} \cN}{}^{\cM} 
\phi_{\mathpzc{n}}{}^{\cN}\ ,
\la{g2.1}
\end{align}
where the diagonal brackets denote the projection from $\cR_1\otimes adj$ onto $\cR_{-1}$ and where
\be
[t_{\mathpzc{m}},e_{\cN}]=t_{\mathpzc{m}\cN}{}^{\cP} e_{\cP}\ ,
\la{g2.2}
\ee
with $t_{\mathpzc{m}\cN}{}^{\cP} $ denoting the generators of $\gg$ in the representation $\cR_1$.
In order to gauge the theory we have to promote some of the abelian gauge fields to non-abelian ones. This is done with the help of the embedding tensor; we define the gauge field to be $\cA:= [A_1,\Th]$. Written out in more detail
\be
\cA=[A_1,\Th] = A_1^\cM \Th_\cM{}^{\mathpzc{m}} t_{\mathpzc{m}}\ .
\la{g3}
\ee
This shows that the generators of the Lie algebra $\gg_0$ of the gauge group $G_0\subset G$ are
\be
X_\cM=\Th_\cM{}^{\mathpzc{m}} t_{\mathpzc{m}} =[e_\cM,\Th]\ .
\ee
We demand that the embedding tensor be invariant under $\cD$:
\be
\cD\Th=0\Rightarrow [X_\cM,\Th]=0 \Rightarrow [e_\cM,[\Th,\Th]]=0\ ,
\la{g4}
\ee
where we used the fact that $\Th_\cM{}^{\mathpzc{m}}$ is itself constant. This is assured if we set $[\Th,\Th]=0$ so that the extended algebra is truncated at level minus-one. So there are two constraints on the embedding tensor, the representation constraint specifying how $\cR_{-1}$ sits inside $\cR_1\otimes adj$, and the quadratic constraint that follows from invariance under $\gg_0$. These are, of course, just the standard constraints imposed in gauging, see, for example \cite{deWit:2002vt,deWit:2003hr,deWit:2005hv}.

The extension of $\gg_{\gf}$ to negative levels using spacetime duality rather than symmetry about level zero gives the tensor hierarchy algebra (THA)  \cite{Palmkvist:2011vz}, and the truncated version, which includes the single element $\Th$ at level minus one with $[\Th,\Th]=0$ (and thus no other non-zero elements at any negative levels) 
is very convenient for discussing the gauged hierarchy \cite{Greitz:2013pua}, as we now demonstrate.

The gauge field $\cA$ is at level zero and its field strength $\cF=d\cA + \cA^2$ is given by $\cF=-[F_2,\Th]$. We claim that the following deformed Bianchi identities are consistent if the corresponding ungauged ones are:
\be
\cD F=-\half[F,F] + [F,\Th]-[F_2,\Th]\ ,
\la{g5}
\ee
where the $F$s start at level one as before. The last term in \eq{g5} is necessary because the set of Bianchi identities starts at $\cD F_2=[F_3,\Th]$. 

To check the consistency of \eq{g5} we apply a second $\cD$ to both sides. Since the last term on the right is just $\cF$ it is annihilated. On the left we get 
\be
\cD^2 F=[F,\cF]=-[F,[F_2,\Th]]\ .
\la{g6}
\ee
Applying $\cD$ on the right-hand side we get terms with zero, one and two $\Th$s. The former vanish because they are the same as in the ungauged case, while the $\Th^2$ terms also vanish  because they have the form $[[F,\Th],\Th]\sim [F,[\Th,\Th]]=0$. So we are left with the single $\Th$ terms. These are also easily seen to cancel with the contribution from the right-hand side using the Jacobi identity. We therefore see that the consistent set of Bianchi identities in an ungauged theory can be extended to the gauged case by means of the embedding tensor interpreted as a level minus-one element of the truncated THA. Of course, the formulae for the field-strengths also have to be deformed. In principle one can do this by deriving the explicit expressions for the field-strength forms such that the Bianchi identities are satisfied. However, it is more straightforward to reformulate everything in terms of the truncated THA.

\subsection{All forms together}

It was observed in the original papers \cite{Cremmer:1997ct,Cremmer:1998px} that, if one thinks of the potentials together as a $\gg_{\gf}$-valued form, one can exponentiate to get a formal group element. Given this, one can then derive a Maurer-Cartan equation which is equivalent to the set of Bianchi identities for all of the field-strength forms and which is guaranteed to be consistent.  Let us define $\hat\gg$ to be the Lie superalgebra obtained by appending the level minus-one element $\Th$ to $\gg_{\gf}$ (thus it is the truncated version of the THA discussed above), $\O$ to be the associative superalgebra of forms and $\cU_{\hat\gg}$ the enveloping algebra for $\hat\gg$. The forms we are interested in take can be considered to be elements of $\O_{\hat\gg}:=\O\otimes\hat\gg$, while if we exponentiate we get objects that lie in $\O\otimes \cU_{\hat\gg}$. As stated above, we take odd forms to anti-commute with odd elements of $\hat\gg$. We consider $d$ to be a skew-derivation that acts from the right while there is another one, $L_{\Th}$, that takes the graded commutator of an object with $\Th$. Both of these square to zero (as $[\Th,\Th]=0$) and they anti-commute with each other. This means that $d_{\Th}:=d+L_{\Th}$ is also nilpotent.

Let us first consider the ungauged case where the extension by $\Th$ is irrelevant. The Bianchi identities written in the form $dF=-\half[F,F]$ can be considered to be a Maurer-Cartan equation that is solved by setting
\be
F=d e^A e^{-A}\ .
\la{g7}
\ee
We can also find the gauge transformations given by the odd parameter $\L:=\sum_{\ell\geq 1} \L_{\ell-1}$, where the individual parameters $\L_{\ell-1}$ are $(\ell-1)$-form parameters at level $\ell$. If we set $Z:=\d e^A\, e^{-A}$ then we find that $F$ is gauge-invariant if $dZ +[Z,F]=0$. This equation is easily seen to be solved by
\be
Z=d\L+[\L,F]\ .
\la{g8}
\ee

We can generalise this to the gauged case reasonably straightforwardly. We define $A$ as before but then put
\be
F':=d_{\Th} e^A\, e^{-A}\ .
\la{g9}
\ee
Since $d_{\Th}^2=0$, we again have a Maurer-Cartan equation
\be
d_{\Th} F' + F'^2=0\ .
\la{g10}
\ee
$F'$ in this case has a level-zero component which is just $\cA=[A_1,\Th]$, so $F'=F+\cA$, where $F$ is the sum of all the field-strength forms starting at level one. It is not difficult to show that the Maurer-Cartan equation \eq{g10} is equivalent to the gauged Bianchi identities \eq{g5}. Since $F^2=\half [F,F]$, we have
\begin{align}
0=d_{\Th} F' + F'^2&=d_{\Th} F + d_{\Th} \cA + \half[F+\cA,F+\cA]\nn\w1
&=\cD F+\half[F,F] +[F,\Th] + d_{\Th}\cA + \cA^2\nn\w1
&=\cD F+\half[F,F] +[F,\Th] +\cF +[\cA,\Th]\nn\w1
&=\cD F+\half[F,F] +[F,\Th]-[F_2,\Th]\ ,
\la{g11}
\end{align}
where we used the facts that $\cF=-[F_2,\Th]$ and $[\cA,\Th]=0$. This confirms the claim.

We can define the gauge transformations in a similar way to the ungauged case, Setting $Z=\d e^A\, e^{-A}$, we find
\begin{align}
\d F&=[F,[\L_0,\Th]]\ ,\la{19}\w1
\d\cA&=[d\L_0 +[\L_0,[A_1,\Th]],\Th]\ .
\la{g12}
\end{align}
where
\be
Z=d_{\Th}\L -[\L_0,\Th] +[\L,F']\ .
\la{g13}
\ee
We can identify $[\L_0,\Th]$ as the gauge parameter for $\gg_0$, so that the equations \eq{g12} show that the gauge field $\cA$ and the field-strength forms have the correct transformations.  The first few levels in the field-strengths and gauge transformations were given explicitly in \cite{Greitz:2013pua} and shown to agree with the formulae previously derived by a Noether-type method in, for example, \cite{deWit:2002vt,deWit:2003hr,deWit:2005hv}.

\subsection{Inclusion of scalars}

It is straightforward to include the scalar fields into the picture. This is most simply accomplished by use of the matrix $\cV\in G$, rather than by fixing a gauge for the R-symmetry group $H$. Recall that the duality group $G$
acts rigidly on $\cV$ to the right, while the local R-symmetry group $H$ acts on the left, $\cV\rightarrow h^{-1}\cV g$. If we now set
\be
\F= d(\cV e^A)\,e^{-A}\cV^{-1}\ ,
\la{g14}
\ee
then clearly $d\F+\F^2=0$.  The Maurer-Cartan form $\F$ can be rewritten as
\be
\F= d\cV \cV^{-1} + \cV F\cV^{-1}\ .
\la{g15}
\ee
Now $d\cV \cV^{-1}=P+Q$, where $Q$ is the composite connection for $\gh$,  the Lie algebra of $H$, 
while $P$, which takes its values in the quotient of $\gg$ by $\gh$, can be considered as the one-form field-strength tensor for the scalar fields.
Note that $\F$ is invariant under $G$, so that we can consider $\cV F\cV^{-1}:=\tilde F$ to be the field-strength forms in the $H$-basis. The Maurer-Cartan equation for $\F$ then gives
\begin{align}
R+ DP+ P^2&=0\ ,\nn\w1
D\tilde F + \tilde F^2 + [\tilde F,P]&=0\ ,
\la{g16}
\end{align}
where $R=dQ+Q^2$ is the $\gh$-curvature and $D$ the $\gh$-covariant derivative.

The above discussion is applicable in the ungauged case. To include the scalars in the gauged case we put
\begin{align} 
\F&=\Th +d_{\Th} (\cV e^A)\,e^{-A}\cV^{-1}\nn\w1
&=\cD\cV \cV^{-1} +\cV(\Th + F)\cV^{-1}\nn\w1
&=\cP+\cQ +\tilde\Th+\tilde F\black \ , 
\la{g17}
\end{align}
where $\cP,\cQ$ are the gauged counterparts of $P,Q$.
The  extra $\Th$-term on the first line is necessary in order to obtain
$\cV\Th\cV^{-1}$ on the second line.\footnote{ This dressed version of the embedding tensor, $\tilde\Th$,  is the original one, known as the T-tensor
\cite{deWit:1981eq,deWit:2002vt,deWit:2003hr}.\black}
Conjugation with $\cV$ converts $\Th$ and $F$ from the $G$-basis to the $H$-basis as indicated on the third line.
The $\cA$ gauge-field in $\cD$ acting on the scalars comes from the level-zero term in $d_{\Th}e^A e^{-A}$. It is not difficult to show that $\F$ satisfies a standard Maurer-Cartan equation $d\F+\F^2=0$. Written out it gives
\begin{align}  
R+D\cP+\cP^2&=  - [\tilde F_2,\Th]= \cV\cF\cV^{-1}\nn\ ,\w1
D\tilde F+ \tilde F^2 +[\tilde F,\cP] + [\tilde F_{\ell\geq2},\tilde \Th]&= 0\nn\ ,\w1
D\tilde\Th + [\tilde\Th,\cP]&=0\ ,
\la{g18}
\end{align}
where $D=d+\cQ$ is the $\gh$-covariant derivative for the gauged theory, and $R=d\cQ +\cQ^2$.  Note that the tilded quantities do not transform under $G$ and, as a result, are also invariant under $G_0$.

\subsection{Curvature deformations}

In the ungauged theory one has local $H$ and rigid $G$ symmetries, but  if the former is included in the superspace structure group, the components of the torsion and curvature tensors do not transform under $G$. In the gauged theory formulated as above there are local $G_0$ and $H$ symmetries. Duality symmetry is lost, although it is still there formally due to the spurionic nature of the embedding tensor. In the geometrical sector of the theory the components of the torsion and curvature tensors transform covariantly under $H$, but are formally invariant under $G$. Nevertheless there are deformations compared to the ungauged case that start at dimension one (because $\Th$ has dimension one). In order to accommodate these in maximal supergravity theories it is necessary to amend the superspace tensors appropriately, which implies, for $D<10$, that there must be at least partially off-shell extensions of the constraints that were imposed to go on-shell in the ungauged case. In fact, one can see from the first of equations \eq{g18} that there must be dimension-one scalars in the $\gh$-curvature and in $D\cP$ that together fill out the representations of $\gh$ contained in the embedding tensor.

The simplest case is $D=3$ where it is known that imposing the standard flatness constraint on the dimension-zero torsion leads to an off-shell conformal supergravity multiplet \cite{Howe:1995zm}. The embedding tensor is in the ${\bf1}+{\bf3875}$ of $\ge_8$ which decomposes to
${\bf1}+{\bf135}+{\bf1820}+{\bf1920'}$ under $\gh=\gs\go(16)$, where ${\bf1920'}$ is a spinor representation (gamma-traceless vector-spinor).
The ${\bf1820}$ is an $\gs\go(16)$ four-form which is the leading component of the super Cotton tensor, while the ${\bf1}+{\bf135}$ together make up a symmetric two-index tensor. This is not in the conformal supergravity multiplet but is a $\th^2$-component of a scalar superfield that reflects the invariance of the conformal theory under local scale transformations. The ${\bf1920'}$ representation can be found in the dimension-one component of $D\cP$. At dimension one-half one has $P_{\a i I}\propto (\S)_{IJ'} \L_\a^{J'}$, where $\L$ is the physical fermion field, $\a=1,2; \ i=1,\ldots, 16,$ and $I,I'$ are 128-component $Spin(16)$ indices. At dimension one, $D_{\a i} \L_{\b J'}$ can therefore contain a term of the form $\ve_{\a\b}$ multiplied by a Lorentz scalar in the ${\bf1920'}$ representation \cite{Greitz:2011vh}.

The appropriate partially off-shell constraints for $D>3$ are only known for $D=4$ \cite{Howe:1981gz}. In this case the embedding tensor is in the
${\bf912}$ of $\ge_7$ and the dimension-one scalars in the torsion, curvature and $D\cP$ relevant to gauging were identified in \cite{Greitz:2011vh}. 

\subsection{Supersymmetry}

In this subsection we shall demonstrate that the full system of gauged Bianchi identities is compatible with supersymmetry to all orders if the ungauged version is. Writing the Bianchi identities in the form
\be
I=d_{\Th} F' + F'^2\ ,
\la{g19}
\ee
and applying  $d_{\Th}$ to this we find, after a short calculation,
\be
\cD I=[F,I]-[I,\Th]\ .
\la{g20}
\ee
In the ungauged case, we had $dI=[F,I]$, \ie $dI=0$ mod $I$, so that we could examine each identity in sequence. This meant that if we had solved up to level $\ell=k$, \ie all $I_{\ell+2}$ up this level $k$ are satisfied, then at the next level, we could use $dI_{k+3}=0$ in order to facilitate the analysis of its different $(p,q)$-form components. In the gauged case, however, the second term on the right of \eq{g20} means that a little more care is required. 

We can arrange the non-zero Bianchi identities in an array:
\bea
\ell=1&&\ I_3: \qquad{\phantom {I_{0,4}}}\qquad I_{0,3}\qquad I_{1,2} \qquad I_{2,1}\qquad I_{3,0}       \nn\w1
\ell=2&&\ I_4: \qquad I_{0,4}\qquad I_{1,3}\qquad I_{2,2} \qquad I_{3,1}\qquad I_{4,0}\nn\w1
\ell=3&&\ I_5: \qquad I_{1,4}\qquad I_{2,3}\qquad I_{3,2} \qquad I_{4,1}\qquad I_{5,0}\nn\w1
\ell=4&&\ I_6: \qquad I_{2,4}\qquad I_{3,3}\qquad I_{4,2}\qquad I_{5,1}\qquad I_{6,0} \nn\w1
\ell=5&&\ I_7: \qquad I_{3,4}\qquad I_{4,3}\qquad\ \  : \qquad \ \ \ :\qquad \ \ \ \,:\nn\w1
\ell=6&&\ I_8: \qquad I_{4,4}\qquad \ \ :\qquad \ \ \ \ : \qquad \ \ \ :\qquad \ \ \ \,:
\la{g21}
\eea
and so on, where the columns correspond to dimension zero, one-half, one, three-halves and two respectively. The idea is to solve the set of identities for $I_{p,q}$ starting at $p=0$, work through all of the $q$s and then go on to $p=1$ and so on. In other words, starting at the top unsolved row we solve for the next value of $p$ and then work downwards on left diagonal keeping $p$ fixed but increasing $q$ in a stepwise fashion. For example, if we have solved for $p=0,1$ we then solve $I_{2,1}, I_{2,2}, I_{2,3}, I_{2,4}$ and then $I_{3,0}, I_{3,1},\ldots$ and so on. 

For maximal supergravity in $3\leq D\leq 9$, the claim is that, if we have solved $I_{0,3}$ and $I_{0,4}$, then all of the other non-trivial Bianchi identity components can be solved by specifying the non-zero components of the field-strength forms.  The solutions to $I_{0,3}=0$ and $I_{0,4}=0$ are the same as in the ungauged case because the deformation terms involve a mass parameter. In particular $I_{0,4}=0$ will be solvable if the supersymmetry constraint is imposed. Now consider the $(0,4)$-component of \eq{g20}. It is
\be
t_0 I_{1,2}=-[I_{0,4},\Th]\ ,
\la{g22}
\ee
as there is no $I_2$. But since $I_{0,4}=0$, $t_0 I_{1,2}=0\Rightarrow I_{1,2}=t_0 J_{2,0}$. Thus $I_{1,2}=0$  if we set $J_{2,0}=0$ which just allows us to identify $F_{2,0}$. We now move on to $I_{1,3}$. We have
\be
t_0 I_{1,3}=[F_{0,2},I_{0,3}] -[I_{0,5},\Th]\ .
\la{g23}
\ee
The first term on the right vanishes because we have assumed that $I_{0,3}=0$, while the second term vanishes because $I_{0,5}=0$ identically in supergravity on dimensional grounds. So $t_0 I_{1,3}=0\Rightarrow I_{1,3}=t_0 J_{2,1}$, so we can solve this identity by setting $J_{2,1}=0$ \ie by finding $F_{2,1}$. Proceeding in this way, we find that all of the Bianchis are solvable if $I_{0,3}=I_{0,4}=0$. The difference with the ungauged case lies with the $\Th$ term on the right of \eq{g20}, but because $\Th$ has dimension one, this term has no effect if we solve the identities in the above order. We therefore conclude that the full set of gauged Bianchi identities is consistent with supersymmetry, and that there is therefore no need to make any explicit checks of the supersymmetry transformations.\footnote{For the $D=10$ IIA case, the Romans massive deformation \cite{Romans:1985tz} was discussed from an algebraic point of view in \cite{Lavrinenko:1999xi}. This can be viewed as a simple example of the truncated THA formalism. All the forms, including the OTT ones, were discussed in a superspace framework in \cite{Greitz:2011da}.}

The above analysis can be extended to the half-maximal case,
although the cohomology is more involved as we saw in the ungauged case.

\section{Concluding remarks}

In this paper we have presented a detailed analysis of the forms in (half-)maximal supergravity theories in dimensions $D\geq 3$. The use of superspace methods has allowed us to prove that all of the forms are consistent with supersymmetry and to directly construct Lie superalgebras from the associated Bianchi identities. We then showed how these form algebras could be extended to Borcherds superalgebras in the non-gauged case, by adding negative levels symmetrically about level zero, and to (truncated)  tensor hierarchy algebras in the gauged case, by including a level minus-one generator, related to the embedding tensor, in a way that is natural from the point of view of spacetime duality. In the latter case the Maurer-Cartan form associated with the formal group obtained from the tensor hierarchy algebra leads to a very simple description of the hierarchy of tensor gauge fields for gauged supergravity theories.

We have shown that the Borcherds superalgebra $\cB$ can be obtained 
from an associated Kac-Moody algebra $\cA$ by assigning V-degrees to the simple roots.
A positive V-degree of a simple root specifies both the form degree of a corresponding generating form, and to which node in an
additional ``gravity line''
we shall connect the corresponding node in the Dynkin diagram of $\cA$ 
when we extend it to a larger Kac-Moody algebra $\cC$.
Of course, neither of these interpretations is possible for negative V-degrees,
but nevertheless we saw that it is natural to allow simple roots of V-degree $-1$, when we consider both $\cB$ and $\cC$
as subalgebras of a unifying Borcherds superalgebra $\cD$. This calls for a deeper understanding of the V-degrees (and also of the map $w$,
through which they enter
in the relation between $\cA$ and $\cB$), and raises the question whether the simple root of V-degree $-1$ can be associated to 
the embedding tensor. The answer might be given by a construction of the tensor hierarchy algebra similar to the construction of the Borcherds superalgebra, or optimally, a unified construction of both. However, such a construction still remains to be found.

The fact that there are infinite-dimensional algebras that arise naturally in supergravity theories raises the question of whether they continue to be relevant in the presence of string theory corrections, and if so, in what way. In this paper we have shown that hierarchies of forms extend naturally beyond the spacetime limit, and that even without higher-order corrections, there can be non-zero field-strength forms with degree $(D+2)$. In \cite{Greitz:2012vp} it was argued that one could expect there to be higher-degree forms that will be turned on in the presence of first-order $\a'$ corrections in half-maximal $D=3$ theories. So if these algebras remain relevant one would certainly expect there to be non-zero forms with higher and higher degrees as the powers of $\a'$ are increased. It might also be that the algebras themselves are deformed by $\a'$ corrections. We certainly know that this happens in the presence of anomalies, for example in $D=11$ \cite{Duff:1995wd}, or in the heterotic and type I string theories  in $D=10$, but it might be the case that such corrections are more commonplace. Moreover, in the presence of non-perturbative effects, one expects the continuous duality symmetries to be replaced by discrete ones, although this in itself does not rule out the possibility that these extended algebras remain relevant. 
\\\\
\noindent
{\bf Acknowledgments}\\\\
\noindent
This project was initially set up with Jesper Greitz. We thank him for his contributions and for the many discussions we have had with him on this topic over the past few years. The work of JP is supported by NSF grant PHY-1214344.

\appendix

\section{Maximal form spectrum} \label{maxreps-app}

Below we list, up to $\ell=D+1$, the representations $\cR_\ell$ of the duality group $\gg=\ge_{11-D}$ that the potential forms of degree $\ell$ transform under in $D$-dimensional maximal supergravity. As we have discussed in the paper, these can be obtained by decomposing the corresponding Borcherds superalgebra $\cB$ (see Table \ref{maxtable}) with respect to the subalgebra $\gg=\ge_{11-D}$. The multiplicities of the representations in the tables are equal to one if not written out explicitly.

\noindent
$D=9:$
\setlength{\arraycolsep}{7.13pt}
\renewcommand{\arraystretch}{1.5}
\begin{flalign*}
\begin{array}{|c|cc|c|c|c|c|cc|cc|cc|cc|cccc|c}
\hline
\ell=k_0+k_{0'} 
& \multicolumn{2}{c|}{1} & 2 &3 & 4&5&\multicolumn{2}{c|}{6}&\multicolumn{2}{c|}{7}&\multicolumn{2}{c|}{8}&\multicolumn{2}{c|}{9}&\multicolumn{4}{c|}{10}&\\
\hline
k_{0} & 1&0&1&2&2&3&4&3&4&4&5&4&5&5&6&6&5&5&\\
k_{0'} &0&1&1&1&2&2&2&3&3&3&3&4&4&4&4&4&5&5&\\
\hline
\cR_\ell &{\bf 2}&{\bf 1}&{\bf 2}&{\bf 1}&{\bf 1}&{\bf 2}&{\bf 1}&{\bf 2}&{\bf 1}
&{\bf 3}&{\bf 2}&{\bf 3}&{\bf 2}&{\bf 4}&{\bf 1}&{\bf 3}&{\bf 2}&{\bf 4}&\\
\hline
\renewcommand{\arraystretch}{1.0}
\begin{array}{c} \text{\,Dynkin}\\ \text{\,labels} \end{array}
& 
\begin{picture}(5,20)(0,2)
\put(0,0){${\scriptstyle 1}$}
\end{picture}& 
\begin{picture}(5,20)(0,2)
\put(0,0){${\scriptstyle 0}$}
\end{picture}&
\begin{picture}(5,20)(0,2)
\put(0,0){${\scriptstyle 1}$}
\end{picture}&
\begin{picture}(5,20)(0,2)
\put(0,0){${\scriptstyle 0}$}
\end{picture}&
\begin{picture}(5,20)(0,2)
\put(0,0){${\scriptstyle 0}$}
\end{picture}&
\begin{picture}(5,20)(0,2)
\put(0,0){${\scriptstyle 1}$}
\end{picture}&
\begin{picture}(5,20)(0,2)
\put(0,0){${\scriptstyle 0}$}
\end{picture}&
\begin{picture}(5,20)(0,2)
\put(0,0){${\scriptstyle 1}$}
\end{picture}&
\begin{picture}(5,20)(0,2)
\put(0,0){${\scriptstyle 0}$}
\end{picture}&
\begin{picture}(5,20)(0,2)
\put(0,0){${\scriptstyle 2}$}
\end{picture}&
\begin{picture}(5,20)(0,2)
\put(0,0){${\scriptstyle 1}$}
\end{picture}&
\begin{picture}(5,20)(0,2)
\put(0,0){${\scriptstyle 2}$}
\end{picture}&
\begin{picture}(5,20)(0,2)
\put(0,0){${\scriptstyle 1}$}
\end{picture}&
\begin{picture}(5,20)(0,2)
\put(0,0){${\scriptstyle 3}$}
\end{picture}&
\begin{picture}(5,20)(0,2)
\put(0,0){${\scriptstyle 0}$}
\end{picture}
&
\begin{picture}(5,20)(0,2)
\put(0,0){${\scriptstyle 2}$}
\end{picture}
&
\begin{picture}(5,20)(0,2)
\put(0,0){${\scriptstyle 1}$}
\end{picture}
&
\begin{picture}(5,20)(0,2)
\put(0,0){${\scriptstyle 3}$}
\end{picture}
&
\\
\hline
\text{multiplicity}& 1&1&1&1&1&1&1&1&1&1&1&1&2&1&1&2&1&1&\\
\hline
\end{array} &&
\end{flalign*}

\noindent
$D=8:$
\setlength{\arraycolsep}{5.35pt}
\renewcommand{\arraystretch}{1.5}
\begin{flalign*}
\begin{array}{|c|c|c|c|c|c|cc|cc|ccc|ccc|c}
\hline
\ell & 1 & 2&3 & 4&5&\multicolumn{2}{c|}{6}&\multicolumn{2}{c|}{7}&\multicolumn{3}{c|}{8}&\multicolumn{3}{c|}{9}&\\
\hline
\cR_\ell{}^{(A_1)} & {\bf 2}&{\bf 1}&{\bf 2}&{\bf 1}&{\bf 2}&{\bf 3}&{\bf 1}&{\bf 2}&{\bf 2}&{\bf 1}&{\bf 3}&{\bf 1}&{\bf 2}&{\bf 2}&{\bf 2}&\\
\cR_\ell{}^{(A_2)} &{\bf 3}&\overline{\bf 3}&{\bf 1}&{\bf 3}&\overline{\bf 3}&{\bf 1}&{\bf 8}&{\bf 3}&\overline{\bf 6}
&\overline{\bf 3}&\overline{\bf 3}&{\bf 15}&{\bf 1}&{\bf 8}&\overline{\bf 10}&\\
\hline
\renewcommand{\arraystretch}{1.0}
\begin{array}{c} \text{\,Dynkin}\\ \text{\,labels} \end{array}
& 
\begin{picture}(15,20)(0,2)
\put(5,0){${\scriptstyle 1}$}
\put(10,0){${\scriptstyle 0}$}
\put(0,8){${\scriptstyle 1}$}
\end{picture}& 
\begin{picture}(15,20)(0,2)
\put(5,0){${\scriptstyle 0}$}
\put(10,0){${\scriptstyle 1}$}
\put(0,8){${\scriptstyle 0}$}
\end{picture}&
\begin{picture}(15,20)(0,2)
\put(5,0){${\scriptstyle 0}$}
\put(10,0){${\scriptstyle 0}$}
\put(0,8){${\scriptstyle 1}$}
\end{picture}&
\begin{picture}(15,20)(0,2)
\put(5,0){${\scriptstyle 1}$}
\put(10,0){${\scriptstyle 0}$}
\put(0,8){${\scriptstyle 0}$}
\end{picture}&
\begin{picture}(15,20)(0,2)
\put(5,0){${\scriptstyle 0}$}
\put(10,0){${\scriptstyle 1}$}
\put(0,8){${\scriptstyle 1}$}
\end{picture}&
\begin{picture}(15,20)(0,2)
\put(5,0){${\scriptstyle 0}$}
\put(10,0){${\scriptstyle 0}$}
\put(0,8){${\scriptstyle 2}$}
\end{picture}&
\begin{picture}(15,20)(0,2)
\put(5,0){${\scriptstyle 1}$}
\put(10,0){${\scriptstyle 1}$}
\put(0,8){${\scriptstyle 0}$}
\end{picture}&
\begin{picture}(15,20)(0,2)
\put(5,0){${\scriptstyle 1}$}
\put(10,0){${\scriptstyle 0}$}
\put(0,8){${\scriptstyle 1}$}
\end{picture}&
\begin{picture}(15,20)(0,2)
\put(5,0){${\scriptstyle 0}$}
\put(10,0){${\scriptstyle 2}$}
\put(0,8){${\scriptstyle 1}$}
\end{picture}&
\begin{picture}(15,20)(0,2)
\put(5,0){${\scriptstyle 0}$}
\put(10,0){${\scriptstyle 1}$}
\put(0,8){${\scriptstyle 0}$}
\end{picture}&
\begin{picture}(15,20)(0,2)
\put(5,0){${\scriptstyle 0}$}
\put(10,0){${\scriptstyle 1}$}
\put(0,8){${\scriptstyle 2}$}
\end{picture}&
\begin{picture}(15,20)(0,2)
\put(5,0){${\scriptstyle 1}$}
\put(10,0){${\scriptstyle 2}$}
\put(0,8){${\scriptstyle 0}$}
\end{picture}&
\begin{picture}(15,20)(0,2)
\put(5,0){${\scriptstyle 0}$}
\put(10,0){${\scriptstyle 0}$}
\put(0,8){${\scriptstyle 1}$}
\end{picture}&
\begin{picture}(15,20)(0,2)
\put(5,0){${\scriptstyle 1}$}
\put(10,0){${\scriptstyle 1}$}
\put(0,8){${\scriptstyle 1}$}
\end{picture}&
\begin{picture}(15,20)(0,2)
\put(5,0){${\scriptstyle 0}$}
\put(10,0){${\scriptstyle 3}$}
\put(0,8){${\scriptstyle 1}$}
\end{picture}
&\\
\hline
\text{mult.}& 1&1&1&1&1&1&1&1&1&2&1&1&2&2&1&\\
\hline
\end{array} &&
\end{flalign*}

\noindent
$D=7:$
\setlength{\arraycolsep}{5.99pt}
{\renewcommand{\arraystretch}{1.5}
\begin{flalign*}
\begin{array}{|c|c|c|c|c|c|cc|ccc|cccc|c}
\hline
\ell & 1 & 2&3 & 4&5&\multicolumn{2}{c|}{6}&\multicolumn{3}{c|}{7}&\multicolumn{4}{c|}{8}& \\
\hline
\cR_\ell & \overline{\bf 10}&{\bf5}&\overline{\bf 5}&{\bf 10}&{\bf 24}&\overline{\bf 15}&{\bf 40}&{\bf 5}&\overline{\bf 45}&{\bf 70}
&\overline{\bf 5}&{\bf 45}&\overline{\bf 70}&{\bf 105}&\\
\hline
\renewcommand{\arraystretch}{1.0}
\begin{array}{c} \text{\,Dynkin}\\ \text{\,labels} \end{array}& 
\begin{picture}(15,20)(0,2)
\put(0,0){${\scriptstyle 1}$}
\put(5,0){${\scriptstyle 0}$}
\put(10,0){${\scriptstyle 0}$}
\put(0,8){${\scriptstyle 0}$}
\end{picture}& 
\begin{picture}(15,20)(0,2)
\put(0,0){${\scriptstyle 0}$}
\put(5,0){${\scriptstyle 0}$}
\put(10,0){${\scriptstyle 1}$}
\put(0,8){${\scriptstyle 0}$}
\end{picture}&
\begin{picture}(15,20)(0,2)
\put(0,0){${\scriptstyle 0}$}
\put(5,0){${\scriptstyle 0}$}
\put(10,0){${\scriptstyle 0}$}
\put(0,8){${\scriptstyle 1}$}
\end{picture}&
\begin{picture}(15,20)(0,2)
\put(0,0){${\scriptstyle 0}$}
\put(5,0){${\scriptstyle 1}$}
\put(10,0){${\scriptstyle 0}$}
\put(0,8){${\scriptstyle 0}$}
\end{picture}&
\begin{picture}(15,20)(0,2)
\put(0,0){${\scriptstyle 0}$}
\put(5,0){${\scriptstyle 0}$}
\put(10,0){${\scriptstyle 1}$}
\put(0,8){${\scriptstyle 1}$}
\end{picture}&
\begin{picture}(15,20)(0,2)
\put(0,0){${\scriptstyle 0}$}
\put(5,0){${\scriptstyle 0}$}
\put(10,0){${\scriptstyle 0}$}
\put(0,8){${\scriptstyle 2}$}
\end{picture}&
\begin{picture}(15,20)(0,2)
\put(0,0){${\scriptstyle 0}$}
\put(5,0){${\scriptstyle 1}$}
\put(10,0){${\scriptstyle 1}$}
\put(0,8){${\scriptstyle 0}$}
\end{picture}&
\begin{picture}(15,20)(0,2)
\put(0,0){${\scriptstyle 0}$}
\put(5,0){${\scriptstyle 0}$}
\put(10,0){${\scriptstyle 1}$}
\put(0,8){${\scriptstyle 0}$}
\end{picture}&
\begin{picture}(15,20)(0,2)
\put(0,0){${\scriptstyle 0}$}
\put(5,0){${\scriptstyle 1}$}
\put(10,0){${\scriptstyle 0}$}
\put(0,8){${\scriptstyle 1}$}
\end{picture}&
\begin{picture}(15,20)(0,2)
\put(0,0){${\scriptstyle 0}$}
\put(5,0){${\scriptstyle 0}$}
\put(10,0){${\scriptstyle 2}$}
\put(0,8){${\scriptstyle 1}$}
\end{picture}&
\begin{picture}(15,20)(0,2)
\put(0,0){${\scriptstyle 0}$}
\put(5,0){${\scriptstyle 0}$}
\put(10,0){${\scriptstyle 0}$}
\put(0,8){${\scriptstyle 1}$}
\end{picture}&
\begin{picture}(15,20)(0,2)
\put(0,0){${\scriptstyle 1}$}
\put(5,0){${\scriptstyle 0}$}
\put(10,0){${\scriptstyle 1}$}
\put(0,8){${\scriptstyle 0}$}
\end{picture}&
\begin{picture}(15,20)(0,2)
\put(0,0){${\scriptstyle 0}$}
\put(5,0){${\scriptstyle 0}$}
\put(10,0){${\scriptstyle 1}$}
\put(0,8){${\scriptstyle 2}$}
\end{picture}&
\begin{picture}(15,20)(0,2)
\put(0,0){${\scriptstyle 0}$}
\put(5,0){${\scriptstyle 1}$}
\put(10,0){${\scriptstyle 2}$}
\put(0,8){${\scriptstyle 0}$}
\end{picture}
&\\
\hline
\end{array} &&
\end{flalign*}
}

\noindent
$D=6:$
\setlength{\arraycolsep}{5.37pt}
{\renewcommand{\arraystretch}{1.5}
\begin{flalign*}
\begin{array}{|c|c|c|c|c|c|ccc|cccc|c}
\hline
\ell &1 & 2&3 & 4&5&\multicolumn{3}{c|}{6}&\multicolumn{4}{c|}{7}&
\\
\hline
\cR_\ell & {\bf 16}_c&{\bf10}&{\bf 16}_s&{\bf45}&{\bf 144}_s&{\bf 10}&{\bf 126}_s&{\bf 320}&{\bf 16}_s & {\bf 144}_c&{\bf 560}_s&{\bf 720}_s&\\
\hline
\renewcommand{\arraystretch}{1.0}
\begin{array}{c} \text{\,Dynkin}\\ \text{\,labels} \end{array}&
\begin{picture}(20,20)(0,2)
\put(0,0){${\scriptstyle 1}$}
\put(5,0){${\scriptstyle 0}$}
\put(10,0){${\scriptstyle 0}$}
\put(15,0){${\scriptstyle 0}$}
\put(5,8){${\scriptstyle 0}$}
\end{picture}& 
\begin{picture}(20,20)(0,2)
\put(0,0){${\scriptstyle 0}$}
\put(5,0){${\scriptstyle 0}$}
\put(10,0){${\scriptstyle 0}$}
\put(15,0){${\scriptstyle 1}$}
\put(5,8){${\scriptstyle 0}$}
\end{picture}& 
\begin{picture}(20,20)(0,2)
\put(0,0){${\scriptstyle 0}$}
\put(5,0){${\scriptstyle 0}$}
\put(10,0){${\scriptstyle 0}$}
\put(15,0){${\scriptstyle 0}$}
\put(5,8){${\scriptstyle 1}$}
\end{picture}& 
\begin{picture}(20,20)(0,2)
\put(0,0){${\scriptstyle 0}$}
\put(5,0){${\scriptstyle 0}$}
\put(10,0){${\scriptstyle 1}$}
\put(15,0){${\scriptstyle 0}$}
\put(5,8){${\scriptstyle 0}$}
\end{picture}& 
\begin{picture}(20,20)(0,2)
\put(0,0){${\scriptstyle 0}$}
\put(5,0){${\scriptstyle 0}$}
\put(10,0){${\scriptstyle 0}$}
\put(15,0){${\scriptstyle 1}$}
\put(5,8){${\scriptstyle 1}$}
\end{picture}&
\begin{picture}(20,20)(0,2)
\put(0,0){${\scriptstyle 0}$}
\put(5,0){${\scriptstyle 0}$}
\put(10,0){${\scriptstyle 0}$}
\put(15,0){${\scriptstyle 1}$}
\put(5,8){${\scriptstyle 0}$}
\end{picture}&
\begin{picture}(20,20)(0,2)
\put(0,0){${\scriptstyle 0}$}
\put(5,0){${\scriptstyle 0}$}
\put(10,0){${\scriptstyle 0}$}
\put(15,0){${\scriptstyle 0}$}
\put(5,8){${\scriptstyle 2}$}
\end{picture}&
\begin{picture}(20,20)(0,2)
\put(0,0){${\scriptstyle 0}$}
\put(5,0){${\scriptstyle 0}$}
\put(10,0){${\scriptstyle 1}$}
\put(15,0){${\scriptstyle 1}$}
\put(5,8){${\scriptstyle 0}$}
\end{picture}&
\begin{picture}(20,20)(0,2)
\put(0,0){${\scriptstyle 0}$}
\put(5,0){${\scriptstyle 0}$}
\put(10,0){${\scriptstyle 0}$}
\put(15,0){${\scriptstyle 0}$}
\put(5,8){${\scriptstyle 1}$}
\end{picture}&
\begin{picture}(20,20)(0,2)
\put(0,0){${\scriptstyle 1}$}
\put(5,0){${\scriptstyle 0}$}
\put(10,0){${\scriptstyle 0}$}
\put(15,0){${\scriptstyle 1}$}
\put(5,8){${\scriptstyle 0}$}
\end{picture}&
\begin{picture}(20,20)(0,2)
\put(0,0){${\scriptstyle 0}$}
\put(5,0){${\scriptstyle 0}$}
\put(10,0){${\scriptstyle 1}$}
\put(15,0){${\scriptstyle 0}$}
\put(5,8){${\scriptstyle 1}$}
\end{picture}&
\begin{picture}(20,20)(0,2)
\put(0,0){${\scriptstyle 0}$}
\put(5,0){${\scriptstyle 0}$}
\put(10,0){${\scriptstyle 0}$}
\put(15,0){${\scriptstyle 2}$}
\put(5,8){${\scriptstyle 1}$}
\end{picture}&\\
\hline
\end{array} &&
\end{flalign*}
}

\noindent
$D=5:$
\setlength{\arraycolsep}{5.07pt}
{\renewcommand{\arraystretch}{1.5}
\begin{flalign*}
\begin{array}{|c|c|c|c|c|cc|ccccc|c}
\hline
\ell & 1 & 2&3 &4& \multicolumn{2}{c|}{5}&\multicolumn{5}{c|}{6}&\\
\hline
\cR_\ell &\overline{\bf 27}&{\bf27}&{\bf 78}&{\bf351}&{\bf 27}&{\bf 1728}&{\bf 1}&{\bf 78}&{\bf 650}&{\bf 2430}&{\bf 5824}&\\
\hline
\renewcommand{\arraystretch}{1.0}
\begin{array}{c} \text{\,Dynkin}\\ \text{\,labels} \end{array}&
\begin{picture}(25,20)(0,2)
\put(0,0){${\scriptstyle 1}$}
\put(5,0){${\scriptstyle 0}$}
\put(10,0){${\scriptstyle 0}$}
\put(15,0){${\scriptstyle 0}$}
\put(20,0){${\scriptstyle 0}$}
\put(10,8){${\scriptstyle 0}$}
\end{picture}& 
\begin{picture}(25,20)(0,2)
\put(0,0){${\scriptstyle 0}$}
\put(5,0){${\scriptstyle 0}$}
\put(10,0){${\scriptstyle 0}$}
\put(15,0){${\scriptstyle 0}$}
\put(20,0){${\scriptstyle 1}$}
\put(10,8){${\scriptstyle 0}$}
\end{picture}& 
\begin{picture}(25,20)(0,2)
\put(0,0){${\scriptstyle 0}$}
\put(5,0){${\scriptstyle 0}$}
\put(10,0){${\scriptstyle 0}$}
\put(15,0){${\scriptstyle 0}$}
\put(20,0){${\scriptstyle 0}$}
\put(10,8){${\scriptstyle 1}$}
\end{picture}& 
\begin{picture}(25,20)(0,2)
\put(0,0){${\scriptstyle 0}$}
\put(5,0){${\scriptstyle 0}$}
\put(10,0){${\scriptstyle 0}$}
\put(15,0){${\scriptstyle 1}$}
\put(20,0){${\scriptstyle 0}$}
\put(10,8){${\scriptstyle 0}$}
\end{picture}& 
\begin{picture}(25,20)(0,2)
\put(0,0){${\scriptstyle 0}$}
\put(5,0){${\scriptstyle 0}$}
\put(10,0){${\scriptstyle 0}$}
\put(15,0){${\scriptstyle 0}$}
\put(20,0){${\scriptstyle 1}$}
\put(10,8){${\scriptstyle 0}$}
\end{picture}&
\begin{picture}(25,20)(0,2)
\put(0,0){${\scriptstyle 0}$}
\put(5,0){${\scriptstyle 0}$}
\put(10,0){${\scriptstyle 0}$}
\put(15,0){${\scriptstyle 0}$}
\put(20,0){${\scriptstyle 1}$}
\put(10,8){${\scriptstyle 1}$}
\end{picture}&\begin{picture}(25,20)(0,0)
\put(0,0){${\scriptstyle 0}$}
\put(5,0){${\scriptstyle 0}$}
\put(10,0){${\scriptstyle 0}$}
\put(15,0){${\scriptstyle 0}$}
\put(20,0){${\scriptstyle 0}$}
\put(10,8){${\scriptstyle 0}$}
\end{picture}& 
\begin{picture}(25,20)(0,2)
\put(0,0){${\scriptstyle 0}$}
\put(5,0){${\scriptstyle 0}$}
\put(10,0){${\scriptstyle 0}$}
\put(15,0){${\scriptstyle 0}$}
\put(20,0){${\scriptstyle 0}$}
\put(10,8){${\scriptstyle 1}$}
\end{picture}& 
\begin{picture}(25,20)(0,2)
\put(0,0){${\scriptstyle 1}$}
\put(5,0){${\scriptstyle 0}$}
\put(10,0){${\scriptstyle 0}$}
\put(15,0){${\scriptstyle 0}$}
\put(20,0){${\scriptstyle 1}$}
\put(10,8){${\scriptstyle 0}$}
\end{picture}& 
\begin{picture}(25,20)(0,2)
\put(0,0){${\scriptstyle 0}$}
\put(5,0){${\scriptstyle 0}$}
\put(10,0){${\scriptstyle 0}$}
\put(15,0){${\scriptstyle 0}$}
\put(20,0){${\scriptstyle 0}$}
\put(10,8){${\scriptstyle 2}$}
\end{picture}&
\begin{picture}(25,20)(0,2)
\put(0,0){${\scriptstyle 0}$}
\put(5,0){${\scriptstyle 0}$}
\put(10,0){${\scriptstyle 0}$}
\put(15,0){${\scriptstyle 1}$}
\put(20,0){${\scriptstyle 1}$}
\put(10,8){${\scriptstyle 0}$}
\end{picture}&
\\
\hline
\end{array} &&
\end{flalign*}
}

\noindent
$D=4:$
\setlength{\arraycolsep}{6.08pt}
{\renewcommand{\arraystretch}{1.5}
\begin{flalign*}
\begin{array}{|c|c|c|c|cc|cccc|c}
\hline
\ell & 1 & 2&3 & \multicolumn{2}{c|}{4}&\multicolumn{4}{c|}{5}&
\\
\hline
\cR_\ell & {\bf56}&{\bf133}&{\bf912}&{\bf133}&{\bf8645}&{\bf 56}&{\bf 912}&{\bf 6480}&{\bf 86184}&\\
\hline
\renewcommand{\arraystretch}{1.0}
\begin{array}{c} \text{\,Dynkin}\\ \text{\,labels} \end{array}&
\begin{picture}(30,20)(0,2)
\put(0,0){${\scriptstyle 1}$}
\put(5,0){${\scriptstyle 0}$}
\put(10,0){${\scriptstyle 0}$}
\put(15,0){${\scriptstyle 0}$}
\put(20,0){${\scriptstyle 0}$}
\put(25,0){${\scriptstyle 0}$}
\put(15,8){${\scriptstyle 0}$}
\end{picture}& 
\begin{picture}(30,20)(0,2)
\put(0,0){${\scriptstyle 0}$}
\put(5,0){${\scriptstyle 0}$}
\put(10,0){${\scriptstyle 0}$}
\put(15,0){${\scriptstyle 0}$}
\put(20,0){${\scriptstyle 0}$}
\put(25,0){${\scriptstyle 1}$}
\put(15,8){${\scriptstyle 0}$}
\end{picture}& 
\begin{picture}(30,20)(0,2)
\put(0,0){${\scriptstyle 0}$}
\put(5,0){${\scriptstyle 0}$}
\put(10,0){${\scriptstyle 0}$}
\put(15,0){${\scriptstyle 0}$}
\put(20,0){${\scriptstyle 0}$}
\put(25,0){${\scriptstyle 0}$}
\put(15,8){${\scriptstyle 1}$}
\end{picture}& 
\begin{picture}(30,20)(0,2)
\put(0,0){${\scriptstyle 0}$}
\put(5,0){${\scriptstyle 0}$}
\put(10,0){${\scriptstyle 0}$}
\put(15,0){${\scriptstyle 0}$}
\put(20,0){${\scriptstyle 0}$}
\put(25,0){${\scriptstyle 1}$}
\put(15,8){${\scriptstyle 0}$}
\end{picture}& 
\begin{picture}(30,20)(0,2)
\put(0,0){${\scriptstyle 0}$}
\put(5,0){${\scriptstyle 0}$}
\put(10,0){${\scriptstyle 0}$}
\put(15,0){${\scriptstyle 0}$}
\put(20,0){${\scriptstyle 1}$}
\put(25,0){${\scriptstyle 0}$}
\put(15,8){${\scriptstyle 0}$}
\end{picture}&
\begin{picture}(30,20)(0,2)
\put(0,0){${\scriptstyle 1}$}
\put(5,0){${\scriptstyle 0}$}
\put(10,0){${\scriptstyle 0}$}
\put(15,0){${\scriptstyle 0}$}
\put(20,0){${\scriptstyle 0}$}
\put(25,0){${\scriptstyle 0}$}
\put(15,8){${\scriptstyle 0}$}
\end{picture}& 
\begin{picture}(30,20)(0,2)
\put(0,0){${\scriptstyle 0}$}
\put(5,0){${\scriptstyle 0}$}
\put(10,0){${\scriptstyle 0}$}
\put(15,0){${\scriptstyle 0}$}
\put(20,0){${\scriptstyle 0}$}
\put(25,0){${\scriptstyle 0}$}
\put(15,8){${\scriptstyle 1}$}
\end{picture}& 
\begin{picture}(30,20)(0,2)
\put(0,0){${\scriptstyle 1}$}
\put(5,0){${\scriptstyle 0}$}
\put(10,0){${\scriptstyle 0}$}
\put(15,0){${\scriptstyle 0}$}
\put(20,0){${\scriptstyle 0}$}
\put(25,0){${\scriptstyle 1}$}
\put(15,8){${\scriptstyle 0}$}
\end{picture}& 
\begin{picture}(30,20)(0,2)
\put(0,0){${\scriptstyle 0}$}
\put(5,0){${\scriptstyle 0}$}
\put(10,0){${\scriptstyle 0}$}
\put(15,0){${\scriptstyle 0}$}
\put(20,0){${\scriptstyle 0}$}
\put(25,0){${\scriptstyle 1}$}
\put(15,8){${\scriptstyle 1}$}
\end{picture}&\\
\hline
\end{array} &&
\end{flalign*}
}

\noindent
$D=3:$
\setlength{\arraycolsep}{5.51pt}
{\renewcommand{\arraystretch}{1.5}
\begin{flalign*}
\begin{array}{|c|c|cc|ccc|c}
\hline
\ell &1 & \multicolumn{2}{c|}{2} & \multicolumn{3}{c|}{3}&\qquad\qquad \qquad 4\quad  \qquad\quad\!\!\!\!\!\\
\hline
\cR_\ell & {\bf248}&{\bf1}&{\bf3875}&{\bf248}&{\bf3875}&{\bf147250}&\multirow{2}{*}{\it $\ $ see below}\\
\cline{1-7}
\renewcommand{\arraystretch}{1.0}
\begin{array}{c} \text{\,Dynkin}\\ \text{\,labels} \end{array}&
\begin{picture}(35,20)(0,2)
\put(0,0){${\scriptstyle 1}$}
\put(5,0){${\scriptstyle 0}$}
\put(10,0){${\scriptstyle 0}$}
\put(15,0){${\scriptstyle 0}$}
\put(20,0){${\scriptstyle 0}$}
\put(25,0){${\scriptstyle 0}$}
\put(30,0){${\scriptstyle 0}$}
\put(20,7){${\scriptstyle 0}$}
\end{picture}& 
\begin{picture}(35,20)(0,2)
\put(0,0){${\scriptstyle 0}$}
\put(5,0){${\scriptstyle 0}$}
\put(10,0){${\scriptstyle 0}$}
\put(15,0){${\scriptstyle 0}$}
\put(20,0){${\scriptstyle 0}$}
\put(25,0){${\scriptstyle 0}$}
\put(30,0){${\scriptstyle 0}$}
\put(20,7){${\scriptstyle 0}$}
\end{picture}& 
\begin{picture}(35,20)(0,2)
\put(0,0){${\scriptstyle 0}$}
\put(5,0){${\scriptstyle 0}$}
\put(10,0){${\scriptstyle 0}$}
\put(15,0){${\scriptstyle 0}$}
\put(20,0){${\scriptstyle 0}$}
\put(25,0){${\scriptstyle 0}$}
\put(30,0){${\scriptstyle 1}$}
\put(20,7){${\scriptstyle 0}$}
\end{picture}& 
\begin{picture}(35,20)(0,2)
\put(0,0){${\scriptstyle 1}$}
\put(5,0){${\scriptstyle 0}$}
\put(10,0){${\scriptstyle 0}$}
\put(15,0){${\scriptstyle 0}$}
\put(20,0){${\scriptstyle 0}$}
\put(25,0){${\scriptstyle 0}$}
\put(30,0){${\scriptstyle 0}$}
\put(20,7){${\scriptstyle 0}$}
\end{picture}& 
\begin{picture}(35,20)(0,2)
\put(0,0){${\scriptstyle 0}$}
\put(5,0){${\scriptstyle 0}$}
\put(10,0){${\scriptstyle 0}$}
\put(15,0){${\scriptstyle 0}$}
\put(20,0){${\scriptstyle 0}$}
\put(25,0){${\scriptstyle 0}$}
\put(30,0){${\scriptstyle 1}$}
\put(20,7){${\scriptstyle 0}$}
\end{picture}& 
\begin{picture}(35,20)(0,2)
\put(0,0){${\scriptstyle 0}$}
\put(5,0){${\scriptstyle 0}$}
\put(10,0){${\scriptstyle 0}$}
\put(15,0){${\scriptstyle 0}$}
\put(20,0){${\scriptstyle 0}$}
\put(25,0){${\scriptstyle 0}$}
\put(30,0){${\scriptstyle 0}$}
\put(20,7){${\scriptstyle 1}$}
\end{picture}
&\\
\hline
\end{array} &&
\end{flalign*}
}

\noindent
$D=3$ {\it (continued)} : 
\setlength{\arraycolsep}{10.73pt}
{\renewcommand{\arraystretch}{1.5}
\begin{flalign*}
\begin{array}{|c|cccccc|c}
\hline
\ell & \multicolumn{6}{c|}{4}& \\
\hline
\cR_\ell & 
{\bf 248}&{\bf 3875}& 
{\bf 30380}&{\bf 147250}&{\bf 779247}&{\bf 6696000}& \\
\hline
\renewcommand{\arraystretch}{1.0}
\begin{array}{c} \text{\,Dynkin}\\ \text{\,labels} \end{array}&
\begin{picture}(35,20)(0,2)
\put(0,0){${\scriptstyle 1}$}
\put(5,0){${\scriptstyle 0}$}
\put(10,0){${\scriptstyle 0}$}
\put(15,0){${\scriptstyle 0}$}
\put(20,0){${\scriptstyle 0}$}
\put(25,0){${\scriptstyle 0}$}
\put(30,0){${\scriptstyle 0}$}
\put(20,7){${\scriptstyle 0}$}
\end{picture}& 
\begin{picture}(35,20)(0,2)
\put(0,0){${\scriptstyle 0}$}
\put(5,0){${\scriptstyle 0}$}
\put(10,0){${\scriptstyle 0}$}
\put(15,0){${\scriptstyle 0}$}
\put(20,0){${\scriptstyle 0}$}
\put(25,0){${\scriptstyle 0}$}
\put(30,0){${\scriptstyle 1}$}
\put(20,7){${\scriptstyle 0}$}
\end{picture}& 
\begin{picture}(35,20)(0,2)
\put(0,0){${\scriptstyle 0}$}
\put(5,0){${\scriptstyle 1}$}
\put(10,0){${\scriptstyle 0}$}
\put(15,0){${\scriptstyle 0}$}
\put(20,0){${\scriptstyle 0}$}
\put(25,0){${\scriptstyle 0}$}
\put(30,0){${\scriptstyle 0}$}
\put(20,7){${\scriptstyle 0}$}
\end{picture}& 
\begin{picture}(35,20)(0,2)
\put(0,0){${\scriptstyle 0}$}
\put(5,0){${\scriptstyle 0}$}
\put(10,0){${\scriptstyle 0}$}
\put(15,0){${\scriptstyle 0}$}
\put(20,0){${\scriptstyle 0}$}
\put(25,0){${\scriptstyle 0}$}
\put(30,0){${\scriptstyle 0}$}
\put(20,7){${\scriptstyle 1}$}
\end{picture}& 
\begin{picture}(35,20)(0,2)
\put(0,0){${\scriptstyle 1}$}
\put(5,0){${\scriptstyle 0}$}
\put(10,0){${\scriptstyle 0}$}
\put(15,0){${\scriptstyle 0}$}
\put(20,0){${\scriptstyle 0}$}
\put(25,0){${\scriptstyle 0}$}
\put(30,0){${\scriptstyle 1}$}
\put(20,7){${\scriptstyle 0}$}
\end{picture}& 
\begin{picture}(35,20)(0,2)
\put(0,0){${\scriptstyle 0}$}
\put(5,0){${\scriptstyle 0}$}
\put(10,0){${\scriptstyle 0}$}
\put(15,0){${\scriptstyle 0}$}
\put(20,0){${\scriptstyle 0}$}
\put(25,0){${\scriptstyle 1}$}
\put(30,0){${\scriptstyle 0}$}
\put(20,7){${\scriptstyle 0}$}
\end{picture}
&\\
\hline
\text{multiplicity}&2&1&2&1&1&1\\
\hline
\end{array} &&
\end{flalign*}
}
\section{Half-maximal Bianchi identities} \label{bianchi-app}
\subsection*{Universal sector}

As we saw in section 3, there is a universal set of forms in half-maximal theories with $D>4$. This consists of the two-form field-strengths $F_2^\cM$, their duals $F_{D-2}^\cM$, the three-form field-strength of the supergravity two-form potential, $F_3$, and its dual $F_{D-3}$, as well as the higher-degree forms they generate.

The basic Bianchi identities are, for $D>4$:
\begin{align}
dF_2^\cM&=0\ , \qquad \ dF_3=F_2\cdot F_2\ , \nn\w1
dF_{D-3}&=0\ , \ \ \ \   dF_{D-2}^\cM= F_{D-3} F_2^\cM \ .
\la{4.6}
\end{align}
At level $(D-2)$ we have
\begin{align}
dF_{D-1}&=F_{D-2}\cdot F_2-F_{D-3} F_3\ ,\nn\w1
dF_{D-1}^{\cM\cN}&=2F_{D-2}^{[\cM} F_2^{\cN]}\ ,
\la{4.7}
\end{align}
at level $(D-1)$  
\begin{align}
dF_D^\cM&=F_{D-1}^{\cM\cN}F_{2\cN}-F_{D-2}^\cM F_3 + F_{D-1} F_2^\cM\ , \nn\w1
dF_D^{\cM\cN\cP}&=3 F_{D-1}^{[\cM\cN}F_2^{\cP]}\ ,
\la{4.8}
\end{align}
at level $D$
\begin{align}
dF_{D+1}&=F_D^\cM F_{2\cM}-F_{D-1} F_3\ ,\nn\w1
dF_{D+1}^{\cM\cN}&=F_D^{\cM\cN\cP} F_{2\cP} + 2F_D^{[\cM} F_2^{\cN]}-F_{D-1}^{\cM\cN} F_3\ ,\nn\w1
dF_{D+1}^{\cM\cN\cP\cQ}&=4 F_D^{[\cM\cN\cP} F_2^{\cQ]}\ ,
\la{4.10a}
\end{align}
and at level $(D+1)$,
\begin{align}
dF_{D+2}^{\cM}&=F_{D+1}^{\cM\cN} F_{2\cN}+F_{D+1} F_2^\cM -F_{D}^\cM F_3\ ,\nn\w1
dF_{D+2}^{\cM\cN\cP}&=F_{D+1}^{\cM\cN\cP\cQ} F_{2\cQ} +3F_{D+1}^{[\cM\cN} F_{2}^{\cP]} -F_{D}^{\cM\cN\cP} F_3\ ,\nn\w1
dF_{D+2}^{\cM\cN\cP\cQ\cR}&=5 F_{D+1}^{[\cM\cN\cP\cQ} F_2^{\cR]}\ .
\la{4.10b}
\end{align}

In the following sections we give the Bianchi identities for all of the additional forms in $D>4$ as well as those for $D=4$. The representations (although not the Bianchi identities) up to level $D$ were given in \cite{Bergshoeff:2007vb} (for $n=0$). Here we include the first OTT level $(D+1)$ as well. In the lists of representations the Young tableaux are taken to be irreducible, \ie traces removed, whereas in the Bianchi identities the tableaux occurring as subscripts denote symmetry type and include traces. We distinguish such tableaux by hats.

\subsection*{$D=8,9$}

In these two dimensions there are additional OTT forms at level $(D+1)$. In $D=9$ we have
\be
dF_{11}=F_6 F_6\ ,
\la{4.10b1}
\ee
while in $D=8$ we have
\be
dF_{10}^\cM=F_5 F_6^\cM\ .
\ee

\subsection*{$D=7$}

The full set of extra forms in $D=7$ is given by a singlet at level $(D-1)=6$, a vector at level $D=7$ and a singlet and two-form at level $(D+1)=8$. The Bianchi identities for these forms are
\begin{align} 
dF_7&= F_4 F_4 \nn\w1
dF_8^\cM&= F_7 F_2^\cM-F_5^\cM F_4\nn\w1
dF_9&=F_8\cdot F_2-F_7 F_3 + F_6 F_4\nn\w1
dF_9^{\cM\cN}&=2 F_8^{[\cM} F_2^{\cN]} + F_5^\cM F_5^\cN\ .
\la{4.10c}
\end{align}
%
\subsection*{$D=6a$}

In $D=6a$, $F_{D-3}$ is another three-form which we denote $F'_3$, with $dF'_3=0$. The additional forms consist of a vector at level $(D-1)=5$, forms in the representations $1+1+\Yboxdim{5pt}\yng(1,1)+\
\yng(2)$ at level $D=6$, and in the representations $4.\Yboxdim{5pt}\yng(1)+2.\tiny\yng(1,1,1)+\tiny\yng(2,1)$ at level $(D+1)=7$, the tableaux being taken as irreducible, \ie traces removed. The Bianchi identities are:
\begin{align}
d {F'}_6^\cM&=F'_3 F_4^\cM\nn\w1
d {F'}_7&=F'_6\cdot F_2 + F_5 F'_3\nn\w1
d{F'}_7^{\cM\cN}&=2{F'}_6^{[\cM} F_2^{\cN]} + F_5^{\cM\cN} {F'}_3\nn\w1
d\tilde F_7^{\cM\cN}&=2{F'}_6^{(\cM} F_2^{\cN)}-F_4^\cM F_4^\cN\nn\w1
d{F'}_8^{\cM\cN\cP}&=a( {F'}_7^{[\cM\cN} F_2^{\cP]}-F_5^{[\cM\cN} F_4^{\cP]}) + b(F_6^{\cM\cN\cP} {F'}_3+3F_5^{[\cM\cN} F_4^{\cP]})\nn\w1
dF_8^{\cM\cN,\cP}&= ({F'}_7^{\cM\cN} F_2^{\cP}-2\tilde F_7^{\cP[\cM} F_2^{\cN]}-F_5^{\cM\cN}F_4^\cP)_{\widehat{\Yboxdim{4pt}\yng(2,1)}} 
\nn\w1
d{F'}_8^{\cM}&=F'^\cM_6 F'_3 \nn\w1
d {F''}_8^\cM&= a\left({F'}_7 F-2^\cM + {F'}_7^\cM\cN F_{2\cN} + F_6^\cM {F'}_3 - {F'}_6^\cM F_3\right)\nn\w1
&+b\left(\tilde F_7^{\cM\cN} F_{2\cN} +{F'}_7^{\cM\cN} F_{2\cN} + F_6^\cM {F'}_3 -2 {F'}_6^\cM F_3 + F_5 F_4^\cM\right)\nn\w1
& + c\left({F'}_7F_2^\cM + 2{F''}_7^{\cM\cN} F_{2\cN} + F_6^\cM F'_3-2{F'}_6^\cM F_3+ F_5 F_4^\cM -F_5^{\cM\cN} F_{4\cN}\right)\ .
\la{4.10d}
\end{align}
Here, the (double) primes are used to denote new forms in representations that are already present in the standard set, and the tilde to distinguish a symmetric two-index representation (not traceless). The mixed symmetry three-index $F$ is not traceless, and there are three solutions to the last Bianchi identity, corresponding to the parameters $a,b,c$, one of which is the trace of the mixed symmetry one. There are thus four new vectors at level seven as well as the vector in the universal set. 

\subsection*{$D=5$}

In five dimensions $F_{D-3}=F_2$, so that all the forms are generated by the level-one set in this case. The new forms in $D=5$ are: at level $(D-1)=4$, $\Yboxdim{5pt}\yng(1,1)$; at level $D=5$, $2.\Yboxdim{5pt}\yng(1) + \yng(2,1)+ \yng(1,1,1)$\,; and at level $(D+1)=6$, $\Yboxdim{5pt}\yng(2,2)+2.\yng(1,1,1,1)+\yng(2,1,1) +\yng(2,1) +2.\yng(2)+ 3.\,\tiny\yng(1,1)+\tiny\yng(1) + 3.1$.
The Bianchi identities for the new forms are:

\begin{align}
dF_5^{\cM\cN}&=F_4^{\cM\cN} F_2-F_3^\cM F_3^\cN\nn\w1
dF_6^{\cM\cN\cP}&=2 F_5^{[\cM\cN}F_2^{\cP]}- F_5^{\cM\cN\cP} + F_4^{[\cM\cN} F_3^{\cP]}\nn\w1
dF_6^{\cM\cN,\cP}&=\left(F_5^{\cM\cN} F_2^{\cP} - F_4^{\cM\cN} F_3^\cP\right)_{\widehat{\Yboxdim{4pt}\yng(2,1)}}\nn\w1
dF_6^\cM&=F_5^{\cM\cN} F_{2\cN} -2F_5^\cM F_2+ F_4^{\cM\cN} F_{3\cN} + 2 F_4 F_3^\cM\nn\w1
dF_7^{\cM\cN\cP,\cQ}&=\left(F_6^{\cM\cN\cP} F_2^\cQ-3F_6^{[\cM\cN,|\cQ|}F_2^{\cP]}+F_5^{\cM\cN\cP} F_3^\cQ\right)_{\widehat{\Yboxdim{4pt}\yng(2,1,1)}}\nn\w1
dF_7^{\cM\cN\cP\cQ}&=a\left(F_6^{\cM\cN\cP\cQ} F_2-4 F_5^{[\cM\cN\cP} F_3^{\cQ]}+3 F_4^{[\cM\cN} F_4^{\cP\cQ]}\right)\nn\w1
&+b\left(F_6^{[\cM\cN\cP} F_2^{\cQ]} +F_5^{[\cM\cN\cP} F_3^{\cQ]} -F_4^{[\cM\cN} F_4^{\cP\cQ]}\right)\nn\w1
dF_7^{\cM\cN,\cP\cQ}&=\left(2 F_6^{\cM\cN,[\cP} F_2^{\cQ]}+ 2F_6^{\cP\cQ,[\cM} F_2^{\cN]}+F_4^{\cM\cN} F_4^{\cP\cQ}\right)_{\widehat{\Yboxdim{4pt}\yng(2,2)}}\nn\w1
dF_7^{\cM\cN,\cP}&=\left(F_6^{\cM\cN,\cP} F_2-F_5^{\cM\cN}F_3^\cP\right)_{\widehat{\Yboxdim{4pt}\yng(2,1)}}\nn\w1
d\tilde F_7^{\cM\cN}&=a\left(F_6^{(\cM} F_2^{\cN)} + 2 F_5^{(\cM} F_3^{\cN)}+ F_4^{\cM\cP} F^\cN_{4\  \cP} +  F_2^{\cN)}\right)\nn\w1
&+b\left(F^{\cP(\cM}_{6\ \ \ \ ,\cP} F_6^{\cP(\cM,\cN)} F_{2\cP}-\half F_4^{\cM\cP} F^\cN_{4\  \cP} -F^{\cP(\cM}_{6\ \ \ \ ,\cP} F_2^{\cN)}\right)\nn\w1
dF_7^{\cM\cN}&=a\left(4F_6^{[\cM} F_2^{\cN]}+ 3F_6^{\cM\cN\cP} F_{2\cP} +5F_6^{\cM\cN} F_2-\right. \nn\w1 &\qquad\ \  \left.  2F_5^{[\cM} F_3^{\cN]} -2F_5^{\cM\cN\cP} 
F_{3\cP}-2 F_5^{\cM\cN} F_3 -
3F_4^{\cM\cN} F_4\right)\nn\w1
&+b\left(4F^{[\cM|\cP|}_{6\ \ \ \ \ ,\cP}  F_2^{\cN]} + 3F_6^{\cM\cN\cP} F_{2\cP} +F_6^{\cM\cN} F_2 - \right.\nn\w1
&\qquad\ \ \left. 2 F_5^{[\cM} F_3^{\cN]} + 2F_5^{\cM\cN\cP} 
F_{3\cP}-2F_5^{\cM\cN} F_3 + F_4^{\cM\cN} F_4\right)\nn\w1
&+c\left(2F_6^{\cM\cN,\cP} F_{2\cP} + F_6^{\cM\cN\cP} F_{2\cP} + F_6^{\cM\cN} F_2-2 F_5^{[\cM} F_3^{\cN]}-2 F_5^{\cM\cN} F_3 + F_4^{\cM\cN} F_4\right)\nn\w1
dF_7&=a\left(F_6 F_2-F_5^\cM F_{3\cM}-\frac{1}{4}F_4\cdot F_4-\half F_4 F_4\right)\nn\\
&+b\left(F_6\cdot F_2+ 2F_5\cdot F_3 +\frac{3}{4} F_4\cdot F_4\right)\nn\\
&+c\left(F^{\cM\cP}_{6\ \ \ ,\cP} F_{2\cM} -\frac{1}{4} F_4\cdot F_4\right)\ .
\end{align}
Only one of the three singlets in the last Bianchi identity is independent, the other two being related to the trace of $\tilde F_7^{\cM\cN}$ and the double-trace of $F_7^{\cM\cN,\cP\cQ}$. The single seven-form in the vector representation of the duality group is the trace of $F_7^{\cM\cN,\cP}$.

\subsection*{$D=4$}

In $D=4$ the supergravity two-form can be dualised to a second scalar which goes together with the dilaton in the coset $U(1)\bsh SL(2,\bbR)$. The duality group is therefore $SL(2,\bbR) \xz SO(k,n+k)$, and the forms carry indices for both groups. As in $D=5$ all of the forms are generated by the level-one set which consists of two-form field-strengths $F_2^{\cM r}$, where $r=1,2$ is an $SL(2,\bbR)$ doublet index. At level three the forms fall into the  $(SL(2,\bbR),SO(k,n+k))$  representations 	$(2,\Yboxdim{5pt}\yng(1))+(2,\Yboxdim{5pt}\yng(1,1,1))$. At level four the representations are $({\bf 3},\Yboxdim{5pt}\yng(1,1,1,1))+({\bf 3},\Yboxdim{5pt}\yng(2,1))+({\bf 3},{\bf 1})+({\bf 1},\Yboxdim{5pt}\yng(2,1,1))
+2.({\bf 1},\Yboxdim{5pt}\yng(1,1))$. Up to level four the Bianchi identities are:
\begin{align}
dF_2^{\cM r}&= 0\phantom{4F_4^{[\cM\cN\cP(r} F_2^{\cQ] s)}}  \nn\w1
dF_3^{\cM\cN}&= F^\cM_{2\ \  r} F_2^{\cN r}\nn\w1
dF_3^{rs}&=F_2^{\cM r} F_{2\cM}{}^s\nn\w1
dF_4^{\cM r}&= F_3^{\cM\cN} F_{2\cN}{}^r -F_3^{rs} F^\cM_{2\ \ s}\nn\w1
dF_4^{\cM\cN\cP r}&= 3F_3^{[\cM\cN}F_2^{\cP] r}\nn\w1
dF_5^{\cM\cN\cP\cQ,rs}&= 4F_4^{[\cM\cN\cP(r} F_2^{\cQ] s)} \nn\w1
dF_5^{\cM\cN\cP,\cQ}&=(F_4^{\cM\cN\cP r} F^\cQ_{2\ \ r}+ 3F_3^{[\cM\cN} F_3^{\cP]\cQ})_{\widehat{\Yboxdim{4pt}\yng(2,1,1)}}\nn\w1
dF_5^{\cM\cN,rs}&=2 F_4^{[\cM(r} F_2^{\cN]s)}+ F_4^{\cM\cN\cP(r} F_{2\cP}{}^{s)}-F_3^{\cM\cN} F_3^{rs}\nn\w1
dF_5^{\cM\cN}&=2 F_4^{[\cM r} F^{\cN]}_{2\ \ r}-F_3^{\cM\cP} F^\cN_{3\ \ \cP}\nn\w1
dF_5^{rs}&= 2F_4^{\cM(r} F_{2\cM}{}^{s)}+ F_3^{rt} F^s_{3\ t}\ .
\end{align}
The second five-form in the $({\bf 1},\Yboxdim{5pt}\yng(1,1))$ representation is the trace of the mixed-symmetry five-form.
At level five the  forms in the $4$ of $SL(2,\bbR)$ fall into the representations $\Yboxdim{5pt}\yng(1)+\tiny\yng(1,1,1)+\tiny\yng(1,1,1,1,1)$ of $SO(k,n+k)$ and have the following consistent Bianchi identities:
\begin{align}
dF_6^{\cM rst}&=2F_5^{\cM\cN(rs} F_{2\cN}{}^{t)}+ F_5^{(rs} F_2^{\cM t)}- F_4^{\cM(r} F_3^{st)}\nn\w1
dF_6^{\cM\cN\cP}&=F_5^{\cM\cN\cP\cQ,(rs} F_{2\cQ}{}^{t)}+ 3F_5^{[\cM\cN(rs}F_2^{\cP] t)}-F_4^{\cM\cN\cP(r} F_3^{st)}\nn\w1
dF_6^{\cM\cN\cP\cQ\cR rst}&=F_5^{[\cM\cN\cP\cQ,(rs} F_2^{\cR] t)}\ .
\end{align}
There are also forms in the ${\bf 2}$ of $SL(2,\bbR)$. Their $SO(k,n+k)$ representations are $3.\Yboxdim{5pt}\yng(1)+3.\tiny\yng(1,1,1)+\tiny\yng(1,1,1,1,1)$ and $2.\Yboxdim{5pt}\yng(2,1)+ \tiny\yng(2,1,1,1)+ \tiny\yng(2,2,1)$. The Bianchi identities for the first set are:

\begin{align}
dF_6^{\cM r}&=a\left(\! F_5^{\cM\cN,rs} F_{2\cN s}+\half F_5^{rs} F^\cM_{2\ \ s} -\frac{3}{2}F_4^{\cN r} F^\cM_{3\ \ \cN}- F_4^{\cM s} F^r_{3\  s}-\frac{3}{4} F_4^{\cM\cN\cP r} F_{3\cN\cP}\right)\nn\w1
\! &+b\left(F_5^{\cM\cN} F_{2\cN}{}^r +F_5^{rs} F^\cM_{2\ \ s}-F_4^{\cN r} F^\cM_{3\ \ \cN}-\half F_4^{\cM s} F^r_{3\ s}\right)\nn\w1
&+c\left(F^{\cM\cN\cP}_{5\ \ \ \ \ ,\cP} F_{2\cN}{}^r+\half F_4^{\cM\cN\cP r} F_{3\cN\cP}\right)\nn\w1
dF_6^{\cM\cN\cP r}&=a\!\left(\! F_5^{[\cM\cN}\! F_2^{\cP]r}\!+\! \frac{2}{3}F_5^{\cM\cN\cP\cQ rs} F_{2\cQ s}
 +F^{\cQ[\cM\cN}_{5\qquad,\cQ} F_2^{\cP] r} -\right. \nn\w1
&\ \ \left. \qquad\frac{1}{3}F_4^{\cM\cN\cP s} F^r_{3 s}\!-\! F_4^{[\cM r} F_3^{\cN\cP]}-\! F_4^{\cM\cN|\cQ| r}F^{\cP]}_{3\ \ \cQ}\! \right)\nn\w1
&+b\! \left(\! F_5^{[\cM\cN rs} F^{\cP]}_{2\ s}\! \! -\! \! \frac{5}{3} F_5^{\cM\cN\cP\cQ rs} F_{2\cQ s}\! -\! \! 3 F^{\cQ[\cM\cN}_{5\qquad,\cQ} F_2^{\cP] r} + \right. \nn\w1
&\qquad \left. \ \ \frac{5}{6}F_4^{\cM\cN\cP s} F^r_{3 s}\! \! +\! \! \frac{3}{2}F_4^{[\cM r} F_3^{\cN\cP]}\! \!  +\! \! \frac{3}{2}F_4^{\cM\cN|\cQ| r}F^{\cP]}_{3\ \ \cQ}\right)\nn\w1
&+c\left(-\frac{3}{2}F_5^{\cM\cN\cP\cQ rs} F_{2\cQ s}-3 F^{\cQ[\cM\cN}_{5\qquad,\cQ} F_2^{\cP] r}+\frac{3}{2}F_4^{[\cM\cN|\cQ| r}F^{\cP]}_{3\ \ \cQ}\right)\nn\w1
dF_6^{\cM\cN\cP\cQ\cR r}&= F_5^{[\cM\cN\cP\cQ rs} F^{\cR]}_{2\ s}+ 3F_4^{[\cM\cN\cP r} F_3^{\cQ\cR]}\ ,
\end{align}
while those for the second set are:
\begin{align}
dF_6^{\cM\cN,\cP r}&=a \left(F_5^{\cM\cN\cQ,\cP}F^r_{2\cQ}-F^{\cM\cN\cQ}_{5\qquad ,\cQ} F_2^{\cP r}+ F_4^{\cM\cN\cQ r} 
F_{3\cQ}{}^\cP\right)_{\widehat{\Yboxdim{5pt}\yng(2,1)}}\qquad \qquad\qquad \qquad\qquad\  \nn\w1
&+b\left(F_5^{\cM\cN rs} F^\cP_{2\ s} -\half F^{\cM\cN\cQ}_{5\qquad ,\cQ} F_2^{\cP r}-\half F_5^{\cM\cN}F_2^{\cP r}-F_4^{\cP r} F_3^{\cM\cN}+\right.\nn\w1
&\qquad \left. F_4^{\cM\cN\cQ r} F_{3\cQ}{}^\cP\right)_{\widehat{\Yboxdim{5pt}\yng(2,1)}}\nn\w1
dF_6^{\cM\cN\cP\cQ,\cR r}&=\left(3F_5^{\cM\cN\cP\cQ rs} F^\cR_{2\ s}-8F_5^{[\cM\cN\cP,|\cR|} F_2^{\cQ] r}+ 4F_4^{[\cM\cN\cP r} F_3^{\cQ]\cR}\right)_{\widehat{\Yboxdim{5pt}\yng(2,1,1,1)}} \nn\w1
dF_6^{\cM\cN\cP,\cR\cS r}&=\left(2 F_5^{\cM\cN\cP,[\cR} F_2^{\cS] r}-F_4^{\cM\cN\cP r} F_3^{\cR\cS}\right)_{\widehat{\Yboxdim{5pt}\yng(2,2,1)}}\ .
\end{align}
Although the traces in these representations are non-zero no additional ones to those listed in the text are present.

\subsection*{$D=3$}

Finally, in $D=3$, the vectors can be dualised to scalars. This implies that the level-one forms should be in the adjoint representation of the duality group rather than the vector representation, \ie we have $F_2^{\cM\cN}$ in the $\Yboxdim{5pt}\yng(1,1)$ representation;  the entire set of forms is generated from these. Up to level three the consistent Bianchi identities are:
\begin{align}
dF_2^{\cM\cN}&= 0\nn\w1
dF_3^{\cM\cN\cP\cQ}&= 3F_2^{[\cM\cN} F_2^{\cP\cQ]}\nn\w1
d\tilde F_3^{\cM\cN}&= F_2^{\cM \cP} F^\cN_{2\ \ \cP}\nn\w1
d F_4^{\cM\cN\cP\cQ\cR,\cS}&= (5F_3^{[\cM\cN\cP\cQ} F_2^{\cR]\cS})_{\widehat{\Yboxdim{4pt}\yng(2,1,1,1,1)}}\nn\w1
dF_4^{\cM\cN\cP,\cQ}&=\left(F^{\cM\cN\cP}_{3\qquad \cR} F_2^{\cQ\cR}-3\tilde F_3^{\cQ[\cM} F_2^{\cN\cP}\right)_{\widehat{\Yboxdim{4pt}
\yng(2,1,1)}}\nn  \w1
d\tilde F_4^{\cM\cN}&=2\tilde F^{(\cM}_{3\ \ \cP} F_2^{\cN)\cP}\ .
\end{align}

The tableaux subscripts here represent symmetry type and include the trace representations $\Yboxdim{5pt}
\yng(1,1,1,1)$ and $\Yboxdim{4pt}
\yng(1,1)$ respectively. The last form is symmetric traceless. The representations for level four are given in \cite{Greitz:2012vp}.

\section{Extended superspace}

In this appendix we reformulate maximal supergravity theories for $3\leq D\leq9$ in extended superspaces, that is, superspaces with additional even co-ordinates that correspond to ``central'' charges in the supersymmetry algebras. The number of these charges is equal to the dimension of the $\cR_1$ representation.\footnote{$D=4,N=8$ supergravity was formulated this way in \cite{Howe:1980th,Howe:1981gz}, see also \cite{Siegel:1980bp}.} The additional co-ordinates will be denoted $y^{\cM}$, and we shall assume that the structure group for the extended superspace is still the product of the relevant spin group and R-symmetry group $H$. The basis forms will be denoted $E^{\unA}:=(E^A, E^{\cA})$, where $\cA$ denotes the representation $\cR_1$ considered as a representation of $H$. In this space we have torsion and curvature but no additional forms, at least for the moment. We make the following assumptions:
\begin{enumerate}
\item $T^A$ and $R_A{}^B$ are unchanged from the standard superspace.
\item The non-zero components of $T^{\cA}$ are $T_{AB}{}^\cC$ and $T_{A\cB}{}^\cC$.
\item All fields are annihilated by $\nab_{\cA}$.
\end{enumerate}
In addition the curvature $R_{\cA}{}^{\cB}$ is the same as the R-symmetry part of $R_A{}^B$, \ie the R-symmetry curvature in the representation $\cR_1$. The key new Bianchi identity is 
\be
DT^{\cA}= E^{\cB} R_{\cB}{}^\cA\ .
\la{.1}
\ee
The $(ABC)$ component of this is
\be
\sum_{(ABC)}\nab_A T_{BC}{}^\cD +T_{AB}{}^E T_{EC}{}^\cD + T_{AB}{}{}^\cE T_{\cE C}{}^\cD=0\ ,
\la{.2}
\ee
where the sum is graded cyclic. Defining $F^\cA=\half E^C E^B T_{BC}{}^\cA$ and the one-forms $P_\cA{}^{\cB}= E^C T_{C\cA}{}^\cB$, we see that this can be rewritten as
\be
D F^\cA = F^\cB P_\cB{}^{\cA}
\label{.3}
\ee
in ordinary superspace. The $(AB\cC)$ component of \eq{.1} 
is
\be
2\nab_{[A} T_{B]\cC}{}^\cD + T_{AB}{}^E T_{E\cC}{}^\cD - 2T_{[A|\cC|}{}^\cE T_{ B]\cE}{}^\cD=R_{AB,\cC}{}^\cD\ .
\label{.4}
\ee
To interpret these equations we recall that the scalars in conventional superspace are given by a matrix $\cV$ that transforms under global $G$ and local $H$ transformations by $\cV\rightarrow h^{-1} \cV g$. The Maurer-Cartan form $d\cV {\cV}^{-1}$ splits into an $\gh$-valued component $Q$, which we identify with the internal part of the connection, and a quotient $\gh\bsh \gg$-valued component $P$. The Maurer-Cartan equation, resolved into isotropy and quotient algebra components, reads
\be
R=-P^2\ ,\qquad DP=0\ ,
\label{.5}
\ee
where $R$ is the $\gh$ curvature. Let us take $\cV$ to be an element of $G$ in the $\cR_1$ representation, so in indices we write $\cV_\cA{}^\cM$.  We then have $D P_\cA{}^\cB=0$ and
\be
R_\cA{}^\cB= -P_\cA{}^\cC P_\cC{}^\cB\ .
\label{.6}
\ee
The two-form field-strength $F^\cM$ obeys a trivial Bianchi identity, but if we define $F^\cA:=F^\cM ({\cV}^{-1})_\cM{}^\cA$ we recover \eq{.3} while \eq{.4} can be rewritten as
\be
R_\cA{}^\cB+DP_\cA{}^\cB +P_\cA{}^\cC P_\cC{}^\cB=0\ .
\la{.7}
\ee
This is equivalent to the two equations in \eq{.5} because $R$ is $\gh$-valued and $P$ takes its values in the quotient. If we let $\unM$ denote all of the coordinate indices in the extended superspace we can see that the ``sehrsupervielbein'' $E_{\unM}{}^{\unA}$ has the form
\be
E_{\unM}{}^{\unA}=\left(\barr{cc} E_M{}^A & A_M{}^\cM E_\cM{}^\cA\\
0 & E_\cM{}^\cA\earr\right)\ ,
\la{.8}
\ee
where we identify $E_\cM{}^\cA$ with $({\cV}^{-1})_\cM{}^\cA$, and where $A_M{}^\cM$ is the level-one potential.

\section{Borcherds and contragredient Lie superalgebras} \label{borcherdsapp}

In this appendix we give the general definitions of Borcherds and Kac-Moody
(super)algebras
as special cases of contragredient Lie (super)algebras (the definitions in the literature vary slightly). 
We explain how general Borcherds and contragredient Lie superalgebras
are constructed from their Cartan matrices,
and introduce the V-diagrams, which in turn completely specify the Cartan matrices that we consider in this paper.

For any real $(r \times r)$ matrix $B_{IJ}$ ($I,J=1,2,\ldots,r$) and any subset $S$ of the set
$R=\{1,2,\ldots,r\}$ one can construct a
Lie superalgebra $\tilde{\cB}$ generated by elements 
$e_I$, $f_I$ and $h_I$
modulo the Chevalley relations
\begin{align} \label{chev-rel-1}
[h_I,e_J]&=B_{IJ}e_J\ , & [h_I,f_J]&=-B_{IJ}f_J\ , & [e_I,f_J]&=\delta_{IJ}h_J\ ,
\end{align}
where the Chevalley generators $e_I$ and $f_I$ are both odd if $I \in S$, and both even otherwise.
It follows that the Cartan elements $h_I=[e_I,f_I]$ are all even and commute with each other, $[h_I,h_J]=0$, spanning an abelian {\it Cartan subalgebra}
$\cH \subset \tilde{\cB}$.
The {\it contragredient Lie superalgebra} $\cB$ of {\it rank} $r$ 
associated to the {\it Cartan matrix} 
$B_{IJ}$ is then obtained by factoring out
from $\tilde{\cB}$ the maximal ideal $\cI$ that intersects $\cH$ trivially, $\cB=\tilde{\cB}/\cI$ \cite{Kac77A,Kac77B}.
The contragredient Lie superalgebra $\cB$ is a {\it Borcherds superalgebra} 
if the Cartan matrix $B_{IJ}$
is non-degenerate and symmetric
with non-positive off-diagonal entries
such that
$2B_{IJ}/B_{II}\in\mathbb{Z}$ if $\cB_{II}>0$, and furthermore $2B_{IJ}/B_{II}\in2\mathbb{Z}$ if in addition
$I\in S$ \cite{Ray95,Ray,Wakimoto}.\footnote{Borcherds superalgebras are also known as Borcherds-Kac-Moody (BKM) superalgebras or
generalised Kac-Moody (GKM) superalgebras.}
A Borcherds superalgebra $\cB$ is a {\it Kac-Moody superalgebra} if $B_{II}>0$ for all $I \in R$.

If $S$ is empty the Lie superalgebras defined here reduce to their Lie algebra analogues:
contragredient Lie algebras, Borcherds algebras and Kac-Moody algebras \cite{Kac68B,Borcherds,Kac}. 
From the Cartan matrix $A_{IJ}$ of a Kac-Moody algebra $\cA$ we can obtain another
matrix with all diagonal entries equal to 2, by multiplying
the row $I$ in $A_{IJ}$ by $2/A_{II}$. The resulting matrix defines a contragredient Lie algebra isomorphic to $\cA$ but unlike $A_{IJ}$ it is not necessarily symmetric.
Usually this matrix is referred to as the Cartan matrix of a Kac-Moody algebra $\cA$, and the matrix $A_{IJ}$ that we here 
call the Cartan matrix of $\cA$ is then
called the {\it symmetrized} Cartan matrix.
Accordingly, the Borcherds or Kac-Moody (super)algebras that we consider are assumed to be
{\it symmetrizable}.

The Kac-Moody algebras $\cA$ that we consider furthermore have Cartan matrices such that
\begin{align}
\text{min}\,\bigg\{\,\Big|\,2\frac{A_{IJ}}{A_{II}}\,\Big|\,,\,\Big|\,2\frac{A_{JI}}{A_{JJ}}\,\Big|\,\bigg\}
\end{align}
is equal to either zero or one.
Any such Cartan matrix can be described (up to isomorphisms of $\cA$)
by a Dynkin diagram consisting of $r$ nodes (where $r$ is the rank of $\cA$) and 
\begin{align}
\text{max}\,\bigg\{\,\Big|\,2\frac{A_{IJ}}{A_{II}}\,\Big|\,,\,\Big|\,2\frac{A_{JI}}{A_{JJ}}\,\Big|\,\bigg\}
\end{align}
lines connecting node $I$ and node $J$, with an arrow pointing at node $I$ if $A_{II} < A_{JJ}$.
In addition to these conventional rules, we also distinguish between the two cases $A_{II}=1$ and $A_{II}=2$,
the only cases that appear for Cartan matrices of Kac-Moody algebras in this paper, by painting node $I$ black if $A_{II}=1$,
and keep it white if $A_{II}=2$.\footnote{This should not be confused with the use of black nodes in for example
\cite{HenryLabordere:2002dk,Bergshoeff:2007vb,Henneaux:2010ys,Palmkvist:2011vz,Kleinschmidt:2013em}.}
Up to isomorphisms, this painting does not give any more information about the Kac-Moody algebra $\cA$ than what is already given by
the number of lines between the nodes and the direction of the arrows.
However, it fixes the overall normalization of each indecomposable block of $A_{IJ}$, which is important when we label the nodes by V-degrees and consider the
Dynkin diagram as a V-diagram of a Borcherds superalgebra, as will be explained below.

If $\cB$ is a Borcherds superalgebra, then the ideal $\cI$ of $\tilde{\cB}$ is generated by the Serre relations, which are
\begin{align} \label{serre-rel1}
(\ad\, e_I)^{1-{2B_{IJ}}/{B_{II}}} (e_J) = (\ad\, f_I)^{1-{2B_{IJ}}/{B_{II}}} (f_J) &=0
\end{align}
for $B_{II}>0$ and $I \neq J$, and
\begin{align} \label{serre-rel2}
[e_I,\,e_J] = [f_I,\,f_J] = 0
\end{align}
for $B_{IJ}=0$ (including the case $I=J$). The condition that the integers $2B_{IJ}/B_{II}$ be
even for $B_{II}>0$ and $I \in S$ is needed for (\ref{serre-rel1}) to generate an ideal that intersects the Cartan subalgebra $\cH$ trivially in that case.

A nonzero element $\beta$ in the dual space $\cH^\ast$ of the Cartan subalgebra $\cH$ of a contragredient Lie superalgebra
$\cB$ is a {\it root} if there is a nonzero element
$e_\beta$ in $\cB$, such that $[h,\,e_\beta]=\beta(h) e_\beta$ for all $h \in \cH$. This element $e_\beta$ is then the corresponding {\it root vector}.
The Cartan matrix $B_{IJ}$ defines a basis of $\cH^\ast$, consisting of {\it simple roots} $\beta_I$, by
$\beta_I(h_J)=B_{IJ}$.
Thus in particular the Chevalley generators $e_I$ and $f_I$ are root vectors corresponding to the simple root $\beta_I$ and its negative $-\beta_I$,
respectively. Accordingly the simple roots can be divided into even and odd ones.
If the Cartan matrix $B_{IJ}$ is symmetric it also defines an inner product in $\cH^\ast$, given by $(\beta_I,\beta_J)=B_{IJ}$,
so that the diagonal value $B_{II}$ is the length squared of the simple root $\beta_I$.

Any contragredient Lie superalgebra $\cB$ that we consider in this paper has a symmetric Cartan matrix $B_{IJ}$ and is equipped with a map
$v:R\to \mathbb{Z}$ such that the subset $S$ and its complement in $R$ are mapped to odd and even integers, respectively.
The map $v$ induces a {\it consistent $\mathbb{Z}$-grading} of $\cB$ given by $e_I \in \cB_{v(I)}$ and $f_I \in \cB_{-v(I)}$.
This is a decomposition of $\mathcal{B}$ into a direct sum of subspaces
$\mathcal{B}_k$ for all integers $k$ such that $[\mathcal{B}_i,\mathcal{B}_j]\subseteq\mathcal{B}_k$, and $\mathcal{B}_k$ consists of
even or odd elements if $k$ is an even or odd integer, respectively.
The map $v$ also induces a linear map $v:\cH^\ast \to \mathbb{Z}$ given by $v(\beta_I)=v(I)$.
Following \cite{Henneaux:2010ys,Kleinschmidt:2013em}
we call the integer $v(I)$ the {\it V-degree} of the simple root
$\beta_I$ and we denote it simply by $v_I$. Furthermore,
the Cartan matrix $B_{IJ}$ of the contragredient Lie superalgebra $\cB$ can in all cases that we consider be obtained from 
the Cartan matrix $A_{IJ}$ of
a corresponding Kac-Moody algebra $\cA$ of the same rank $r$,
equipped with the same map $v:R \to \mathbb{Z}$.
The Cartan matrices of $\cB$ and $\cA$ are then related by
\begin{align} \label{fromatob}
B_{IJ}=A_{IJ}-w(v_I,v_J)\ ,
\end{align}
where $w$ is a symmetric map $w:\mathbb{Z} \times \mathbb{Z} \to \mathbb{Z}$ defined by $w(a,b)=a(b+1)$ for
$0 \leq a \leq b$ and $w(-a,b)=w(a,-b)=-w(a,b)$.
For example, in the simply-laced case, simple roots
of V-degree $\pm 1$ are odd null roots of $\cB$, whereas simple roots of V-degree
zero remain even and of length squared equal to two when we go from $\cA$ to $\cB$. If only one of the simple roots has a nonzero V-degree,
then the off-diagonal entries of the Cartan matrix remain unchanged.
If the nonzero V-degrees are all positive (or all negative) then the off-diagonal entries of the Cartan matrix remain non-positive, and
$\cB$ is a Borcherds superalgebra.

According to (\ref{fromatob}) the Borcherds superalgebra $\cB$ can be completely specified by
drawing the Dynkin diagram of $\cA$ and labelling the nodes by the V-degrees of the corresponding simple roots.
We call the result a {\it V-diagram} of $\cB$.
It describes $\cB$ very efficiently, but one must bear in mind that two disconnected nodes in the V-diagram 
actually correspond to a nonzero off-diagonal entry in the Cartan matrix of $\cB$ if both V-degrees are nonzero.

A fundamental Weyl reflection with respect to a simple root $\beta_J$ of a contragredient Lie algebra $\cB$ with nonzero length,
$(\beta_J,\beta_J)\neq0$,
is a linear transformation of $\cH^\ast$ given by
\begin{align}
\beta_I \mapsto 2\frac{(\beta_I,\beta_J)}{(\beta_J,\beta_J)}\beta_J = 2\frac{B_{IJ}}{B_{JJ}}\beta_J
\end{align}
for any simple root $\beta_I$.
It is easy to see that a fundamental Weyl reflection is indeed a reflection and thus preserves the inner product in $\cH^\ast$ and leaves the Cartan matrix of $\cB$
invariant.
If $(\beta_J,\beta_J)\neq0$ there is no fundamental Weyl reflection with respect to $\beta_J$, but if in addition
$\beta_J$ is an odd root, 
then there is a
{\it generalised Weyl transformation} \cite{Frappat, Leites} given by 
\begin{align} \label{cases}
\beta_I &\mapsto
\begin{cases}
-\beta_I &\text{if $I=J$}\ ,\\
\beta_I+\beta_J &\text{if  $I\neq J$ and $B_{IJ}\neq0$}\ ,\\
\beta_I &\text{if $I\neq J$ and $B_{IJ}=0$}\ .
\end{cases}
\end{align}
for the simple roots, and extended to the whole of $\cH^\ast$ by linearity. The generalised Weyl transformation does not preserve the inner product in
$\cH^\ast$, but maps the basis of simple roots to another one, corresponding to a different Cartan matrix.
In particular, if we start with a Cartan matrix of a Borcherds superalgebra, and the second case in (\ref{cases}) appears,
then the new Cartan matrix will have positive off-diagonal entries and thus no longer satisfy the conditions for a Cartan matrix
of a Borcherds superalgebra (but still those for a Cartan matrix
of a contragredient Lie superalgebra).
As before we can describe the different Cartan matrices by V-diagrams.
Any node labelled by V-degree $v_J=1$ or $v_J=-1$ corresponds to an odd null root
$\beta_J$, and thus
to a generalised Weyl transformation. Since we consider $v$ as linear map from $\cH^\ast$ to the integers, the generalised Weyl transformation also changes $v_J=v(\beta_J)$ to $v(-\beta_J)=-v_J$. This is how the different V-diagrams of the Borcherds superalgebra $\cD$ in Table
\ref{maxtable} are obtained from each other.


\raggedright

\end{document}